\documentclass[10pt,journal,compsoc]{IEEEtran}
\usepackage{multirow}
\usepackage{tikz}
\usepackage{tikz-qtree}
\usepackage{pgfplots}
\usetikzlibrary{patterns}
\usepackage{subfigure}
\usepackage{url}
\usetikzlibrary{matrix}
\usepgfplotslibrary{groupplots}
\pgfplotsset{compat=newest}
\usepackage{tablefootnote}
\usepackage{comment}
\usepackage{makecell}
\usepackage{enumitem}
\usepackage{booktabs}
\usepackage{amsmath,amssymb}

\ifCLASSOPTIONcompsoc
  \usepackage[nocompress]{cite}
\else
  \usepackage{cite}
\fi
\ifCLASSINFOpdf
\else
\fi
\begin{document}
%
\title{DaisyRec 2.0: Benchmarking Recommendation \\for Rigorous Evaluation}
%
%
%
%

\author{
        Zhu Sun,
        Hui Fang, 
        Jie Yang,
        Xinghua Qu,
        Hongyang Liu, 
        Di Yu,\\
        Yew-Soon Ong,~\IEEEmembership{Fellow,~IEEE,}
        and~Jie Zhang
\IEEEcompsocitemizethanks{
\IEEEcompsocthanksitem Zhu Sun is with Institute of High Performance Computing and Centre for Frontier AI Research, A*STAR, Singapore. E-mail:sunzhuntu@gmail.com

\IEEEcompsocthanksitem Hui Fang (corresponding author) is with Shanghai University of Finance and Economics, China. E-mail: fang.hui@mail.shufe.edu.cn
\IEEEcompsocthanksitem Jie Yang is with Delft University of Technology, the Netherlands. E-mail: j.yang-3@tudelft.nl
\IEEEcompsocthanksitem Xinghua Qu is with ByteDance AI Lab, Singapore. E-mail:quxinghua17@gmail.com
\IEEEcompsocthanksitem Hongyang Liu is with Yanshan University, China. E-mail: hyliu767289@gmail.com
\IEEEcompsocthanksitem Di Yu is with Singapore Management University, Singapore. E-mail: yudi201909@gmail.com
\IEEEcompsocthanksitem Yew-Soon Ong is with Nanyang Technological University, Singapore. E-mail: asysong@ntu.edu.sg. He is
also with A*STAR Centre for Frontier AI Research, Singapore. E-mail: Ong\_Yew\_Soon@hq.a-star.edu.sg
\IEEEcompsocthanksitem Jie Zhang is with Nanyang Technological University, Singapore. E-mail: zhangj@ntu.edu.sg
}
}

\IEEEtitleabstractindextext{%
\begin{abstract}

Recently, one critical issue looms large in the field of recommender systems -- there are no effective benchmarks for rigorous evaluation -- which consequently leads to unreproducible evaluation and unfair comparison. We, therefore, conduct studies from the perspectives of practical theory and experiments, aiming at benchmarking recommendation for rigorous evaluation. Regarding the theoretical study, a series of hyper-factors affecting recommendation performance throughout the whole evaluation chain are systematically summarized and analyzed via an exhaustive review on 141 papers published at eight top-tier conferences within 2017-2020. We then classify them into model-independent and model-dependent hyper-factors, and different modes of rigorous evaluation are defined and discussed in-depth accordingly. For the experimental study, we release DaisyRec 2.0 library by integrating these hyper-factors to perform rigorous evaluation, whereby a holistic empirical study is conducted to unveil the impacts of different hyper-factors on recommendation performance. 
Supported by the theoretical and experimental studies, we finally create benchmarks for rigorous evaluation by proposing standardized procedures and providing performance of ten state-of-the-arts across six evaluation metrics on six datasets as a reference for later study. Overall, our work sheds light on the issues in recommendation evaluation, provides potential solutions for rigorous evaluation, and lays foundation for further investigation.
\end{abstract}

\begin{IEEEkeywords}
Recommender Systems, Reproducible Evaluation, Fair Comparison, Benchmarks, Standardized Procedures
\end{IEEEkeywords}}

\maketitle
\IEEEpeerreviewmaketitle

\section{Introduction}
\IEEEPARstart{W}{ith} the advent of the big data era, we are flooded by the exponentially increased information on the Internet.  To 
ease the severe information overload problem~\cite{sun2019research}, 
recommender systems have been extensively studied in academia and widely applied in industry across different domains, such as e-commerce (e.g., Amazon, Tmall), location-based social networks (e.g., Foursquare, Yelp), multi-media (e.g., Netflix, Spotify), and so forth. With a massive amount of recommendation approaches being proposed, one critical issue has attracted much attention from researchers in the field of recommender systems: there are few effective benchmarks for evaluation~\cite{said2014comparative,said2014rival,sun2020we}, which, consequently, leads to unreproducible evaluation and unfair comparison. 
As indicated by the recent study~\cite{rendle2019difficulty}, results for baselines that have been used in numerous publications over the past five years are suboptimal; with a careful setup, the baselines even outperform the reported results of any newly proposed method. This is in alignment with another latest study~\cite{dacrema2019we}, which discovers that the recent proposed deep learning models (DLMs) can be defeated by comparably simple baselines, such as MostPop and ItemKNN~\cite{sarwar2001item} with fine-tuned parameters. These findings initiate an extremely heated discussion on the evaluation of recommendation methods and inspire us to deeply consider the underlying barriers that hinder the rigorous evaluation in recommendation. 

\begin{figure}[t]
    \centering
    \includegraphics[width=0.5\textwidth]{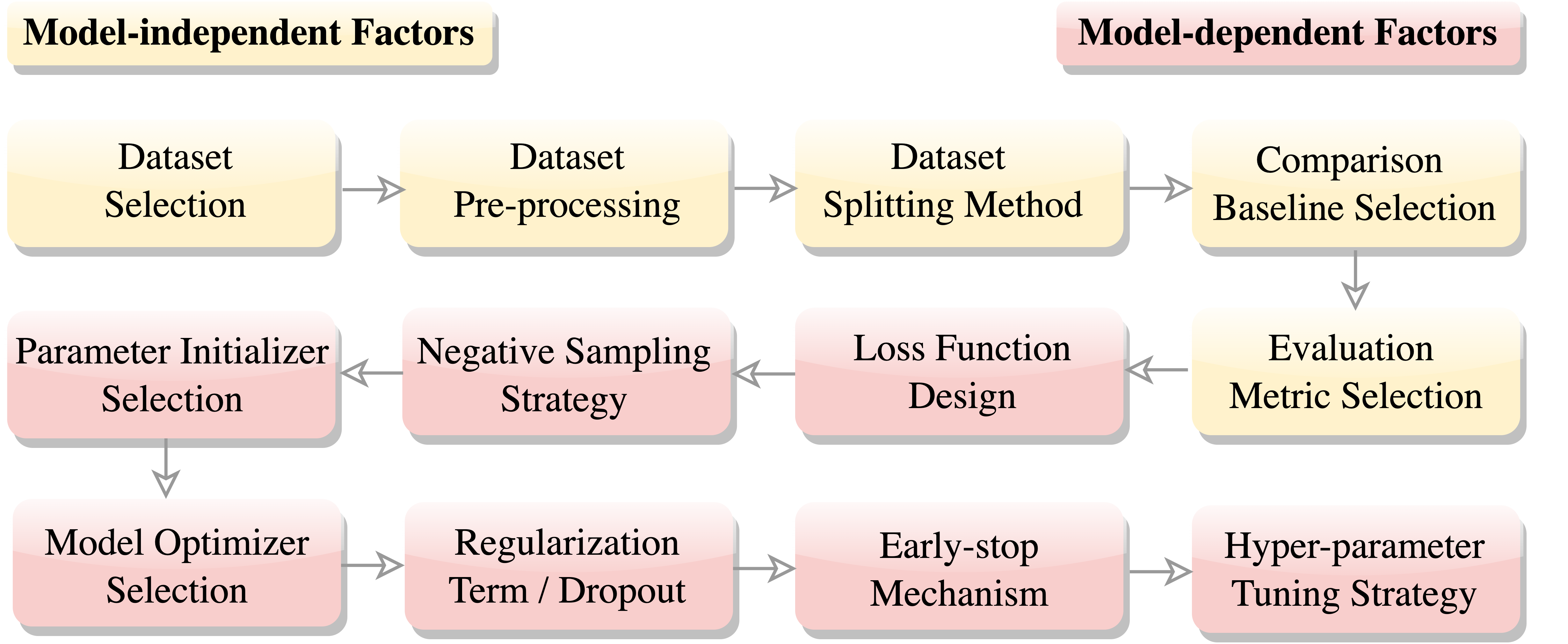}
    \vspace{-0.15in}
    \caption{Hyper-factors within the whole recommendation evaluation chain.}
    \label{fig:evaluation_chain}
    \vspace{-0.15in}
\end{figure}

%
\begin{table*}[t]
\scriptsize
\centering
\addtolength{\tabcolsep}{1pt}
\caption{Summary of the collected papers.}\label{tab:collected_papers}
\vspace{-0.15in}
    \begin{tabular}{|l|c|l|l|l|l|}
    \specialrule{.15em}{.05em}{.05em}
    \multirow{2}{*}{\textbf{Venue}} & \multirow{2}{*}{\textbf{No.}} &\multicolumn{4}{c|}{\textbf{Reference}} \\\cline{3-6}
    & &\textbf{2017} &\textbf{2018} &\textbf{2019} &\textbf{2020}\\
    \specialrule{.05em}{.05em}{.05em}
    \specialrule{.05em}{.05em}{.05em}
    \multirow{2}{*}{AAAI} &\multirow{2}{*}{22} &\multirow{2}{*}{\cite{sun2017exploiting,li2017ermma}} &\multirow{2}{*}{\cite{yu2018walkranker,wang2018collaborative,do2018coupled}} &\multirow{2}{*}{\cite{zhang2019hierarchical,wang2019camo,lin2019non,li2019zero,liu2019discrete,deng2019deepcf,hu2019hers,wang2019explainable}} &\cite{shen2020peia,zhu2020knowledge,chen2020efficientA,guo2020leveraging,li2020symmetric,xu2020multi,chen2020fast}\\
    &&&&&\cite{le2020stochastically,wang2020setrank}\\
    \specialrule{.05em}{.05em}{.05em}
    \multirow{2}{*}{CIKM} &\multirow{2}{*}{19}&\multirow{2}{*}{\cite{zhang2017joint,le2017indexable,pei2017interacting}}&\multirow{2}{*}{\cite{mei2018attentive,wang2018ripplenet,tran2018regularizing}}&\multirow{2}{*}{\cite{ma2019dbrec,kang2019candidate,xu2019relation}} &\cite{chang2020learning,chen2020tgcn,lee2020news,sun2020multi,kang2020rrd,chuang2020tpr}\\
    &&&&&\cite{xu2020commerce,wang2020disenhan,xian2020cafe,yuan2020exploring}\\
    \specialrule{.05em}{.05em}{.05em}
    {IJCAI} &{20} &{\cite{zhao2017learning,sun2017mrlr,xue2017deep}} &{\cite{liu2018dynamic,wang2018matrix,zhao2018plastic,ding2018improving,cheng2018delf,liu2018discrete}} &\cite{xin2019cfm,jiang2019convolutional,guo2019discrete,xu2019learning,zhang2019quaternion,fan2019deep,wang2019unified} &{\cite{han2020contextualized,xie2020internal,chen2020neural,liu2020hypernews}}\\
    \specialrule{.05em}{.05em}{.05em}
    KDD &12 &\cite{li2017collaborative} &\cite{zhu2018learning,christakopoulou2018local,hu2018leveraging} &\cite{wang2019kgat,tang2019akupm,zhao2019intentgc,wang2019graph,chen2019lambdaopt} &\cite{jin2020efficient,ma2020probabilistic,ji2020dual}\\
    \specialrule{.05em}{.05em}{.05em}
    RecSys &14 &\cite{otunba2017mpr,rafailidis2017learning} &\cite{valcarce2018robustness,sun2018recurrent,zheng2018spectral} &\cite{ouyang2019asymmetric,liu2019deep,nikolakopoulos2019personalized,costa2019collective,frolov2019hybridsvd,elahi2019variational} &\cite{zhang2020content,liu2020kred,zhou2020tafa}\\
    \specialrule{.05em}{.05em}{.05em}
    \multirow{2}{*}{SIGIR} &\multirow{2}{*}{22} &\multirow{2}{*}{\cite{chen2017attentive}} &\multirow{2}{*}{\cite{he2018adversarial,ebesu2018collaborative,xu2018graphcar,canamares2018should,wang2018streaming}}
    &\multirow{2}{*}{\cite{chen2019bayesian,wang2019neural,wu2019noise,xin2019relational}} &\cite{wang2020ckan,chen2020jointly,gong2020attentional,hansen2020content,he2020lightgcn,shi2020beyond}\\
    &&&&&\cite{tai2020mvin,wu2020joint,chae2020ar,wang2020disentangled,zou2020neural,sun2020neighbor}\\
    \specialrule{.05em}{.05em}{.05em}
    \multirow{2}{*}{WSDM} &\multirow{2}{*}{16} &\multirow{2}{*}{\cite{zhao2017multi}} &\multirow{2}{*}{\cite{zhang2018discrete,jiang2018recommendation,niu2018neural}} &\multirow{2}{*}{\cite{ma2019gated,nikolakopoulos2019recwalk,chen2019social,liu2019spiral}} &\cite{steck2020admm,li2020adversarial,liu2020end,gu2020hierarchical,wang2020key,sun2020lara}\\
    &&&&&\cite{zamani2020learning,shenbin2020recvae}\\
    \specialrule{.05em}{.05em}{.05em}
    WWW &16 &\cite{hsieh2017collaborative,he2017neural} &\cite{yu2018aesthetic,liang2018variational} &\cite{wang2019multi,ma2019jointly,cao2019unifying,tran2019signed,wang2019knowledge,chen2019collaborative} &\cite{khawar2020learning,liu2020deep,javari2020weakly,wang2020personalized,tan2020learning,chen2020efficientW}\\
    \specialrule{.05em}{.05em}{.05em}
    Total &141 &15 &28 &43 &55\\
    \specialrule{.15em}{.05em}{.05em}
    \end{tabular}
    \vspace{-0.1in}
\end{table*}

As a matter of fact, there are a number of \textit{hyper-factors} which may affect the recommendation performance throughout the whole evaluation chain, and their best settings are unknown. They can be broadly classified into two types as depicted in Figure~\ref{fig:evaluation_chain}, namely model-independent and model-dependent hyper-factors. The former refers to the hyper-factors that are isloated from the model design and optimization process (e.g., dataset and comparison baseline selection); whilst the latter indicates the ones involved in the model development and parameter optimization procedure (e.g., loss function design and regularization terms). According to this categorization, three main aspects may inherently lead to such non-rigorous evaluation.
\begin{itemize}[leftmargin=*]
    \item \textbf{Diverse settings on model-independent hyper-factors}. With being prominent in different platforms, there are diverse recommendation datasets (i.e., dataset selection) in various application domains shown in Table~\ref{tab:summary_datasets} (Appendix). Taking the movie domain as an example, the datasets vary from MovieLens (ML), Netflix to Amazon-Movie, etc. Even for the same dataset, it may have different versions with different sizes covering different durations, such as, ML-100K/1M/10M/20M/25M/Lastest. Different researchers may choose different datasets across different domains based on their requirements, and only report results on their selected datasets; meanwhile different settings on other model-independent hyper-factors (e.g., dataset pre-processing and splitting strategies) may generate entirely different recommendation performance.
    \item \textbf{Diverse settings on model-dependent hyper-factors}. There are different choices for the model-dependent hyper-factors. For instance, the loss function could be either point-wise (square error loss~\cite{koren2009matrix} and cross-entropy loss~\cite{he2017neural}) or pair-wise (log loss~\cite{rendle2009bpr}, hinge loss~\cite{zhao2018plastic} and top-1 loss~\cite{hidasi2016parallel}); different types of model optimizers are also available, ranging from stochastic gradient descent (SGD) to adaptive moment estimation (Adam).  The recommendation results may vary a lot with different settings on these model-dependent factors even with fixed settings on model-independent ones.
    \item \textbf{Missing setting details}. Most importantly, a majority of papers do not report details on the settings of either model-independent and model-dependent hyper-factors, such as data processing and parameter settings, which increases the difficulty on evaluation and leads to inconsistent results in reproduction by different researchers, thus heavily aggravating the unreproducible evaluation and unfair comparison issues.  
\end{itemize}

With the above issues in mind, we are seeking at benchmarking recommendation for rigorous (i.e., reproducible and fair) evaluation, thus helping achieve a healthy and sustainable growth of research in this area. Considering the diverse recommendation tasks (e.g., temporal, sequential, diversity, explanation, location, group and cross-domain aware recommendation), we first mainly focus on the \textbf{general top-$N$} recommendation task, which is one of the hottest and most prominent topics in recommendation. 
To this end, we conduct extensive studies from the perspectives of both practical theory and experiments.
\begin{itemize}[leftmargin=*]
    \item Regarding the theoretical study, we conduct an exhaustive review on $141$ papers related to top-$N$ recommendation published in the recent four years (2017-2020) on eight prestigious conferences as representatives, including KDD, SIGIR, WWW, IJCAI, AAAI, RecSys, WSDM and CIKM\footnote{They are most important venues (full names see Table~\ref{tab:conference-name} in Appendix) to accept high-quality recommendation papers and other related conferences and journals will be considered in our future study.}. In doing so, we systematically extract and summarize hyper-factors throughout the whole evaluation chain, and classify them into model-independent and model-dependent factors in Figure~\ref{fig:evaluation_chain}.
    Accordingly, different modes (e.g., relax, strict and mixed) of rigorous evaluation are defined and discussed in-depth, acting as valuable guidance for later study.  
    \item For the experimental study, we release a Python-based recommendation testbed -- DaisyRec 2.0 to integrate the hyper-factors throughout the evaluation chain\footnote{\url{https://github.com/recsys-benchmark/DaisyRec-v2.0}}. Our testbed advances existing libraries (e.g., LibRec~\cite{guo2015librec}, DeepRec~\cite{zhang2019deeprec} and RecBole~\cite{zhao2021recbole}), which mainly aim to implement various state-of-the-art recommenders, in the light of performing rigorous evaluation in recommendation. 
    Based on DaisyRec 2.0, a holistic empirical study has been performed to comprehensively analyze the impact of different hyper-factors on recommendation performance. 
\end{itemize}
%

%
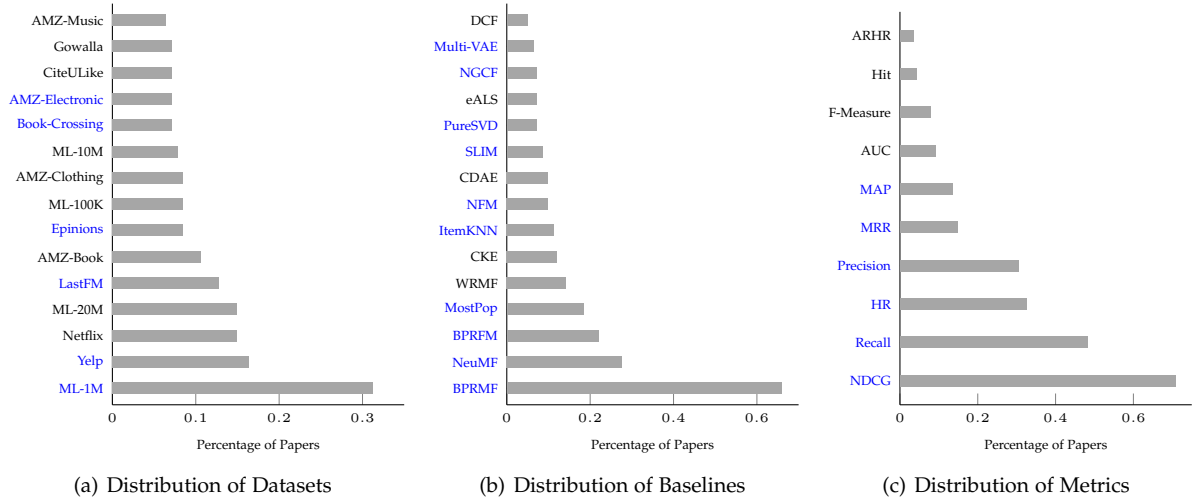
\begin{figure*}
\centering
\subfigure[Distribution of Datasets]{
\begin{tikzpicture}
\begin{axis}[
    xbar stacked,
    legend style={
    legend columns=4,
        at={(xticklabel cs:0.5)},
        anchor=north,
        draw=none
    },
    ytick=data,
    axis y line*=none,
    axis x line*=bottom,
    tick label style={font=\tiny},
    legend style={font=\tiny},
    label style={font=\tiny},
    xlabel={Percentage of Papers},
    xtick={0, 0.1, 0.2, 0.3},
    width=.3\textwidth,
    bar width=1.5mm,
    yticklabels={
    {\textcolor{blue}{ML-1M}}, 
    {\textcolor{blue}{Yelp}}, 
    {Netflix}, 
    {ML-20M}, 
    {\textcolor{blue}{LastFM}}, 
    {AMZ-Book},
    {\textcolor{blue}{Epinions}},
    {ML-100K},
    {AMZ-Clothing},
    {ML-10M},
    {\textcolor{blue}{Book-Crossing}},
    {\textcolor{blue}{AMZ-Electronic}},
    {CiteULike},
    {Gowalla},
    {AMZ-Music},
    },
    xmin=0,
    xmax=0.35,
    area legend,
    y=3.5mm,
    enlarge y limits={abs=0.625},
    visualization depends on=y \as \pgfplotspointy,
    every axis plot/.append style={fill}
]
\addplot[draw=none, fill=gray!70!white, thick] coordinates
  {(0.3120567376,0) (0,1) (0,2) (0,3) (0,4) (0,5) (0,6) (0,7) (0,8) (0,9) (0,10) (0,11) (0,12) (0,13) (0,14)};
\addplot[draw=none,fill=gray!70!white, thick] coordinates
  {(0,0) (0.1631205674,1) (0,2) (0,3) (0,4) (0,5) (0,6) (0,7) (0,8) (0,9) (0,10) (0,11) (0,12) (0,13) (0,14)};
\addplot[draw=none,fill=gray!70!white, thick] coordinates
  {(0,0) (0,1) (0.1489361702,2) (0,3) (0,4) (0,5) (0,6) (0,7) (0,8) (0,9) (0,10) (0,11) (0,12) (0,13) (0,14)};
\addplot[draw=none,fill=gray!70!white, thick] coordinates
  {(0,0) (0,1) (0,2) (0.1489361702,3) (0,4) (0,5) (0,6) (0,7) (0,8) (0,9) (0,10) (0,11) (0,12) (0,13) (0,14)};
\addplot[draw=none,fill=gray!70!white, thick] coordinates
  {(0,0) (0,1) (0,2) (0,3) (0.1276595745,4) (0,5) (0,6) (0,7) (0,8) (0,9) (0,10) (0,11) (0,12) (0,13) (0,14)};
\addplot[draw=none,fill=gray!70!white, thick] coordinates
  {(0,0) (0,1) (0,2) (0,3) (0,4) (0.1063829787,5) (0,6) (0,7) (0,8) (0,9) (0,10) (0,11) (0,12) (0,13) (0,14)};
\addplot[draw=none,fill=gray!70!white, thick] coordinates
  {(0,0) (0,1) (0,2) (0,3) (0,4) (0,5) (0.08510638298,6) (0,7) (0,8) (0,9) (0,10) (0,11) (0,12) (0,13) (0,14)};
\addplot[draw=none,fill=gray!70!white, thick] coordinates
  {(0,0) (0,1) (0,2) (0,3) (0,4) (0,5) (0,6) (0.08510638298,7) (0,8) (0,9) (0,10) (0,11) (0,12) (0,13) (0,14)};
\addplot[draw=none,fill=gray!70!white, thick] coordinates
  {(0,0) (0,1) (0,2) (0,3) (0,4) (0,5) (0,6) (0,7) (0.08510638298,8) (0,9) (0,10) (0,11) (0,12) (0,13) (0,14)};
\addplot[draw=none,fill=gray!70!white, thick] coordinates
  {(0,0) (0,1) (0,2) (0,3) (0,4) (0,5) (0,6) (0,7) (0,8) (0.0780141844,9) (0,10) (0,11) (0,12) (0,13) (0,14)};
\addplot[draw=none,fill=gray!70!white, thick] coordinates
  {(0,0) (0,1) (0,2) (0,3) (0,4) (0,5) (0,6) (0,7) (0,8) (0,9) (0.07092198582,10) (0,11) (0,12) (0,13) (0,14)};
\addplot[draw=none,fill=gray!70!white, thick] coordinates
  {(0,0) (0,1) (0,2) (0,3) (0,4) (0,5) (0,6) (0,7) (0,8) (0,9) (0,10) (0.07092198582,11) (0,12) (0,13) (0,14)};
\addplot[draw=none,fill=gray!70!white, thick] coordinates
  {(0,0) (0,1) (0,2) (0,3) (0,4) (0,5) (0,6) (0,7) (0,8) (0,9) (0,10) (0,11) (0.07092198582,12) (0,13) (0,14)};
\addplot[draw=none,fill=gray!70!white, thick] coordinates
  {(0,0) (0,1) (0,2) (0,3) (0,4) (0,5) (0,6) (0,7) (0,8) (0,9) (0,10) (0,11) (0,12) (0.07092198582,13) (0,14)};
\addplot[draw=none,fill=gray!70!white, thick] coordinates
  {(0,0) (0,1) (0,2) (0,3) (0,4) (0,5) (0,6) (0,7) (0,8) (0,9) (0,10) (0,11) (0,12) (0,13) (0.06382978723,14)};  
\end{axis}  
\end{tikzpicture}
}
\subfigure[Distribution of Baselines]{
\begin{tikzpicture}
\begin{axis}[
    xbar stacked,
    legend style={
    legend columns=4,
        at={(xticklabel cs:0.5)},
        anchor=north,
        draw=none
    },
    ytick=data,
    axis y line*=none,
    axis x line*=bottom,
    tick label style={font=\tiny},
    legend style={font=\tiny},
    label style={font=\tiny},
    xlabel={Percentage of Papers},
    xtick={0, 0.2, 0.4, 0.6},
    width=.3\textwidth,
    bar width=1.5mm,
    yticklabels={
    {\textcolor{blue}{BPRMF}}, 
    {\textcolor{blue}{NeuMF}}, 
    {\textcolor{blue}{BPRFM}}, 
    {\textcolor{blue}{MostPop}}, 
    {WRMF}, 
    {CKE},
    {\textcolor{blue}{ItemKNN}},
    {\textcolor{blue}{NFM}},
    {CDAE},
    {\textcolor{blue}{SLIM}},
    {\textcolor{blue}{PureSVD}},
    {eALS},
    {\textcolor{blue}{NGCF}},
    {\textcolor{blue}{Multi-VAE}},
    {DCF},
    },
    xmin=0,
    xmax=0.7,
    area legend,
    y=3.5mm,
    enlarge y limits={abs=0.625},
    visualization depends on=y \as \pgfplotspointy,
    every axis plot/.append style={fill}
]
\addplot[draw=none, fill=gray!70!white, thick] coordinates
  {(0.6595744681,0) (0,1) (0,2) (0,3) (0,4) (0,5) (0,6) (0,7) (0,8) (0,9) (0,10) (0,11) (0,12) (0,13) (0,14)};
\addplot[draw=none,fill=gray!70!white, thick] coordinates
  {(0,0) (0.2765957447,1) (0,2) (0,3) (0,4) (0,5) (0,6) (0,7) (0,8) (0,9) (0,10) (0,11) (0,12) (0,13) (0,14)};
\addplot[draw=none,fill=gray!70!white, thick] coordinates
  {(0,0) (0,1) (0.219858156,2) (0,3) (0,4) (0,5) (0,6) (0,7) (0,8) (0,9) (0,10) (0,11) (0,12) (0,13) (0,14)};
\addplot[draw=none,fill=gray!70!white, thick] coordinates
  {(0,0) (0,1) (0,2) (0.1843971631,3) (0,4) (0,5) (0,6) (0,7) (0,8) (0,9) (0,10) (0,11) (0,12) (0,13) (0,14)};
\addplot[draw=none,fill=gray!70!white, thick] coordinates
  {(0,0) (0,1) (0,2) (0,3) (0.1418439716,4) (0,5) (0,6) (0,7) (0,8) (0,9) (0,10) (0,11) (0,12) (0,13) (0,14)};
\addplot[draw=none,fill=gray!70!white, thick] coordinates
  {(0,0) (0,1) (0,2) (0,3) (0,4) (0.1205673759,5) (0,6) (0,7) (0,8) (0,9) (0,10) (0,11) (0,12) (0,13) (0,14)};
\addplot[draw=none,fill=gray!70!white, thick] coordinates
  {(0,0) (0,1) (0,2) (0,3) (0,4) (0,5) (0.1134751773,6) (0,7) (0,8) (0,9) (0,10) (0,11) (0,12) (0,13) (0,14)};
\addplot[draw=none,fill=gray!70!white, thick] coordinates
  {(0,0) (0,1) (0,2) (0,3) (0,4) (0,5) (0,6) (0.09929078014,7) (0,8) (0,9) (0,10) (0,11) (0,12) (0,13) (0,14)};
\addplot[draw=none,fill=gray!70!white, thick] coordinates
  {(0,0) (0,1) (0,2) (0,3) (0,4) (0,5) (0,6) (0,7) (0.09929078014,8) (0,9) (0,10) (0,11) (0,12) (0,13) (0,14)};
\addplot[draw=none,fill=gray!70!white, thick] coordinates
  {(0,0) (0,1) (0,2) (0,3) (0,4) (0,5) (0,6) (0,7) (0,8) (0.08510638298,9) (0,10) (0,11) (0,12) (0,13) (0,14)};
\addplot[draw=none,fill=gray!70!white, thick] coordinates
  {(0,0) (0,1) (0,2) (0,3) (0,4) (0,5) (0,6) (0,7) (0,8) (0,9) (0.07092198582,10) (0,11) (0,12) (0,13) (0,14)};
\addplot[draw=none,fill=gray!70!white, thick] coordinates
  {(0,0) (0,1) (0,2) (0,3) (0,4) (0,5) (0,6) (0,7) (0,8) (0,9) (0,10) (0.07092198582,11) (0,12) (0,13) (0,14)};
\addplot[draw=none,fill=gray!70!white, thick] coordinates
  {(0,0) (0,1) (0,2) (0,3) (0,4) (0,5) (0,6) (0,7) (0,8) (0,9) (0,10) (0,11) (0.07092198582,12) (0,13) (0,14)};
\addplot[draw=none,fill=gray!70!white, thick] coordinates
  {(0,0) (0,1) (0,2) (0,3) (0,4) (0,5) (0,6) (0,7) (0,8) (0,9) (0,10) (0,11) (0,12) (0.06382978723,13) (0,14)};
\addplot[draw=none,fill=gray!70!white, thick] coordinates
  {(0,0) (0,1) (0,2) (0,3) (0,4) (0,5) (0,6) (0,7) (0,8) (0,9) (0,10) (0,11) (0,12) (0,13) (0.04964539007,14)};  
\end{axis}  
\end{tikzpicture}
}
\subfigure[Distribution of Metrics]{
\begin{tikzpicture}
\begin{axis}[
    xbar stacked,
    legend style={
    legend columns=4,
        at={(xticklabel cs:0.5)},
        anchor=north,
        draw=none
    },
    ytick=data,
    axis y line*=none,
    axis x line*=bottom,
    tick label style={font=\tiny},
    legend style={font=\tiny},
    label style={font=\tiny},
    xlabel={Percentage of Papers},
    xtick={0, 0.2, 0.4, 0.6},
    width=.3\textwidth,
    bar width=1.5mm,
    yticklabels={
    {\textcolor{blue}{NDCG}}, 
    {\textcolor{blue}{Recall}}, 
    {\textcolor{blue}{HR}}, 
    {\textcolor{blue}{Precision}}, 
    {\textcolor{blue}{MRR}}, 
    {\textcolor{blue}{MAP}},
    {AUC},
    {F-Measure},
    {Hit},
    {ARHR}
    },
    xmin=0,
    xmax=0.75,
    area legend,
    y=5.1mm,
    enlarge y limits={abs=0.625},
    visualization depends on=y \as \pgfplotspointy,
    every axis plot/.append style={fill}
]
\addplot[draw=none, fill=gray!70!white, thick] coordinates
  {(0.7092198582,0) (0,1) (0,2) (0,3) (0,4) (0,5) (0,6) (0,7) (0,8) (0,9)};
\addplot[draw=none,fill=gray!70!white, thick] coordinates
  {(0,0) (0.4822695035,1) (0,2) (0,3) (0,4) (0,5) (0,6) (0,7) (0,8) (0,9)};
\addplot[draw=none,fill=gray!70!white, thick] coordinates
  {(0,0) (0,1) (0.3262411348,2) (0,3) (0,4) (0,5) (0,6) (0,7) (0,8) (0,9)};
\addplot[draw=none,fill=gray!70!white, thick] coordinates
  {(0,0) (0,1) (0,2) (0.304964539,3) (0,4) (0,5) (0,6) (0,7) (0,8) (0,9)};
\addplot[draw=none,fill=gray!70!white, thick] coordinates
  {(0,0) (0,1) (0,2) (0,3) (0.1489361702,4) (0,5) (0,6) (0,7) (0,8) (0,9)};
\addplot[draw=none,fill=gray!70!white, thick] coordinates
  {(0,0) (0,1) (0,2) (0,3) (0,4) (0.134751773,5) (0,6) (0,7) (0,8) (0,9)};
\addplot[draw=none,fill=gray!70!white, thick] coordinates
  {(0,0) (0,1) (0,2) (0,3) (0,4) (0,5) (0.09219858156,6) (0,7) (0,8) (0,9)};
\addplot[draw=none,fill=gray!70!white, thick] coordinates
  {(0,0) (0,1) (0,2) (0,3) (0,4) (0,5) (0,6) (0.0780141844,7) (0,8) (0,9)};
\addplot[draw=none,fill=gray!70!white, thick] coordinates
  {(0,0) (0,1) (0,2) (0,3) (0,4) (0,5) (0,6) (0,7) (0.04255319149,8) (0,9)};
\addplot[draw=none,fill=gray!70!white, thick] coordinates
  {(0,0) (0,1) (0,2) (0,3) (0,4) (0,5) (0,6) (0,7) (0,8) (0.03546099291,9)};
\end{axis}  
\end{tikzpicture}
}
\vspace{-0.1in}
\caption{(a) popularity of the top-15 datasets, where `ML, AMZ' denote MovieLens and Amazon, respectively; (b) popularity of the top-15 baselines; and (c) popularity of the top-10 evaluation metrics. Note that the selected datasets, baselines and metrics in our study are highlighted in blue.}
\label{fig:factor_distribution}
\vspace{-0.15in}
\end{figure*}

Supported by both theoretical and experimental study, we finally create benchmarks by proposing the standardized procedures to help enhance the reproducibility and fairness of evaluation. Meanwhile, the performance of ten well-tuned state-of-the-arts on six widely-used datasets across six metrics is provided as a reference for fair evaluation. Additionally, a number of interesting findings are noted throughout our study, for example, \textbf{(1)} 
the recommendation performance does not necessarily improve with denser datasets; \textbf{(2)} some non-deep learning based baselines, e.g., PureSVD~\cite{cremonesi2010performance} can achieve a better balance between recommendation accuracy and complexity; \textbf{(3)} the
best hyper-parameter settings for one specific metric does not necessarily guarantee optimums w.r.t. other metrics;
\textbf{(4)} although the
objective function with pair-wise log loss generally outperforms
others, different methods may have their best fit objective functions; \textbf{(5)} uniform sampler, though simple, performs better than the popularity based sampler; and \textbf{(6)} different parameter initializers and model optimizers can extensively affect the final recommendation accuracy.  
To sum up, our work sheds lights on the issues in evaluation for recommendation, provides potential solutions for rigorous evaluation, and paves the way for further investigation\footnote{A preliminary report of our work was published at RecSys'20 as a reproducibility paper~\cite{sun2020we}. In this study, we have extended it from two aspects: (1) with regards to theoretical study, we conduct a more in-depth analysis on the hyper-factors throughout the whole evaluation chain by reviewing more latest literature in 2020, whereby several new hyper-factors (e.g., regularization terms, parameter initializers and model optimizers) are further considered; we systematically classify these hyper-factors into model-independent and -dependent ones, whereby different modes of rigorous evaluation are well defined and discussed in-depth; and
(2) for the experimental study, we release DaisyRec 2.0 by further fusing these new hyper-factors and extending existing ones (e.g., more types of loss function designs, negative sampling strategies, data splitting methods and deep learning based baselines). To be more user-friendly, we design a user interface tool for automatic command generation. 
Thereby, more holistic experiments are conducted based on DaisyRec 2.0 to unveil the impacts of different hyper-factors on recommendation performance, where more interesting and insightful observations are gained.
}.

\section{Practical Theory Study}\label{sec:theoretical_study}

\subsection{Hyper-factor Extraction}
As we seek to benchmark recommendation for rigorous evaluation, we first conduct study from the angle of practical theory by an exhaustive literature review, so as to extract and summarize hyper-factors affecting recommendation performance throughout the whole evaluation chain.
In particular, we review papers published in the recent four years (2017-2020) on eight top tier conferences, namely, AAAI, CIKM, IJCAI, KDD, RecSys, SIGIR WSDM and WWW. As a starting point, we mainly focus on recommendation methods for implicit feedback based top-$N$ recommendation, which is one of the hottest topics in recommendation.  
Other tasks (e.g., sequential recommendation)
are remained for future exploration. Specifically, we first search the accepted full paper lists ($8*4=32$) for the eight conferences in the four years. Given our interest and the 32 lists, we only consider papers with titles containing keywords `recommend$\ast$' or `collaborative filtering'. After that, we manually select papers towards top-$N$ recommendation 
adopting ranking metrics (e.g., Precision, Recall) to evaluate the \textit{accuracy} of recommendation. In the end, we obtain a collection of 141 relevant 
papers as listed in Table~\ref{tab:collected_papers}.

By delicately reviewing the collected papers in Table~\ref{tab:collected_papers}, we find that there are a bunch of hyper-factors that may affect the recommendation performance along with the entire evaluation chain. Typically, they can be classified into two types: (1) \textbf{model-independent} factors that are isolated from the model design and optimization process (e.g., dataset and baseline selection); and (2) \textbf{model-dependent} ones involved in the model development and parameter optimization process (e.g., loss function and regularization terms). Figure~\ref{fig:evaluation_chain} illustrates the two types of  hyper-factors along with the whole evaluation chain, starting with the dataset selection and ending with hyper-parameter tuning strategy. Next, we will analyze these hyper-factors one by one. 

\subsection{Analysis on Model-independent Hyper-factors}\label{subsec:independent-factor}
 
\subsubsection{Dataset Selection}\label{subsubsec:dataset}
We find two major issues on the utilized datasets by analyzing the collected papers: (1) domain diversity, i.e., there are massive different datasets within and across various domains, as shown in Table~\ref{tab:summary_datasets}; and
(2) version diversity, i.e., many datasets, though with the same names, may have different versions. For example, we find more than three versions for Yelp, as it has been updated for different rounds of the challenge. By treating different versions as a same dataset, there are $84$ different datasets used in the $141$ papers. Figure~\ref{fig:factor_distribution}(a) shows the dataset popularity, i.e., percentage of papers for the top-15 used datasets. Around $90\%$ of the $141$ papers adopt at least one of the 15 datasets. 
\begin{table}[t]
    \scriptsize
    \centering
    \renewcommand{\arraystretch}{0.9}
    \addtolength{\tabcolsep}{-2.5pt}
    \caption{Statistics of the six selected  datasets.}\label{tab:dataset_statistics}
    \vspace{-0.15in}
    \begin{tabular}{|l|l|l|l|l|l|l|l|l|l|l|l|l|l|l|l}
    \specialrule{.15em}{.05em}{.05em}
    \multicolumn{2}{|c|}{Dataset} &ML-1M &Yelp &LastFM &Epinions &Book-X &AMZe \\
    \specialrule{.05em}{.05em}{.05em}
    \specialrule{.05em}{.05em}{.05em}
    \multirow{4}{*}{\rotatebox[origin=c]{90}{origin}}
    &\#User  &6,038  &1,326,101 &1,892 &22,164   &105,283 &4,201,696 \\
    &\#Item  &3,533 &174,567 &17,632  &296,277 &340,556 &476,002 \\
    &\#Record  &575,281 &5,261,669 &92,834 &922,267  &1,149,780 &7,824,482 \\
    &Density  &2.697e-2 &2.273e-5 &2.783e-3 &1.404e-4 &3.207e-5 &3.912e-6\\
    \specialrule{.05em}{.05em}{.05em}
    \multirow{4}{*}{\rotatebox[origin=c]{90}{5-filter}}
    &\#User  &6,034 &227,109 &1,874 &21,995 &22,072  &253,994\\
    &\#Item  &3,125 &123,985 &2,828 &31,678 &43,748  &145,199\\
    &\#Record  &574,376 &3,419,587 &71,411 &550,117 &623,405 &2,109,869  \\
    &Density  &3.046e-2 &1.214e-4 &1.348e-2 &7.895e-4 &6.456e-4 &5.721e-5 \\
    \specialrule{.05em}{.05em}{.05em}
    \multirow{4}{*}{\rotatebox[origin=c]{90}{10-filter}}
    &\#User &5,950 &96,168 &1,867 &21,111 &12,720 &63,161 \\
    &\#Item  &2,811 &80,351 &1,530 &14,030 &18,318 &85,930 \\
    &\#Record  &571,549  &2,458,153 &62,984 &434,162 &443,196 &949,416 \\
    &Density  &3.412e-2 &3.181e-4 &2.205e-2 &1.466e-3 &1.902e-3 &1.749e-4 \\
    \specialrule{.05em}{.05em}{.05em}
    \multicolumn{2}{|c|}{Timestamp} &${\surd}$ &${\surd}$ &${\times}$ &${\surd}$ &${\times}$ &${\surd}$\\
    \specialrule{.15em}{.05em}{.05em}
    \end{tabular}
    \vspace{-0.1in}
\end{table}

For a practical study, we further delicately select six among them by considering popularity and domain coverage, thus resolving the domain diversity issue. They are \textbf{ML-1M} (Movie), \textbf{Yelp} (LBSNs), \textbf{LastFM} (Music),  \textbf{Epinions} (SNs), \textbf{Book-X} (Book) and \textbf{AMZe} (Consumable), covering $62\%$ papers of the collection. To ease the version diversity issue, we conduct a careful selection by considering the authority and information richness
of data sources, which could benefit the study on diverse recommendation models. 
Specifically, 
we use MovieLens-1M (ML-1M) released by GroupLens\footnote{\url{grouplens.org/datasets/movielens/}};
Yelp was created by Kaggle in 2018\footnote{\url{www.kaggle.com/yelp-dataset/yelp-dataset}}; 
LastFM was released by the 2nd international workshop HetRec 2011~\cite{Cantador:RecSys2011};
Epinions was crawled by~\cite{tang-etal12b} containing timestamp and item category information;
Book-Crossing (Book-X)~\cite{ziegler2005improving} was collected by Cai-Nicolas Ziegler from the Book-Crossing community\footnote{\url{grouplens.org/datasets/book-crossing/}}; Amazon Electronic (AMZe) was released by Julian McAuley\footnote{\url{jmcauley.ucsd.edu/data/amazon/links.html}}.
The statistics of all datasets are listed in Table~\ref{tab:dataset_statistics} and all links for the datasets are available at the homepage of DaisyRec 2.0.

\subsubsection{Dataset Pre-Processing}
There are typically two core steps for the dataset pre-processing, namely {binarization} and {filtering}. 

\smallskip\noindent\textit{Binarization}. As our current study focuses on the implicit feedback, all the datasets with explicit feedback (e.g., ratings or counts) should be binarized into implicit data. 
Let $u \in \mathcal{U}, i \in \mathcal{I}$ denote user $u$ and item $i$; $\mathcal{U, I}$ are user and item sets; and $r_{ui} \in \mathcal{R}$ is the binary feedback of user $u$ over item $i$.
For each user $u$, we transform all her explicit feedback with no less than a threshold (denoted by $r$) into positive feedback ($r_{ui}=1$); otherwise, negative feedback ($r_{ui}=0$).
Different papers may have different settings for $r$ (e.g., $r = 1/2/3/4$). By following the majority studies~\cite{wu2019noise,tang2019akupm,wang2019graph}, we recommend to set $r=4$ for ML-1M, and $r=1$ for the rest datasets.

\smallskip\noindent\textit{Filtering}. The original datasets are generally quite sparse, where some users (items) only interact with few items (users), e.g., less than $5$. To ensure the quality, the filtering strategy is usually adopted to help remove the inactive users and items. By analyzing the paper collection, we have found out that around $57\%$ of papers adopt filtering strategies; while $22\%$ of papers utilize the original datasets; and $21\%$ of papers do not report details on data filtering. 
Among the papers adopting pre-processing strategies, more than $58\%$ of them utilize 5- or 10-filter/core setting~\cite{xu2019relation,wang2019kgat,wang2019neural} on the datasets, which filter out users and items with less than 5 or 10 interactions, respectively. While, others adopt, such as 1-, 2-, 3-, 4-, 20- or 30-filter/core settings. 
Therefore, to check the performance and robustness of different methods w.r.t. various data sparsity levels, besides original datasets, we also take the two most common settings (i.e., \textbf{5-} and \textbf{10-filter}) on all selected datasets, the statistics of which are summarized in Table~\ref{tab:dataset_statistics}. Note that $F$-fiter is different from $F$-core: the former means that users and items are only filtered with less than $F$ interactions in one pass; by contrast, the latter indicates a recursive filtering until all users and items have at least $F$ interactions.

\subsubsection{Dataset Splitting Methods}\label{subsec:data-splitting-method} 
Three types of data splitting methods are mainly used in the collected papers, including \textbf{split-by-ratio} ($69\%$ of the papers), \textbf{leave-one-out} ($21\%$ of the papers) and \textbf{split-by-time} ($6\%$ of the papers). There are also $4\%$ of papers not reporting their data splitting methods. 
In particular, split-by-ratio means that a proportion $\rho$ (e.g., $\rho=80\%$) of the dataset (i.e., user-item interaction records) is treated as training set, and the rest ($1-\rho =20\%$) as test set; leave-one-out refers to that for each user, only one record is kept as test set and the remaining are for training; and split-by-time indicates directly dividing training and test sets by a fixed timestamp, that is, the data before the fixed timestamp is used as training set, and the rest as test set. 

Although $69\%$ of papers adopt split-by-ratio, they are quite different from each other due to: (1) different proportion settings, e.g., $\rho=50\%, 60\%, 70\%, 80\%, 90\%$; (2) global- or user-level split. That is, some globally split the entire records into training and test sets regardless of different users; whilst others split training and test sets on the user basis; and (3) random- or time-aware split. Among papers exploiting split-by-ratio, $87\%$ of papers merely randomly split the data, whereas $13\%$ of papers split the data based on the timestamp, i.e., the earlier (e.g., $\rho=80\%$) records as training and the later ones as test. 
In terms of leave-one-out, the split is generally on the user basis and timestamp is taken into account by $60\%$ of papers. To validate the impact of different data splitting methods, in our study, we compare the recommendation performance with random-/time-aware \textbf{split-by-ratio} at \textbf{global-level} with \textbf{$\rho=80\%$} and random-/time-aware \textbf{leave-one-out} at \textbf{user-level}, and leave split-by-time as our future exploration.  

Besides, to improve the test efficiency, they usually randomly sample a number of negative items (e.g., $neg\_test=99, 100, 999, 1,000$) that are not interacted by each
user, and then rank each test item among the $(neg\_test+1)$ items for recommendation~\cite{ma2019jointly,shi2020beyond,xin2019relational,chuang2020tpr}. To speed up the test process, we randomly sample negative items for each user to ensure her test candidates to be $1,000$, and then rank all test items among the $1,000$ items w.r.t. both split-by-ratio and leave-one-out.  Table~\ref{tab:statistics_train_test_set} depicts the average number of test items for each user on the six datasets across origin, 5- and 10-filter settings w.r.t. split-by-ratio, where all values are smaller than $100$, indicating that $1,000$ test candidates is sufficient to examine the performance of recommenders.

\begin{table}[t]
    \centering
    \scriptsize
    \addtolength{\tabcolsep}{-2pt}
    \renewcommand{\arraystretch}{0.9}
	\caption{Average size of training \& test sets for each user.}
	\label{tab:statistics_train_test_set}
	\vspace{-0.15in}
    \begin{tabular}{|l|l|llllll|}
    \specialrule{.15em}{.05em}{.05em}
    Type &Setting &ML-1M &Yelp &LastFM &Epinions &Book-X &AMZe \\
    \specialrule{.05em}{.05em}{.05em}
    \multirow{3}{*}{\textbf{Train}} &origin &86&4&39&35&10&2\\
    &5-filter &86&13&30&23&23&7\\
    &10-filter &86&21&27&20&28&12\\
    \specialrule{.05em}{.05em}{.05em}
    \multirow{3}{*}{\textbf{Test}} &origin &64&2&10&30&5&2\\
    &5-filter &64&5&8&13&7&3\\
    &10-filter &64&8&7&10&8&5\\
    \specialrule{.15em}{.05em}{.05em}
    \end{tabular}
    \vspace{-0.15in}
\end{table}

\subsubsection{Comparison Baseline Selection}\label{subsubsec:comparison_baselines}
As observed, the compared baselines vary a lot in different collected papers.   
We show the top-15 widely-compared baselines in these papers in Figure~\ref{fig:factor_distribution}(b), covering $98\%$ of papers in total, that is, $98\%$ of the papers consider at least one of the 15 baselines. They can be classified into three types, (1) memory-based methods (MMs): MostPop, ItemKNN~\cite{sarwar2001item}; (2) latent factor methods (LFMs): BPRMF~\cite{rendle2009bpr}, FM~\cite{rendle2010factorization}, WRMF~\cite{hu2008collaborative}, SLIM~\cite{ning2011slim}, PureSVD~\cite{cremonesi2010performance}, eALS~\cite{he2016fast} and DCF~\cite{zhang2016discrete}; and (3) deep learning methods (DLMs):
NeuMF~\cite{he2017neural}, CKE~\cite{zhang2016collaborative}, NeuFM~\cite{he2017neural}, CDAE~\cite{wu2016collaborative}, NGCF~\cite{wang2019neural} and Multi-VAE~\cite{liang2018variational}. 

For a practical study, we ultimately take 10 baselines into account, as highlighted in blue in Figure~\ref{fig:factor_distribution}(b). Specifically, two MMs are considered. \textbf{MostPop} is a non-personalized method and recommends most popular items to all users; and
\textbf{ItemKNN} is a $K$-nearest neighborhood based method recommending items based on item similarity. We adapt it for implicit feedback data by following~\cite{hu2008collaborative}, and adopt cosine similarity. 
In terms of LFMs, \textbf{BPRMF} is selected as the representative of matrix factorization method (WRMF, eALS and DCF are remained for future exploration);
\textbf{BPRFM} (factorization machine) considers the second-order feature interactions between inputs and we train it by optimizing the BPR loss~\cite{rendle2009bpr};
\textbf{SLIM}~\cite{ning2011slim} learns a sparse item similarity matrix by minimizing a constrained reconstruction square loss; and
\textbf{PureSVD} directly performs conventional singular
value decomposition on the user-item implicit interaction matrix, where all the unobserved entries are set as 0.  
Regarding DLMs, \textbf{NeuMF}~\cite{he2017neural} is a state-of-the-art neural method taking advantage of both generalized matrix factorization and multi-layer perceptron (MLP); \textbf{NFM}~\cite{he2017neural} seamlessly combines the linearity of FM in modelling second-order feature interactions and the non-linearity of neural network in modelling
higher-order feature interactions, and we train it by optimizing the BPR loss; \textbf{NGCF}~\cite{wang2019neural} leverages graph neural networks to capture the high-order connectivity in the user-item graph;
and \textbf{Multi-VAE}~\cite{liang2018variational} is a generative model with multinomial likelihood and extends variational autoencoders to collaborative filtering. CKE and CDAE are remained for future study, as CKE involves textual and visual information besides user-item interaction data; both CDAE and Multi-VAE are in the family of autoencoders, while Multi-VAE has proven to be a stronger baseline~\cite{dacrema2019we}. 


\begin{table}[t]
    \centering
    \scriptsize
    \addtolength{\tabcolsep}{-0pt}
    \renewcommand{\arraystretch}{1.05}
	\caption{Objective functions of different baselines.}
	\label{tab:summary_objective_functions}
	\vspace{-0.15in}
    \begin{tabular}{|l|l|lll|}
    \specialrule{.15em}{.05em}{.05em}
    Method &\multicolumn{1}{c|}{Origin} &\multicolumn{3}{c|}{To explore} \\
    \specialrule{.05em}{.05em}{.05em}
    BPRMF &$\mathcal{L}_{pai}$ + $f_{ll}$ &$\mathcal{L}_{poi}$ + $f_{cl}$
    &$\mathcal{L}_{pai}$ + $f_{tl}$
    &$\mathcal{L}_{pai}$ + $f_{hl}$\\
    BPRFM &$\mathcal{L}_{pai}$ + $f_{ll}$
    &$\mathcal{L}_{poi}$ + $f_{cl}$
    &$\mathcal{L}_{pai}$ + $f_{tl}$
    &$\mathcal{L}_{pai}$ + $f_{hl}$\\
    SLIM &$\mathcal{L}_{poi}$ + $f_{sl}$ &--&--&-- \\
    NeuMF &$\mathcal{L}_{poi}$ + $f_{cl}$
    &$\mathcal{L}_{pai}$ + $f_{ll}$
    &$\mathcal{L}_{pai}$ + $f_{tl}$
    &$\mathcal{L}_{pai}$ + $f_{hl}$ \\
    NFM &$\mathcal{L}_{pai}$ + $f_{ll}$
    &$\mathcal{L}_{poi}$ + $f_{cl}$
    &$\mathcal{L}_{pai}$ + $f_{tl}$
    &$\mathcal{L}_{pai}$ + $f_{hl}$ \\
    NGCF &$\mathcal{L}_{pai}$ + $f_{ll}$
    &$\mathcal{L}_{poi}$ + $f_{cl}$ 
    &$\mathcal{L}_{pai}$ + $f_{tl}$
    &$\mathcal{L}_{pai}$ + $f_{hl}$ \\
    Multi-VAE &$\mathcal{L}_{poi}$ + $f_{cl}$ &--&--&--\\
    \specialrule{.15em}{.05em}{.05em}
    \end{tabular}
    \vspace{-0.15in}
\end{table}

\subsubsection{Evaluation Metric Selection}\label{subsubsec:evluation_metrics}
The evaluation metrics vary a lot in different papers in the collection. Figure~\ref{fig:factor_distribution}(c) depicts the popularity of the used evaluation metrics. 
We thus adopt the top-6 evaluation metrics covering $99\%$ of the collected papers. That is to say, $99\%$ of these papers adopt at least one of the six metrics. 
They are \textbf{Precision}, \textbf{Recall}, Mean Average Precision (\textbf{MAP}), Hit Ratio (\textbf{HR}), Mean Reciprocal Rank (\textbf{MRR}) and Normalized Discounted Cumulative Gain (\textbf{NDCG}). In particular, the first four metrics intuitively measure whether a test item is present in the top-$N$ recommendation list, whilst the latter two accounts for the ranking positions of test items. Detailed formulas are given by Table~\ref{tab:evaluation_metrics} (Appendix), where $R(u), T(u)$ represent the recommendation set and the test set for user $u$, respectively; $rel_j=1/0$ indicates whether the item at rank $j$ is in the intersection of $R(u)$ and $T(u)$, i.e., $(R(u) \cap T(u))$; $\delta(x)=1$ if $x$ is true, otherwise 0; and IDCG means the maximum possible DCG through ideal ranking.

\subsection{Analysis on Model-dependent Hyper-factors}\label{subsec:dependent-factor}

\subsubsection{Loss Function Design}\label{subsubsec:objective_function}
Two types of objective functions are widely utilized by the collected papers: \textbf{point-wise} ($55\%$ of the collected papers) and \textbf{pair-wise} ($40\%$ of the collected papers). The former only relies on the accuracy of the prediction of individual preferences; whilst the latter
approximates ranking loss by considering the relative order
of the predictions for pairs of items. Regardless of which one is deployed, it is critical to properly exploit unobserved feedback within the model, as merely considering the observed feedback fails to account for the fact that feedback is not missing at random, thus being not suitable for top-N recommenders~\cite{wu2016collaborative}. 
Let $\mathcal{L}$ denote the objective function of recommendation task. The point- and pair-wise objectives are thus given by:
\begin{equation}\scriptsize
\begin{aligned}
    &\mathcal{L}_{poi} = \sum\nolimits_{(u,i)\in \mathcal{\widetilde{O}}}{f(r_{ui}, \hat{r}_{ui})} + \lambda\Omega(\mathbf{\Theta}); \\ &\mathcal{L}_{pai} = \sum\nolimits_{(u,i,j)\in \mathcal{\widetilde{O}}}{f(r_{uij}, \hat{r}_{uij})} + \lambda\Omega(\mathbf{\Theta}),
\end{aligned}
\end{equation}
where $\mathcal{\widetilde{O}} = \{\mathcal{O}^+ \cup \mathcal{O}^-\}$ is the augmented dataset with the unobserved user-item set $\mathcal{O}^- =\{(u,j)\vert r_{uj}=0\}$ in addition to the observed user-item set $\mathcal{O}^+ = \{(u,i)\vert r_{ui}=1\}$; $f(\cdot)$ is the loss function; $r_{ui}, \hat{r}_{ui}$ are the observed and estimated feedback of user $u$ on item $i$, respectively; $(u,i,j)$ is the triple meaning that $u$ prefers positive item $i$ to negative item $j$;
$r_{uij} = r_{ui} - r_{uj}, \hat{r}_{uij} = \hat{r}_{ui} - \hat{r}_{uj}$; $\Omega(\mathbf{\Theta})$ is the regularization term, whose impact is illustrated in Section~\ref{subsubsec:over-fitting}; and $\mathbf{\Theta}$ is the set of model parameters.

W.r.t. the loss function $f(\cdot)$, point-wise objective usually adopts square loss and cross-entropy (CE) loss, whereas pair-wise objective generally employs log loss, top-1 loss and hinge loss:
\begin{equation}
\scriptsize
\begin{aligned}
  &\mathcal{L}_{poi} =
    \begin{cases}
      \text{Square Loss} & f_{sl}=\frac{1}{2}(r_{ui}-\hat{r}_{ui})^2\\
      \text{CE Loss}  & f_{cl}=-r_{ui}log(\hat{r}_{ui}) - (1-r_{ui})log(1-\hat{r}_{ui})
    \end{cases} \\
    &\mathcal{L}_{pai} =
    \begin{cases}
      \text{Log Loss} & f_{ll} = -log(\sigma(\hat{r}_{uij}))\\
      \text{Top-1 Loss}  & f_{tl} = \sigma (-\hat{r}_{uij}) + \sigma(\hat{r}_{uj}^2)\\
      \text{Hinge Loss}  & f_{hl} = max (0, 1-\hat{r}_{uij}).
    \end{cases}
\end{aligned}
\end{equation}
Table~\ref{tab:summary_objective_functions} shows the original objectives used by BPRMF, BPRFM, SLIM, NeuMF, NFM, NGCF and Multi-VAE.
Besides, we vary different objectives on these baselines to further examine their respective impacts.
Note that MostPop, ItemKNN and PureSVD do not have objective functions; 
we did not consider the square loss, as it is more suitable for rating prediction task instead of ranking problem~\cite{rendle2009bpr};
Multi-VAE cannot be easily adapted with different objective functions; and we did not study the impacts of different objectives on SLIM due to its high complexity and low scalability, which will be discussed in Section~\ref{subsec:baseline_selection}.

\subsubsection{Negative Sampling Strategies}
As pointed out in Section~\ref{subsubsec:objective_function}, properly exploiting the unobserved feedback (i.e., negative samples) helps learn users' relative preferences and benefits more accurate top-$N$ recommendation. This can be further supported by the fact that around $65\%$ of the collected papers consider the unobserved feedback when designing objective functions regardless of point- and pair-wise ones. However, it is not practical to leverage all unobserved feedback in large volume, as most users only provide feedback for a small number of items. Negative sampling is, therefore, adopted to balance the efficiency and effectiveness. It is noteworthy that we follow majority studies~\cite{he2017neural,sun2018recurrent,wang2019kgat,wang2019multi} and directly treat the unobserved feedback as negative feedback. There may be different explanations behind the unobserved feedback~\cite{zhao2018interpreting}, but we leave it for further exploration. 

There are various kinds of negative sampling strategies. Specifically, \textbf{uniform sampler}~\cite{he2017neural}, where all unobserved items of each user are sampled with an equal probability, has been adopted by almost all papers in the collection. To better study the impact of negative sampling, we additionally consider item popularity-based samplers, which have also been adopted in recommendation~\cite{he2016fast,xu2019relation}.  \textbf{Low-popularity sampler} refers to that for each user, her unobserved items with a lower popularity are sampled with a higher probability. This is based on the assumption that
a user is less likely to prefer less popular items. \textbf{High-popularity sampler} is opposite to the low-popularity sampler, where the unobserved items of each user with a higher popularity are more likely to be sampled. The rationale behind is that if a user provides no feedback for a quite popular item favored by a large number of users, it indicates that she may be not into this item. Moreover, we also compare two types of hybrid samplers by leveraging both uniform and popularity samplers, namely \textbf{uniform+low-popularity sampler} and \textbf{uniform+high-popularity sampler}. In these cases, half of the unobserved items are sampled via the uniform sampler, while the rest half are sampled via popularity samplers.

\subsubsection{Parameter Initializer Selection}
There are normally a set of learnable parameters (e.g. user/item representation matrix and the network weights) for the recommendation models, ranging from early LFMs to recently emerged DLMs. A proper parameter initializer will assist in a faster model convergence and better model performance. Specifically, the core of LFMs is to learn accurate user and item representations, which are generally initialized based on either a \textbf{Uniform distribution} in the range of $(0, a)$ or a \textbf{Normal/Gaussian distribution} with zero mean and a variance of $\sigma^2$~\cite{zhang2020content,le2020stochastically}, given by,
\begin{equation}
\scriptsize
    \boldsymbol{v}_{uf}/\boldsymbol{v}_{vf} \sim \boldsymbol{U}(0, a);\;\;\;\; 
    \boldsymbol{v}_{uf}/\boldsymbol{v}_{vf} \sim \mathcal{N}(0, \sigma^2), 
\end{equation}
where the common settings are $a=1$; and $\sigma=0.01$.

Regarding DLMs, besides user and item representations, initializing with proper weights helps ensure the network to converge in a reasonable amount of time; otherwise the network loss function does not go anywhere even after hundreds of thousands of iterations. Given too small weights, the variance of the input diminishes as it passes through each layer in the network, and eventually drops to a really low value thus failing to work. Contrarily, a too-large weights leads to exploding gradients. Xavier initialization~\cite{glorot2010understanding} has been proven as an effective fashion, and widely adopted by DLMs in recommendation~\cite{sun2020neighbor,wang2020ckan,wang2020disentangled,sun2020lara}. Typically, the weights are also initialized based on either a Uniform or Normal distribution, defined as\footnote{\url{https://pytorch.org/docs/stable/nn.init.html}},
\begin{equation}
\scriptsize
    \boldsymbol{W}_{ij} \sim \boldsymbol{U}(-\frac{\sqrt{6}}{\sqrt{n_{\text{in}} + n_{\text{out}}}}, \frac{\sqrt{6}}{\sqrt{n_{\text{in}} + n_{\text{out}}}}); \;
    \boldsymbol{W}_{ij} \sim \mathcal{N}(0, \frac{2}{n_{\text{in}} + n_{\text{out}}}),
\end{equation}
where \begin{small}$\boldsymbol{W}_{ij}$\end{small} is the network weight; $n_{\text{in}}, n_{\text{out}}$ are the number of input and output neurons, respectively. 

By analyzing the collected papers, we find that around 59\% of them do not report the parameter initializers. Among the papers mentioning parameter initializers, 36\% of them are based on a Normal distribution; 10\% of them use a Uniform distribution; 18\% of them adopt the Xavier Initialization; 13\% of them employ the pre-trained embeddings (e.g., via BPRMF) for initialization; and the rest 23\% ultilize other methods. 
The impacts of different parameter initializers are investigated in Section~\ref{subsec:parameter-initializer}.

\subsubsection{Model Optimizer Selection}
Opitimizer is used to update model parameters, thus minimizing the loss function; meanwhile loss function acts as guides to the terrain telling optimizer if it is moving in the right direction to reach the bottom of the valley, i.e., global minimum. 
Different optimizers also affect the recommendation performance. By looking into the collected papers, we find that 23\% of them do not report their optimizers. Among the papers mentioning used optimizers, 50\% of them adopt {Adam}~\cite{liang2018variational,wu2016collaborative,he2017neural}; 23\% of them use {SGD}~\cite{rendle2009bpr,rendle2010factorization}; and the rest 27\% employ other optimizers (e.g., ALS~\cite{tran2018regularizing}, AdaGrad~\cite{xin2019relational} and RMSProp~\cite{zheng2018spectral}).

Here, we discuss six commonly-selected optimizers as shown in Table~\ref{tab:optimizer} (Appendix). (1) Gradient Descent (\textbf{GD}) is a first-order optimization algorithm which is dependent on the first order derivative of a loss function. The parameters are then updated in the negative gradient direction to minimize the loss; (2) Stochastic Gradient Descent (\textbf{SGD}) is a variant of GD to update the model’s parameters after computation of loss on each training example, whilst parameters are changed after calculating gradient on the whole dataset by GD; (3) Mini Batch Stochastic Gradient Descent (\textbf{MB-SGD}) is an improvement on both SGD and standard GD, where the dataset is divided into various batches and after every batch, the parameters are updated; (4) Adaptive Gradient (\textbf{AdaGrad}) is an algorithm for gradient-based optimization that adapts the learning rate to the parameters, performing smaller updates
(i.e. low learning rates) for frequently parameters, whilst larger updates (i.e. high learning rates) for infrequent parameters;
(5) Root Mean Square Propagation (\textbf{RMSProp}) is devised to resolve AdaGrad’s radically diminishing learning rates, and divides the learning rate by the average of the exponential decay of squared gradients; 
and (6) Adaptive Moment Estimation (\textbf{Adam}) calculates the adaptive learning rate for each parameter from estimates of first and second moments of the gradients. In addition to the decaying average of past squared gradients like RMSprop, it also keeps a decaying average of past gradients.

\subsubsection{Strategies to Avoid Over-fitting}\label{subsubsec:over-fitting}
In machine learning, different strategies are exploited to combat the issue of over-fitting, which refers to the model over-fits the training data, thus achieving poor performance on the validation or test data. As a matter of fact, the most widely used regularization techniques include regularization terms, dropout and early-stop mechanism. 

\smallskip\noindent\textit{Regularization.} It is generally integrated into the loss function, so as to help avoid over-fitting while training a recommendation model. Two types of terms are mainly adopted, namely $L1$ and $L2$ regularization (i.e. norm). $L1$ norm is also known as Manhattan Distance, which is the most natural way of measure distance between vectors. It is the sum of the magnitudes of the vectors in a space, where all the components of the vector are weighted equally. $L2$ norm is the most popular norm, also known as the Euclidean norm, which is the shortest distance between two points. Different from $L1$ norm, each component of the vector is squared for $L2$ norm, indicating that the outliers have more weighting, so it can skew results. 
The main difference between the $L1$ and $L2$ regularization lies in (1) $L1$ regularization attempts to estimate the median of the data, whereas $L2$ regularization tries to estimate the mean of the data to avoid over-fitting; and (2) $L1$ regularization helps in feature selection by eliminating less important features,
which is helpful given a large number of feature points.

\smallskip\noindent\textit{Dropout.} It has been widely adopted in DLMs to help avoid over-fitting~\cite{srivastava2014dropout}. The key idea is to randomly drop units (along with their connections) from the neural network during training, which prevents units from co-adapting too much. Hence, an extra hyper-parameter, i.e., the probability of retaining a unit $p$, is introduced, controlling the intensity of dropout. For instance, $p = 1$ implies no dropout and low values of $p$ mean more dropout. Smaller $p$ could lead to under-fitting; whereas large $p$ may not produce enough dropout to prevent over-fitting. Typical values of $p$ for hidden units are in the range 0.5 to 0.8~\cite{srivastava2014dropout}.

\smallskip\noindent\textit{Early-stop Mechanism.}
Early-stop is also a form of regularization used to avoid over-fitting. 
A major issue with training recommenders (e.g., LFMs and DLMs) is in the choice of the number of training epochs to use. Too many epochs can lead to over-fitting of the training dataset, whereas too few may result in an under-fit model. Early-stop is a method that allows us to specify an arbitrary large number of training epochs and stop training once the model performance stops improving on a hold out validation dataset. To be more specific, if the validation loss stops decreasing for several epochs in a row, the training stops. 
Through analyzing on the collected papers, only 11\% of them point out early-stop strategy is adopted in their papers. 
In our study, we also investigate the impacts of ealry-stop mechanism on recommendation evaluation.

\subsubsection{Hyper-parameter Tuning Strategies}\label{subsubsec:parameter-tuning}
Hyper-parameter tuning, including validation and searching strategies, plays a vital role in the training process of recommendation approaches, and greatly influences the final recommendation performance. 

\smallskip\noindent\textit{Validation Strategy.} 
Through the paper analysis, we notice that 
more than 33\% of papers directly tune hyper-parameters according to the performance on the test set. That is to say, they use the same data to tune model parameters and evaluate the model performance. Information may thus leak into the model and overfit the historical data. As a matter of fact, besides the training and test sets, an extra validation set should be retained to help tune the hyper-parameters, which is called \textit{nested validation}\footnote{\url{https://scikit-learn.org/stable/auto_examples/model_selection/plot_nested_cross_validation_iris.html}}. With nested validation, the optimal hyper-parameter settings are obtained when the model achieves the best performance on the validation set.
By doing so, the information leak issue is well avoided in the model training and evaluation process. Therefore, in our study, we adopt the nested validation strategy. To be more specific, we further select \textbf{10\%} of records from the training set as the validation set for split-by-ratio; and for leave-one-out, we retain one record from the training set as the validation set to tune hyper-parameters. Finally, the performance on the test set is reported. Due to the computational requirements of certain of baselines, we were unable to search the hyper-parameter space for cross-validation in a reasonable amount of time.

\begin{figure*}[t]
    \centering
    \includegraphics[width=1.0\textwidth]{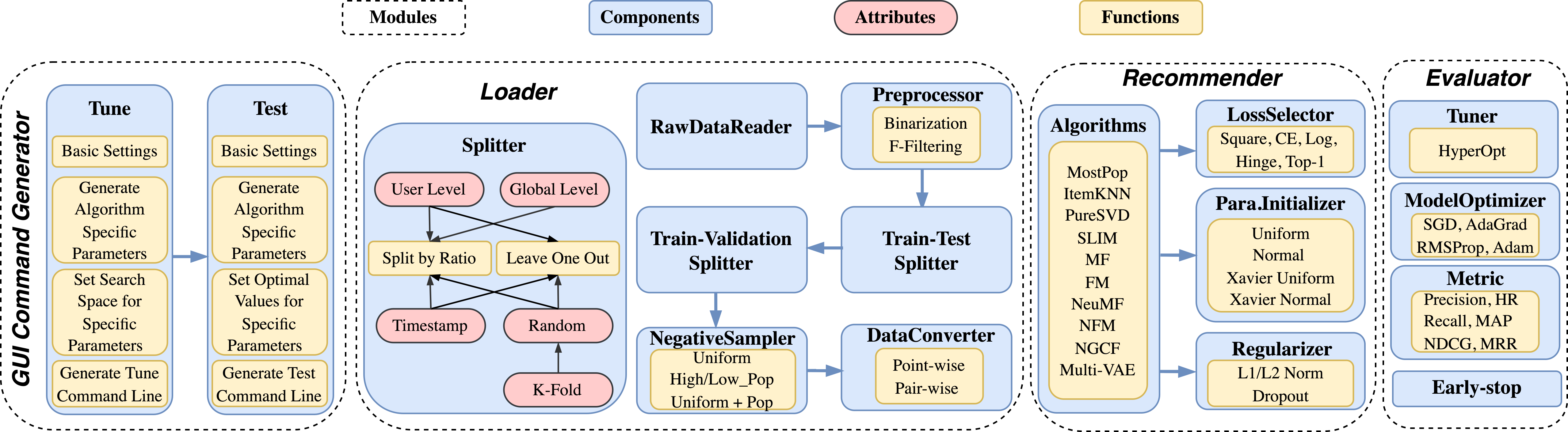}
    \vspace{-0.2in}
    \caption{The overall structure of DaisyRec 2.0, composed of four components, i.e., GUI Command Generator, Loader, Recommender, and Evaluator.}\label{fig:framework}
    \vspace{-0.15in}
\end{figure*}

\smallskip\noindent\textit{Searching Strategy.} 
From our observation, almost all collected papers employ 
the most straightforward and simple method -- \textbf{grid search}~\cite{he2017neural,wang2019kgat} to find out the optimal parameter settings. 
In particular, each hyper-parameter is provided with a set of possible values (i.e., search space) based on the prior-knowledge, and the optimal setting is then obtained by traversing the entire search space. 
Suppose a model has $m$ parameters, where each parameter has an average of $n$ possible values, the model needs to be executed for $n^m$ times to find out optimal settings for all parameters.
Hence, grid search is more suitable for models with less hyper-parameters; otherwise, it may suffer from the combination explosion issue.
To improve the tuning efficiency, other strategies have been introduced. Given the search space of each parameter, \textbf{random search}~\cite{bergstra2012random} randomly chooses trials for a pre-defined times (e.g., 30) instead of traversing the entire search space. It is able to find models that are as good or slightly worse but within a smaller fraction of the computation time. In contrast, \textbf{Bayesian HyperOpt}~\cite{snoek2012practical} is not a brute force but more intelligent technique compared to grid and random search. 
It makes use of information from past trials to inform the next set of hyper-parameters to explore, while not compromising the quality of the results~\cite{dacrema2019we}. Therefore, for each baseline, we adopt Bayesian HyperOpt to perform hyper-parameter optimization on \textbf{NDCG}, which is the most popular metrics as shown in Figure~\ref{fig:factor_distribution}(c); and other metrics are expected to be simultaneously optimized with the optimal results on NDCG.

\subsection{Categorization on Evaluation Modes}\label{subsec:categorization-evaluation-modes}
Based on the model-independent and model-dependent hyper-factors introduced in Sections~\ref{subsec:independent-factor}-\ref{subsec:dependent-factor}, we define four modes of rigorous evaluation as below.
\begin{itemize}[leftmargin=*]
    \item \textbf{Relax Mode} keeps exactly the same settings for all model-independent hyper-factors and follows the original settings as per individual approach for model-dependent hyper-factors. 
    \item \textbf{Hard-strict Mode} keeps exactly the same settings for both model-independent and model-dependent hyper-factors for all approaches.
    \item \textbf{Soft-strict Mode} keeps exactly the same settings for all model-independent hyper-factors and empirically finds out the optimal settings for per individual approach for model-dependent hyper-factors.
    \item \textbf{Mixed Mode} keeps exactly the same settings for all model-independent hyper-factors; while applies hard-strict mode on
    some model-dependent hyper-factors (e.g., the same initializer/optimizer), and relax or soft-strict modes on
    the others (e.g., empirically searching the optimal hyper-parameter settings for different baselines). 
\end{itemize}
Through analysis, we find that most of the collected papers adopt the mixed mode for evaluation~\cite{sun2020neighbor,sun2020multi,wang2020disenhan}, for example, using the same model optimizer and parameter initializer for all approaches; adopting different loss functions as indicated in the original papers; and empirically finding out best parameter settings for all approaches.
Regardless of different modes, it is essential to keep exactly the same settings for all model-independent hyper-factors for a rigorous evaluation. W.r.t. the model-dependent hyper-factors, different modes have their own pros and cons. 
\begin{itemize}[leftmargin=*]
    \item Relax mode can extremely reduce the cost on exploring the optimal performance. However, it relies on the original settings indicated by the individual approach, which are not always available and may lead to a unfair comparison, e.g., one model defeats the others merely because it adopts a different loss function.
    \item Hard-strict mode ensures a fair comparison among different baselines, while it may not always be reasonable to, e.g., have the same settings for all shared hyper-parameters for all baselines, as the optimal hyper-parameter settings for different baselines may vary a lot across different datasets.
    \item Soft-strict mode could help find out the optimal performance for per individual approach, which, however, may be quite expensive due to the large amount of combinations of model-dependent hyper-factors.
    \item Mixed mode would be a better balance among complexity, performance and fairness for per individual approach. For instance, soft-strict mode can be applied regarding, e.g., optimal hyper-parameter settings, to maintain model performance; hard-strict mode can be used for, e.g., parameter initializer and model optimizer, to ensure fair comparison and less exploration complexity. 
\end{itemize}
In summary, \textbf{mixed mode} could be the most practical way for achieving rigorous evaluation in recommendation.

\section{Experimental Study}\label{sec:experiement}

\subsection{Introduction to DaisyRec 2.0}
To support the empirical study, we release a user-friendly Python toolkit named as
\textbf{DaisyRec 2.0}, where Daisy is short for `Multi-\textbf{D}imension f\textbf{AI}rly comp\textbf{A}r\textbf{I}son for recommender \textbf{SY}stem'. 
Different from existing open-source libraries (e.g., LibRec~\cite{guo2015librec}, OpenRec~\cite{yang2018openrec} and 
DeepRec~\cite{zhang2019deeprec}), which mainly aim to reproduce various state-of-the-art recommenders, DaisyRec 2.0 is designed with the goal of performing rigorous evaluation in recommendation by seamlessly accommodating the extracted hyper-factors in Section~\ref{sec:theoretical_study}. 
It is built upon the widely-used deep learning framework Pytorch (\url{pytorch.org}), and Figure~\ref{fig:framework} depicts its overall structure consisting of four modules: GUI Command Generator, Loader, Recommender and Evaluator. 

\begin{figure}[!ht]
    \vspace{-0.1in}
    \centering
    \includegraphics[width=0.48\textwidth]{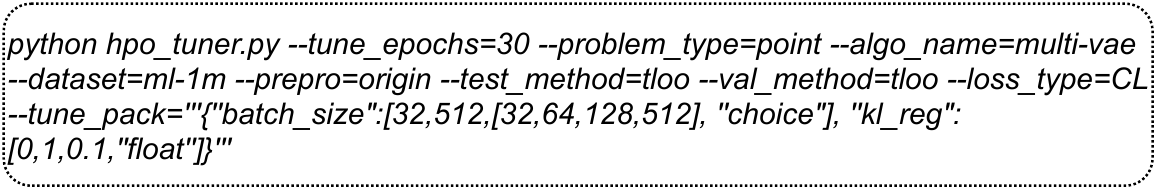}
    \vspace{-0.15in}
    \caption{An example of the generated tune command for Multi-VAE.}
    \label{fig:command}
    \vspace{-0.05in}
\end{figure}

In particular, GUI Command Generator\footnote{\url{http://DaisyRecGuiCommandGenerator.pythonanywhere.com}} is used to help generate tune and test commands in a more user-friendly fashion. Taking the tune command generator as an example, users first need to select values for the basic settings (e.g., algorithm name and dataset) from a drop-down menu. Based on the selected algorithm, it then automatically displays the algorithm-specific parameters (e.g., KL regularization for Multi-VAE). Accordingly, users can select and set the search space for the algorithm-specific parameters. Lastly, it generates the corresponding tune command (shown in Figure~\ref{fig:command}) based on all selected settings, which can be directly copied and pasted into the terminal to run. 

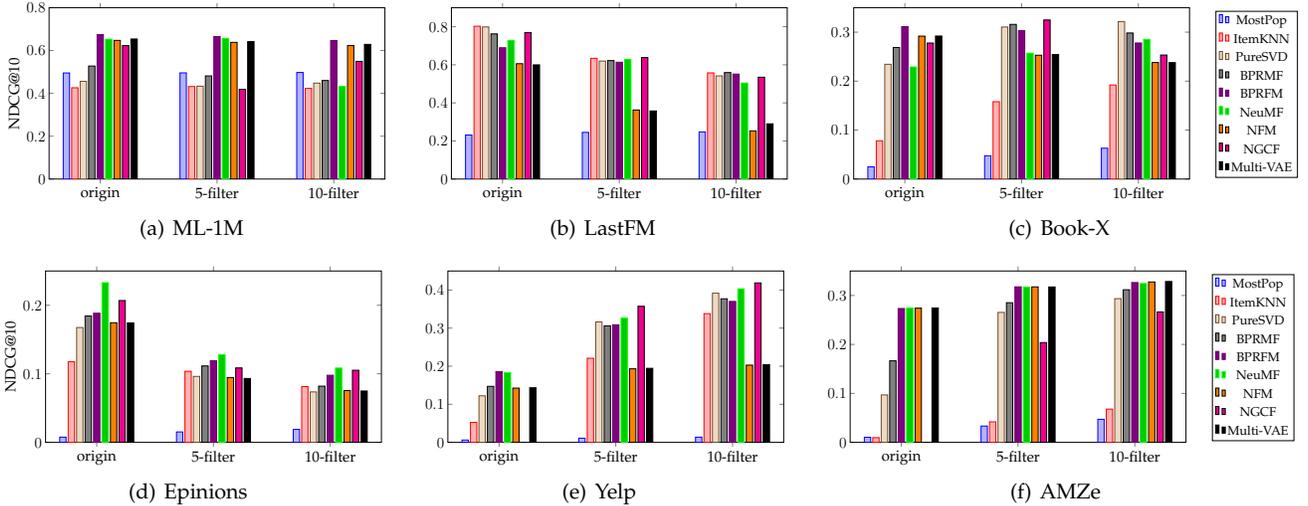
\begin{figure*}[t]
\centering
\subfigure[ML-1M]{
\begin{tikzpicture}[scale=0.4]
\pgfplotsset{%
    width=0.7\textwidth,
    height=0.4\textwidth
}
\begin{axis}[
    ybar,
    bar width=6pt,
    ylabel={NDCG@10},
    ylabel style ={font = \Large},
    xlabel style ={font = \Large},
    enlarge x limits={abs=1.7cm},
    scaled ticks=false,
    tick label style={/pgf/number format/fixed, font=\Large},
    ymin=0, ymax=0.8,
    symbolic x coords={origin, 5-filter, 10-filter},
    xtick=data,
    ytick={0,0.2,0.4,0.6,0.8},
    legend style={at={(0.75,0.98)}, anchor=north,legend columns=2, font=\large},
]
\addplot coordinates {
(origin,0.4952)(5-filter, 0.4957) (10-filter,0.4975)};
\addplot coordinates {
(origin,0.4261)(5-filter, 0.4319) (10-filter,0.4233)};   
\addplot coordinates {
(origin,0.4562)(5-filter, 0.4334) (10-filter,0.4479)};
\addplot coordinates {
(origin,0.5275)(5-filter, 0.4814) (10-filter,0.4601)};
\addplot coordinates {
(origin,0.6745)(5-filter, 0.6651) (10-filter,0.6466)};
\addplot coordinates {
(origin,0.6528)(5-filter, 0.6568) (10-filter,0.4322)};
\addplot[fill=orange] coordinates {
(origin,0.6473)(5-filter, 0.6381) (10-filter,0.6232)};
\addplot[fill=magenta] coordinates {
(origin,0.6230)(5-filter, 0.4186) (10-filter,0.5494)};
\addplot[fill=black] coordinates {
(origin,0.6538)(5-filter, 0.6415) (10-filter,0.6287)};
\end{axis}
\end{tikzpicture}}
\hspace{0.1in}
\subfigure[LastFM]{
\begin{tikzpicture}[scale=0.4]
\pgfplotsset{%
    width=0.7\textwidth,
    height=0.4\textwidth
}
\begin{axis}[
    ybar,
    bar width=6pt,
    ylabel style ={font = \Large},
    xlabel style ={font = \Large},
    enlarge x limits={abs=1.7cm},
    scaled ticks=false,
    tick label style={/pgf/number format/fixed, font=\Large},
    ymin=0, ymax=0.9,
    symbolic x coords={origin, 5-filter, 10-filter},
    xtick=data,
    ytick={0,0.2,0.4,0.6,0.8},
    legend style={at={(1.2,0.98)}, anchor=north,legend columns=1, row sep=5pt},
]
\addplot coordinates {
(origin,0.2311)(5-filter, 0.2450) (10-filter,0.2476)};
\addplot coordinates {
(origin,0.8026)(5-filter, 0.6341) (10-filter,0.5577)};   
\addplot coordinates {
(origin,0.7989)(5-filter, 0.6197) (10-filter,0.5418)};
\addplot coordinates {
(origin,0.7624)(5-filter, 0.6222) (10-filter,0.5592)};
\addplot coordinates {
(origin,0.6897)(5-filter, 0.6135) (10-filter,0.5513)};
\addplot coordinates {
(origin,0.7279)(5-filter, 0.6284) (10-filter,0.5044)};
\addplot[fill=orange] coordinates {
(origin,0.6058)(5-filter, 0.3625) (10-filter,0.2532)};
\addplot[fill=magenta] coordinates {
(origin,0.7688)(5-filter, 0.6381) (10-filter,0.5349)};
\addplot[fill=black] coordinates {
(origin,0.5998)(5-filter, 0.3577) (10-filter,0.2901)};
\end{axis}
\end{tikzpicture}}
\hspace{0.1in}
\subfigure[Book-X]{
\begin{tikzpicture}[scale=0.4]
\pgfplotsset{%
    width=0.7\textwidth,
    height=0.4\textwidth
}
\begin{axis}[
    ybar,
    bar width=6pt,
    ylabel style ={font = \Large},
    xlabel style ={font = \Large},
    enlarge x limits={abs=1.7cm},
    scaled ticks=false,
    tick label style={/pgf/number format/fixed, font=\Large},
    ymin=0, ymax=0.35,
    symbolic x coords={origin, 5-filter, 10-filter},
    xtick=data,
    ytick={0,0.1,0.2,0.3,0.4},
    legend style={at={(1.2,0.98)}, anchor=north,legend columns=1, row sep=3.5pt, font=\large},
]
\addplot coordinates {
(origin,0.0249)(5-filter, 0.0476) (10-filter,0.0630)};
\addplot coordinates {
(origin,0.0782)(5-filter, 0.1580) (10-filter,0.1921)};   
\addplot coordinates {
(origin,0.2344)(5-filter, 0.3105) (10-filter,0.3216)};
\addplot coordinates {
(origin,0.2687)(5-filter, 0.3159) (10-filter,0.2983)};
\addplot coordinates {
(origin,0.3115)(5-filter, 0.3033) (10-filter,0.2779)};
\addplot coordinates {
(origin,0.2289)(5-filter, 0.2571) (10-filter,0.2852)};
\addplot[fill=orange] coordinates {
(origin,0.2919)(5-filter, 0.2529) (10-filter,0.2380)};
\addplot[fill=magenta] coordinates {
(origin,0.2778)(5-filter, 0.3249) (10-filter,0.2533)};
\addplot[fill=black] coordinates {
(origin,0.2921)(5-filter, 0.2542) (10-filter,0.2378)};
\legend{MostPop, ItemKNN, PureSVD, BPRMF, BPRFM, NeuMF, NFM, NGCF, Multi-VAE}
\end{axis}
\end{tikzpicture}}
\subfigure[Epinions]{
\begin{tikzpicture}[scale=0.4]
\pgfplotsset{%
    width=0.7\textwidth,
    height=0.4\textwidth
}
\begin{axis}[
    ybar,
    bar width=6pt,
    ylabel={NDCG@10},
    ylabel style ={font = \Large},
    xlabel style ={font = \Large},
    enlarge x limits={abs=1.7cm},
    scaled ticks=false,
    tick label style={/pgf/number format/fixed, font=\Large},
    ymin=0, ymax=0.25,
    symbolic x coords={origin, 5-filter, 10-filter},
    xtick=data,
    ytick={0,0.1,0.2},
    legend style={at={(1.2,0.98)}, anchor=north,legend columns=1, row sep=5pt},
]
\addplot coordinates {
(origin,0.0076)(5-filter, 0.0152) (10-filter,0.0189)};
\addplot coordinates {
(origin,0.1176)(5-filter, 0.1037) (10-filter,0.0813)};   
\addplot coordinates {
(origin,0.1675)(5-filter, 0.0962) (10-filter,0.0736)};
\addplot coordinates {
(origin,0.1844)(5-filter, 0.1115) (10-filter,0.0820)};
\addplot coordinates {
(origin,0.1884)(5-filter, 0.1192) (10-filter,0.0979)};
\addplot coordinates {
(origin,0.2329)(5-filter, 0.1279) (10-filter,0.1084)};
\addplot[fill=orange] coordinates {
(origin,0.1745)(5-filter, 0.0946) (10-filter,0.0756)};
\addplot[fill=magenta] coordinates {
(origin,0.2069)(5-filter, 0.1086) (10-filter,0.1052)};
\addplot[fill=black] coordinates {
(origin,0.1743)(5-filter, 0.0932) (10-filter,0.0749)};
\end{axis}
\end{tikzpicture}}
\hspace{0.1in}
\subfigure[Yelp]{
\begin{tikzpicture}[scale=0.4]
\pgfplotsset{%
    width=0.7\textwidth,
    height=0.4\textwidth
}
\begin{axis}[
    ybar,
    bar width=6pt,
    ylabel style ={font = \Large},
    xlabel style ={font = \Large},
    enlarge x limits={abs=1.7cm},
    scaled ticks=false,
    tick label style={/pgf/number format/fixed, font=\Large},
    ymin=0, ymax=0.45,
    symbolic x coords={origin, 5-filter, 10-filter},
    xtick=data,
    ytick={0,0.1,0.2,0.3,0.4},
    legend style={at={(0.75,0.98)}, anchor=north,legend columns=2},
]
\addplot coordinates {
(origin,0.0061)(5-filter, 0.0109) (10-filter,0.0137)};
\addplot coordinates {
(origin,0.0524)(5-filter, 0.2208) (10-filter,0.3380)};   
\addplot coordinates {
(origin,0.1222)(5-filter, 0.3162) (10-filter,0.3918)};
\addplot coordinates {
(origin,0.1471)(5-filter, 0.3060) (10-filter,0.3770)};
\addplot coordinates {
(origin,0.1859)(5-filter, 0.3088) (10-filter,0.3701)};
\addplot coordinates {
(origin,0.1827)(5-filter, 0.3264) (10-filter,0.4032)};
\addplot[fill=orange] coordinates {
(origin,0.1425)(5-filter, 0.1931) (10-filter,0.2028)};
\addplot[fill=magenta] coordinates {
(origin,0.0)(5-filter, 0.3573) (10-filter,0.4187)};
\addplot[fill=black] coordinates {
(origin,0.1433)(5-filter, 0.1944) (10-filter,0.2041)};
\end{axis}
\end{tikzpicture}}
\hspace{0.1in}
\subfigure[AMZe]{
\begin{tikzpicture}[scale=0.4]
\pgfplotsset{%
    width=0.7\textwidth,
    height=0.4\textwidth
}
\begin{axis}[
    ybar,
    bar width=6pt,
    ylabel style ={font = \Large},
    xlabel style ={font = \Large},
    enlarge x limits={abs=1.7cm},
    scaled ticks=false,
    tick label style={/pgf/number format/fixed, font=\Large},
    ymin=0, ymax=0.35,
    symbolic x coords={origin, 5-filter, 10-filter},
    xtick=data,
    ytick={0,0.1,0.2,0.3,0.4},
    legend style={at={(1.2,0.98)}, anchor=north,legend columns=1, row sep=3.5pt, font=\large},
]
\addplot coordinates {
(origin,0.0104)(5-filter, 0.0333) (10-filter,0.0470)};
\addplot coordinates {
(origin,0.0096)(5-filter, 0.0419) (10-filter,0.0679)};   
\addplot coordinates {
(origin,0.0969)(5-filter, 0.2654) (10-filter,0.2935)};
\addplot coordinates {
(origin,0.1665)(5-filter, 0.2852) (10-filter,0.3117)};
\addplot coordinates {
(origin,0.2734)(5-filter, 0.3177) (10-filter,0.3264)};
\addplot coordinates {
(origin,0.2745)(5-filter, 0.3172) (10-filter,0.3244)};
\addplot[fill=orange] coordinates {
(origin,0.2741)(5-filter, 0.3176) (10-filter,0.3276)};
\addplot[fill=magenta] coordinates {
(origin, 0.0)(5-filter, 0.2038) (10-filter,0.2664)};
\addplot[fill=black] coordinates {
(origin,0.2745)(5-filter, 0.3174) (10-filter,0.3285)};
\legend{MostPop, ItemKNN, PureSVD, BPRMF, BPRFM, NeuMF, NFM, NGCF, Multi-VAE}
\end{axis}
\end{tikzpicture}}
\vspace{-0.1in}
\caption{Performance of baselines w.r.t. time-aware split-by-ratio on the six datasets across origin, 5- and 10-filter settings.
}\label{fig:data-preprocessing}
\vspace{-0.15in}
\end{figure*}

Loader mainly aims to: (1) load and pre-process the dataset; (2) split it into training and test sets based on the selected Splitter; (3) divide validation set from training set by choosing proper Splitter according to Step 2; (4) sample negative items for training by choosing different samplers; and (5) convert the data into the specific format to fit the Recommender. 
Four components are included in 
Recommender, where `Algorithms' implements the ten selected state-of-the-arts in Section~\ref{subsubsec:comparison_baselines} (more recommenders will be implemented); `LossSelector' makes it flexible to change different objective functions for the algorithms; `ParameterInitializer' allows to select different initialization methods (e.g., Xavier uniform distribution); and `Regularizer' provides options for different regularization terms to avoid overfitting (e.g., $L1$ and $L2$).
Evaluator is equipped with `Tuner', `ModelOptimizer', `Metric', and `Early-stop', where `Tuner' helps accomplish hyper-parameter optimization; `ModelOptimizer' provides options for different optimizers; `Metric' implements the classic ranking metrics, e.g., Precision; and `Early-stop' helps further avoid over-fitting.

To sum up, all modules in DaisyRec 2.0 are wrapped friendly for users to deploy, and new algorithms can be easily added into this extensible and adaptable framework. We keep DaisyRec 2.0 updated by adding more features.

\subsection{Analysis on Model-independent Hyper-Factors}

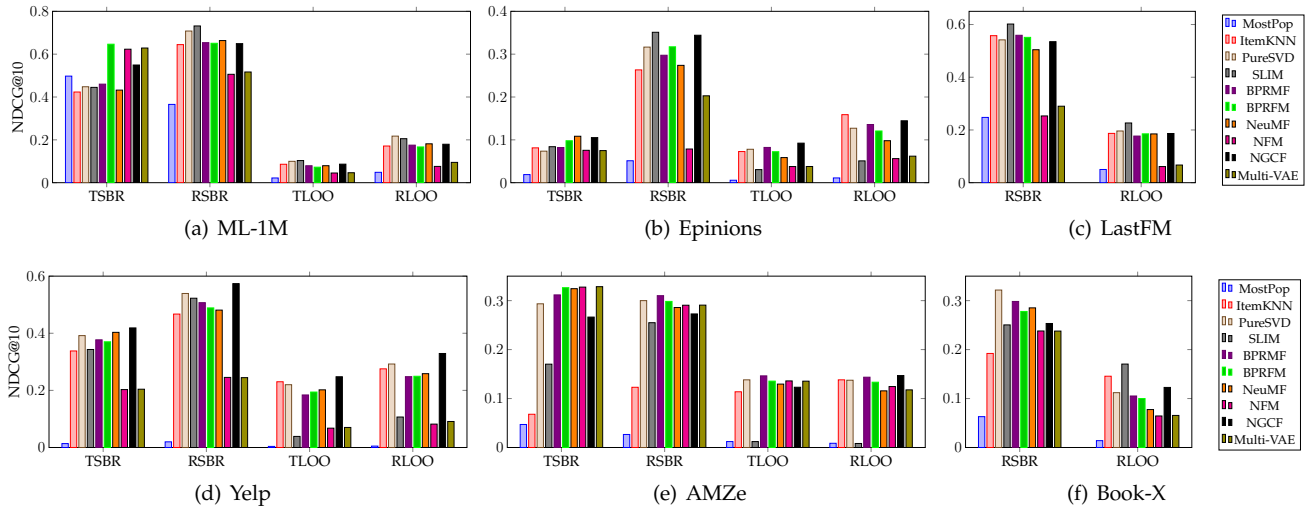
\begin{figure*}[t]
\centering
\subfigure[ML-1M]{
\begin{tikzpicture}[scale=0.4]
\pgfplotsset{%
    width=0.85\textwidth,
    height=0.4\textwidth
}
\begin{axis}[
    ybar,
    bar width=6pt,
    ylabel={NDCG@10},
    ylabel style ={font = \Large},
    xlabel style ={font = \Large},
    enlarge x limits={abs=1.8cm},
    scaled ticks=false,
    tick label style={/pgf/number format/fixed, font=\Large},
    ymin=0, ymax=0.8,
    symbolic x coords={TSBR, RSBR, TLOO, RLOO},
    xtick=data,
    ytick={0,0.2,0.4,0.6,0.8},
    legend style={at={(0.75,0.98)}, anchor=north,legend columns=2},
]
\addplot coordinates {
(TSBR,0.4975) (RSBR,0.3655)
(TLOO, 0.0220) (RLOO,0.0484)};
\addplot coordinates {
(TSBR,0.4233) (RSBR,0.6446)
(TLOO, 0.0860) (RLOO,0.1712)};  
\addplot coordinates {
(TSBR,0.4479) (RSBR,0.7080)
(TLOO, 0.0997) (RLOO,0.2174)};
\addplot coordinates {
(TSBR,0.4450) (RSBR,0.7317)
(TLOO, 0.1033) (RLOO,0.2054)};
\addplot coordinates {
(TSBR,0.4601) (RSBR,0.6537)
(TLOO, 0.0791) (RLOO,0.1753)};
\addplot coordinates {
(TSBR,0.6466) (RSBR,0.6509)
(TLOO, 0.0725) (RLOO,0.1669)};
\addplot[fill=orange] coordinates {
(TSBR,0.4322) (RSBR,0.6635)
(TLOO, 0.0793) (RLOO,0.1815)};
\addplot[fill=magenta] coordinates {
(TSBR,0.6232) (RSBR,0.5060)
(TLOO, 0.0452) (RLOO,0.0760)};
\addplot[fill=black] coordinates {
(TSBR,0.5494) (RSBR,0.6491)
(TLOO, 0.0865) (RLOO,0.1795)};
\addplot[fill=olive] coordinates {
(TSBR,0.6287) (RSBR,0.5168)
(TLOO, 0.0465) (RLOO,0.0948)};
\end{axis}
\end{tikzpicture}}
\subfigure[Epinions]{
\begin{tikzpicture}[scale=0.4]
\pgfplotsset{%
    width=0.85\textwidth,
    height=0.4\textwidth
}
\begin{axis}[
    ybar,
    bar width=6pt,
    ylabel style ={font = \Large},
    xlabel style ={font = \Large},
    enlarge x limits={abs=1.8cm},
    scaled ticks=false,
    tick label style={/pgf/number format/fixed, font=\Large},
    ymin=0, ymax=0.4,
    symbolic x coords={TSBR, RSBR, TLOO, RLOO},
    xtick=data,
    ytick={0,0.1,0.2,0.3,0.4},
    legend style={at={(1.2,0.98)}, anchor=north,legend columns=1, row sep=3pt},
]
\addplot coordinates {
(TSBR,0.0189) (RSBR,0.0512)
(TLOO, 0.0057) (RLOO,0.0111)};
\addplot coordinates {
(TSBR,0.0813) (RSBR,0.2633)
(TLOO, 0.0727) (RLOO,0.1588)};  
\addplot coordinates {
(TSBR,0.0736) (RSBR,0.3165)
(TLOO, 0.0780) (RLOO,0.1271)};
\addplot coordinates {
(TSBR,0.0840) (RSBR,0.3511)
(TLOO, 0.0305) (RLOO,0.0509)};
\addplot coordinates {
(TSBR,0.0820) (RSBR,0.2975)
(TLOO, 0.0823) (RLOO,0.1358)};
\addplot coordinates {
(TSBR,0.0979) (RSBR,0.3172)
(TLOO, 0.0723) (RLOO,0.1207)};
\addplot[fill=orange] coordinates {
(TSBR,0.1084) (RSBR,0.2739)
(TLOO, 0.0586) (RLOO,0.0980)};
\addplot[fill=magenta] coordinates {
(TSBR,0.0756) (RSBR,0.0786)
(TLOO, 0.0377) (RLOO,0.0562)};
\addplot[fill=black] coordinates {
(TSBR,0.1052) (RSBR,0.3442)
(TLOO, 0.0923) (RLOO,0.1445)};
\addplot[fill=olive] coordinates {
(TSBR,0.0749) (RSBR,0.2028)
(TLOO, 0.0378) (RLOO,0.0619)};
\end{axis}
\end{tikzpicture}}
\subfigure[LastFM]{
\begin{tikzpicture}[scale=0.4]
\pgfplotsset{%
    width=0.5\textwidth,
    height=0.4\textwidth
}
\begin{axis}[
    ybar,
    bar width=6pt,
    ylabel style ={font = \Large},
    xlabel style ={font = \Large},
    enlarge x limits={abs=1.8cm},
    scaled ticks=false,
    tick label style={/pgf/number format/fixed, font=\Large},
    ymin=0, ymax=0.65,
    symbolic x coords={RSBR, RLOO},
    xtick=data,
    ytick={0,0.2,0.4,0.6},
    legend style={at={(1.3,0.98)}, anchor=north,legend columns=1, row sep=2pt, font=\large},
]
\addplot coordinates {
(RSBR,0.2476) 
(RLOO,0.0501)};
\addplot coordinates {
(RSBR,0.5577)
(RLOO,0.1870)};  
\addplot coordinates {
(RSBR,0.5418)
(RLOO,0.1961)};
\addplot coordinates {
(RSBR,0.6018)
(RLOO,0.2265)};
\addplot coordinates {
(RSBR,0.5592)
(RLOO,0.1769)};
\addplot coordinates {
 (RSBR,0.5513)
 (RLOO,0.1853)};
\addplot[fill=orange] coordinates {
 (RSBR,0.5044)
 (RLOO,0.1850)};
\addplot[fill=magenta] coordinates {
 (RSBR,0.2532)
 (RLOO,0.0612)};
\addplot[fill=black] coordinates {
 (RSBR,0.5349)
 (RLOO,0.1866)};
\addplot[fill=olive] coordinates {
 (RSBR,0.2901)
 (RLOO,0.0670)};
\legend{MostPop, ItemKNN, PureSVD, SLIM, BPRMF, BPRFM, NeuMF, NFM, NGCF, Multi-VAE}
\end{axis}
\end{tikzpicture}}
\subfigure[Yelp]{
\begin{tikzpicture}[scale=0.4]
\pgfplotsset{%
    width=0.85\textwidth,
    height=0.4\textwidth
}
\begin{axis}[
    ybar,
    bar width=6pt,
    ylabel={NDCG@10},
    ylabel style ={font = \Large},
    xlabel style ={font = \Large},
    enlarge x limits={abs=1.8cm},
    scaled ticks=false,
    tick label style={/pgf/number format/fixed, font=\Large},
    ymin=0, ymax=0.6,
    symbolic x coords={TSBR, RSBR, TLOO, RLOO},
    xtick=data,
    ytick={0,0.2,0.4,0.6},
    legend style={at={(0.75,0.98)}, anchor=north,legend columns=2},
]
\addplot coordinates {
(TSBR,0.0137) (RSBR,0.0198)
(TLOO, 0.0038) (RLOO,0.0049)};
\addplot coordinates {
(TSBR,0.3380) (RSBR,0.4672)
(TLOO, 0.2302) (RLOO,0.2754)};  
\addplot coordinates {
(TSBR,0.3918) (RSBR,0.5393)
(TLOO, 0.2199) (RLOO,0.2923)};
\addplot coordinates {
(TSBR,0.3432) (RSBR,0.5227)
(TLOO, 0.0388) (RLOO,0.1067)};
\addplot coordinates {
(TSBR,0.3770) (RSBR,0.5071)
(TLOO, 0.1840) (RLOO,0.2480)};
\addplot coordinates {
(TSBR,0.3701) (RSBR,0.4888)
(TLOO, 0.1942) (RLOO,0.2495)};
\addplot[fill=orange] coordinates {
(TSBR,0.4032) (RSBR,0.4811)
(TLOO, 0.2019) (RLOO,0.2585)};
\addplot[fill=magenta] coordinates {
(TSBR,0.2028) (RSBR,0.2454)
(TLOO, 0.0677) (RLOO,0.0819)};
\addplot[fill=black] coordinates {
(TSBR,0.4187) (RSBR,0.5738)
(TLOO, 0.2477) (RLOO,0.3292)};
\addplot[fill=olive] coordinates {
(TSBR,0.2041) (RSBR,0.2446)
(TLOO, 0.0702) (RLOO,0.0911)};
\end{axis}
\end{tikzpicture}}
\subfigure[AMZe]{
\begin{tikzpicture}[scale=0.4]
\pgfplotsset{%
    width=0.85\textwidth,
    height=0.4\textwidth
}
\begin{axis}[
    ybar,
    bar width=6pt,
    ylabel style ={font = \Large},
    xlabel style ={font = \Large},
    enlarge x limits={abs=1.8cm},
    scaled ticks=false,
    tick label style={/pgf/number format/fixed, font=\Large},
    ymin=0, ymax=0.35,
    symbolic x coords={TSBR, RSBR, TLOO, RLOO},
    xtick=data,
    ytick={0,0.1,0.2,0.3},
    legend style={at={(1.2,0.98)}, anchor=north,legend columns=1, row sep=2pt, font=\large},
]
\addplot coordinates {
(TSBR,0.0470) (RSBR,0.0266)
(TLOO, 0.0121) (RLOO,0.0086)};
\addplot coordinates {
(TSBR,0.0679) (RSBR,0.1230)
(TLOO, 0.1138) (RLOO,0.1382)};  
\addplot coordinates {
(TSBR,0.2935) (RSBR,0.3000)
(TLOO, 0.1382) (RLOO,0.1374)};
\addplot coordinates {
(TSBR,0.1702) (RSBR,0.2549)
(TLOO, 0.0121) (RLOO,0.0082)};
\addplot coordinates {
(TSBR,0.3117) (RSBR,0.3101)
(TLOO, 0.1463) (RLOO,0.1436)};
\addplot coordinates {
(TSBR,0.3264) (RSBR,0.2984)
(TLOO, 0.1356) (RLOO,0.1334)};
\addplot[fill=orange] coordinates {
(TSBR,0.3244) (RSBR,0.2859)
(TLOO, 0.1295) (RLOO,0.1157)};
\addplot[fill=magenta] coordinates {
(TSBR,0.3276) (RSBR,0.2906)
(TLOO, 0.1360) (RLOO,0.1245)};
\addplot[fill=black] coordinates {
(TSBR,0.2664) (RSBR,0.2728)
(TLOO, 0.1231) (RLOO,0.1469)};
\addplot[fill=olive] coordinates {
(TSBR,0.3285) (RSBR,0.2908)
(TLOO, 0.1355) (RLOO,0.1177)};
\end{axis}
\end{tikzpicture}}
\subfigure[Book-X]{
\begin{tikzpicture}[scale=0.4]
\pgfplotsset{%
    width=0.5\textwidth,
    height=0.4\textwidth
}
\begin{axis}[
    ybar,
    bar width=6pt,
    ylabel style ={font = \Large},
    xlabel style ={font = \Large},
    enlarge x limits={abs=1.8cm},
    scaled ticks=false,
    tick label style={/pgf/number format/fixed, font=\Large},
    ymin=0, ymax=0.35,
    symbolic x coords={RSBR, RLOO},
    xtick=data,
    ytick={0,0.1,0.2,0.3},
    legend style={at={(1.3,0.98)}, anchor=north,legend columns=1, row sep=2pt, font=\large},
]
\addplot coordinates {
 (RSBR,0.0630)
 (RLOO,0.0140)};
\addplot coordinates {
 (RSBR,0.1921)
 (RLOO,0.1456)};  
\addplot coordinates {
 (RSBR,0.3216)
 (RLOO,0.1120)};
\addplot coordinates {
 (RSBR,0.2504)
 (RLOO,0.1704)};
\addplot coordinates {
 (RSBR,0.2983)
 (RLOO,0.1052)};
\addplot coordinates {
 (RSBR,0.2779)
 (RLOO,0.1000)};
\addplot[fill=orange] coordinates {
 (RSBR,0.2852)
 (RLOO,0.0775)};
\addplot[fill=magenta] coordinates {
 (RSBR,0.2380)
 (RLOO,0.0644)};
\addplot[fill=black] coordinates {
 (RSBR,0.2533)
 (RLOO,0.1227)};
\addplot[fill=olive] coordinates {
 (RSBR,0.2378)
 (RLOO,0.0655)};
\legend{MostPop, ItemKNN, PureSVD, SLIM, BPRMF, BPRFM, NeuMF, NFM, NGCF, Multi-VAE}
\end{axis}
\end{tikzpicture}}
\vspace{-0.1in}
\caption{Performance of baselines w.r.t. 10-filter on the six datasets with different data splitting methods.
}\label{fig:data-splitting}
\vspace{-0.15in}
\end{figure*}
\begin{table*}[t]
    \centering
    \addtolength{\tabcolsep}{2pt}
    \scriptsize
	\caption{Baseline comparisons on training time w.r.t. time-aware split-by-ratio on the 10-filter view (seconds).
	}
	\label{tab:complexity_train}
	\vspace{-0.15in}
	\addtolength{\tabcolsep}{-1pt}
    \begin{tabular}{|l|ll|llll|llll|}
    \specialrule{.15em}{.05em}{.05em}
    \multirow{2}{*}{} &\multicolumn{2}{c|}{\textbf{MMs}} &\multicolumn{4}{c|}{\textbf{LFMs}} &\multicolumn{4}{c|}{\textbf{DLMs}}\\
    \cline{2-11}
    &MostPop &ItemKNN &PureSVD &BPRMF &BPRFM &SLIM &NeuMF &NFM &NGCF &Multi-VAE\\
    \specialrule{.05em}{.05em}{.05em}
    ML-1M &0.0048 &22.882 &0.6808 &2286.0 &1847.4 &62.924 &9627.6 &1682.6 &2216.2 &58.197\\
    Lastfm &0.0010 &2.4913 &0.4728 &246.35 &103.35 &5.1933 &95.156 &1178.6 &538.06 &31.807\\
    Book-X &0.0055 &29.013 &3.1499 &506.89 &2129.0 &860.71 &5250.0 &895.42 &6251.6 &101.00\\
    Epinions &0.0070 &25.847 &2.8042 &1497.0 &3525.4 &3283.7 &4265.6 &3387.0 &9453.9 &177.75 \\
    Yelp &0.0314 &244.80 &16.284 &1566.4 &6882.0 &63990 &5931.2 &2459.4 &38351 &1388.6 \\
    AMZe &0.0167 &121.96 &1.7262 &491.56 &1258.4 &18273 &8625.5 &2290.9 &10099 &1497.3 \\
    \specialrule{.15em}{.05em}{.05em}
    \end{tabular}
\end{table*}


\subsubsection{Impacts of Dataset Pre-processing}\label{subsubsec:data_preprocessing}
To study the impacts of pre-processing strategies (origin, 5- and 10-filter), we adopt Bayesian HyperOpt to perform hyper-parameter optimization w.r.t. NDCG$@10$ for each baseline under each view on each dataset for 30 trails~\cite{dacrema2019we}.
We keep original objective functions for each baseline (see Table~\ref{tab:summary_objective_functions}), employ the uniform sampler, and adopt time-aware split-by-ratio at global level ($\rho=80\%$) as the data splitting method. Besides, 10\% of the latest training set is held out as the validation set to tune the hyper-parameters. Once the optimal hyper-parameters are decided, we feed the whole training set to train the final model
and report the performance on the test set.
Figure~\ref{fig:data-preprocessing} depicts the final results, where SLIM is omitted due to its extremely high computational complexity on large-scale datasets, which is unable to complete in a reasonable amount of time; and NGCF on Yelp and AMZe under origin view is also omitted because of the same reason
(see Section~\ref{subsec:baseline_selection}).
Due to the space limitation, we only report the results on NDCG@10.

Overall, three different trends can be observed from the results: (1) the performance of different baselines keeps relatively stable on ML-1M with varied settings; (2) on Book-X, Yelp and AMZe, the performance of all baselines generally climbs up; and (3) an obvious performance drop is observed on Lastfm and Epinions. The most probable explanation is that although the density of the datasets increases (origin $\rightarrow$ 5-filter $\rightarrow$ 10-filter) as shown in Table~\ref{tab:dataset_statistics}, the average length of the training sets for each user keeps stable on ML-1M ($86$); increases on Book-X, Yelp and AMZe; and decreases on Lastfm ($39 \rightarrow 30 \rightarrow 27$) and Epinions ($35 \rightarrow 23 \rightarrow 20$), as depicted by  Table~\ref{tab:statistics_train_test_set}.
The more training data per user, the better a model can be trained, meaning that the more accurate performance can be achieved, and \textit{vice versa}. 

Regarding the performance of different baselines, 
several major findings can be noted as below.
(1) Regarding MMs, MostPop performs the worst in most cases, showing the importance of personalization in recommendation; and ItemKNN is defeated by LFMs and DLMs, indicating the superiority of LFMs and DLMs on effective recommendation. However, on ML-1M, the performance of MostPop exceeds that of ItemKNN, PureSVD and even BPRMF, demonstrating the potential of popularity in effective recommendation; and on Lastfm, ItemKNN achieves the best performance compared with LFMs and DLMs. This implies that, the neighborhood-based idea, though simple, could be absorbed by LFMs and DLMs to further improve the recommendation accuracy~\cite{jannach2017recurrent}.
(2) W.r.t. the three LFMs, BPRMF generally performs better than PureSVD but worse than BPRFM. Although PureSVD is simple -- directly applying conventional sigular value decomposition on the user-item interaction matrix, it sometimes achieves comparable and even better performance in comparison with BPRMF and BPRFM (see LastFM, Book-X and Yelp with 5/10-filter views).
(3) For the four DLMs (i.e., NeuMF, NFM, NGCF and Multi-VAE), their performance varies across different datasets. For instance, NeuMF obtains the highest accuracy on Epinions; NFM is the winner on AMZe; NGCF defeats the others on both LastFM and Yelp; and Multi-VAE achieves extraordinary results on ML-1M. However, they generally perform comparably to BPRFM across all datasets, and sometimes even worse than BPRFM (e.g., ML-1M and Book-X). Besides, on LastFM, they even underperform ItemKNN. This is consistent with the previous findings~\cite{dacrema2019we} that DLMs are not always better than traditional methods with well-tuned parameters, but mostly cost much more in training as verified by Table \ref{tab:complexity_train}.

%


\subsubsection{Impacts of Dataset Splitting Methods}
We now test the impacts of different data splitting methods on the recommendation performance. To this end, we compare \emph{random-} and \emph{time-aware split-by-ratio} (i.e., \textbf{RSBR} vs. \textbf{TSBR}) at global level with $\rho=80\%$ as well as 
\emph{random-} and \emph{time-aware leave-one-out} (i.e., \textbf{RLOO} vs. \textbf{TLOO}) on the 10-filter view. Note that we adopt Bayesian HyperOpt to perform hyper-parameter optimization w.r.t. NDCG@10 for each baseline under each data splitting method on each dataset fro 30 trails. Meanwhile, LastFM and Book-X do not have the timestamp information, so their results on TSBR and TLOO for each baseline are omitted.

Figure~\ref{fig:data-splitting} displays the results of ten baselines on the six datasets. 
First of all, we can clearly observe that the performance of TSBR/SBR (split-by-ratio) is generally better than that of TLOO/LOO (leave-one-out). This could be largely affected by the different settings on the test procedure. To be specific, as introduced in Section~\ref{subsec:data-splitting-method}, to improve the test efficiency, we randomly sample negative items for each user to ensure her test candidates to be $1,000$, and then rank all test items among the $1,000$ items w.r.t. both SBR and LOO. However, the number of positive items in the test set of SBR ($>1$) is normally larger than that of LOO ($=1$), thus leading to a higher accuracy of SBR.
Second, baselines with RSBR/RLOO outperform those with TSBR/TLOO, especially on Epinions. The reason behind is that compared with random-aware split,
time-aware split poses a stronger constraint on the pattern of training and test data, thus increasing the training difficulty. However, this is more close to the real recommendation scenario, which strives to infer future by history. Our study also implies that the empirical results disclosed in previous studies using RSBR might be overestimated compared to those for real-world scenarios.

\begin{figure*}
    \centering
    \subfigure[TSBR]{\includegraphics[width=0.25\textwidth]{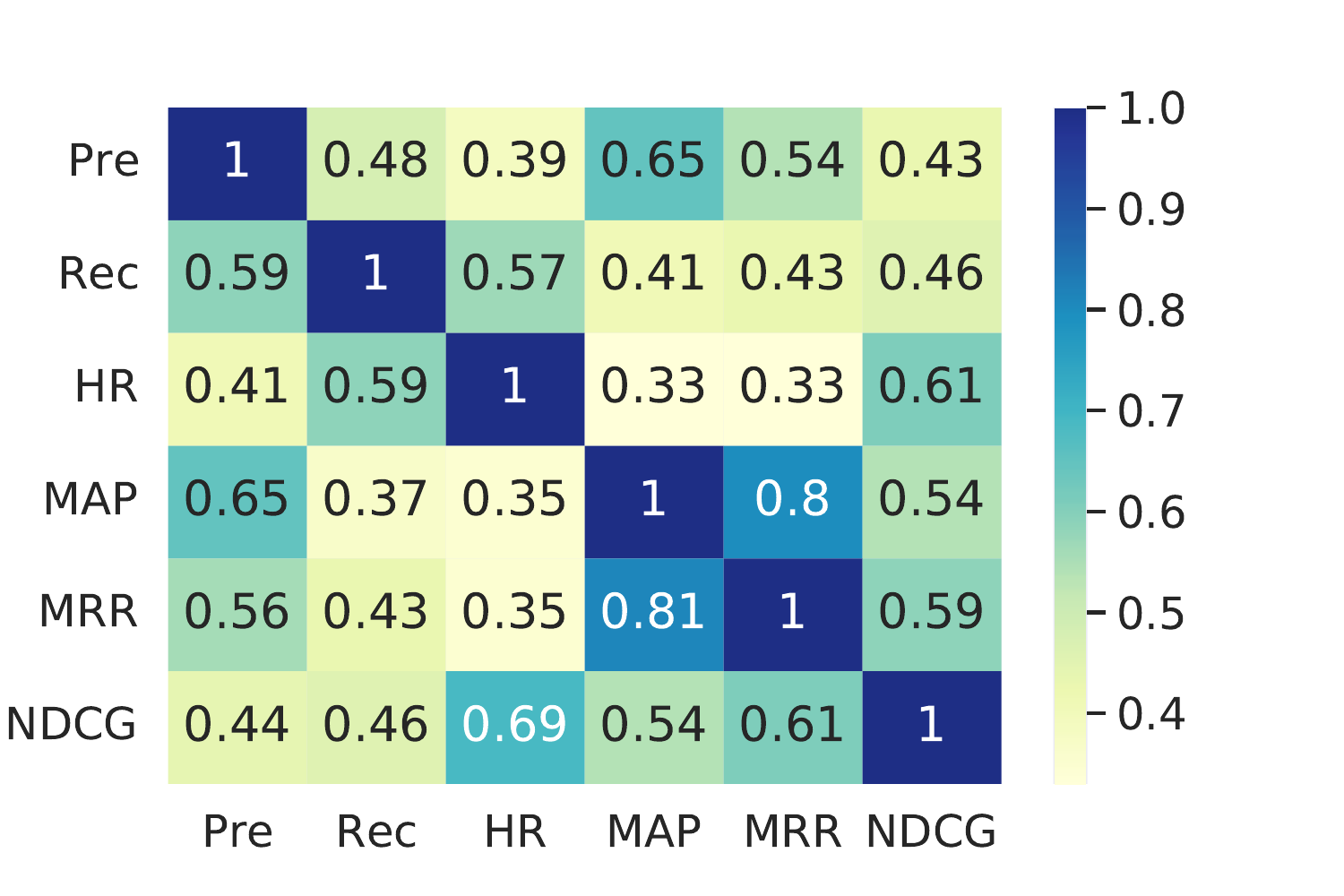}}
    \hspace{-0.1in}
    \subfigure[RSBR]{\includegraphics[width=0.25\textwidth]{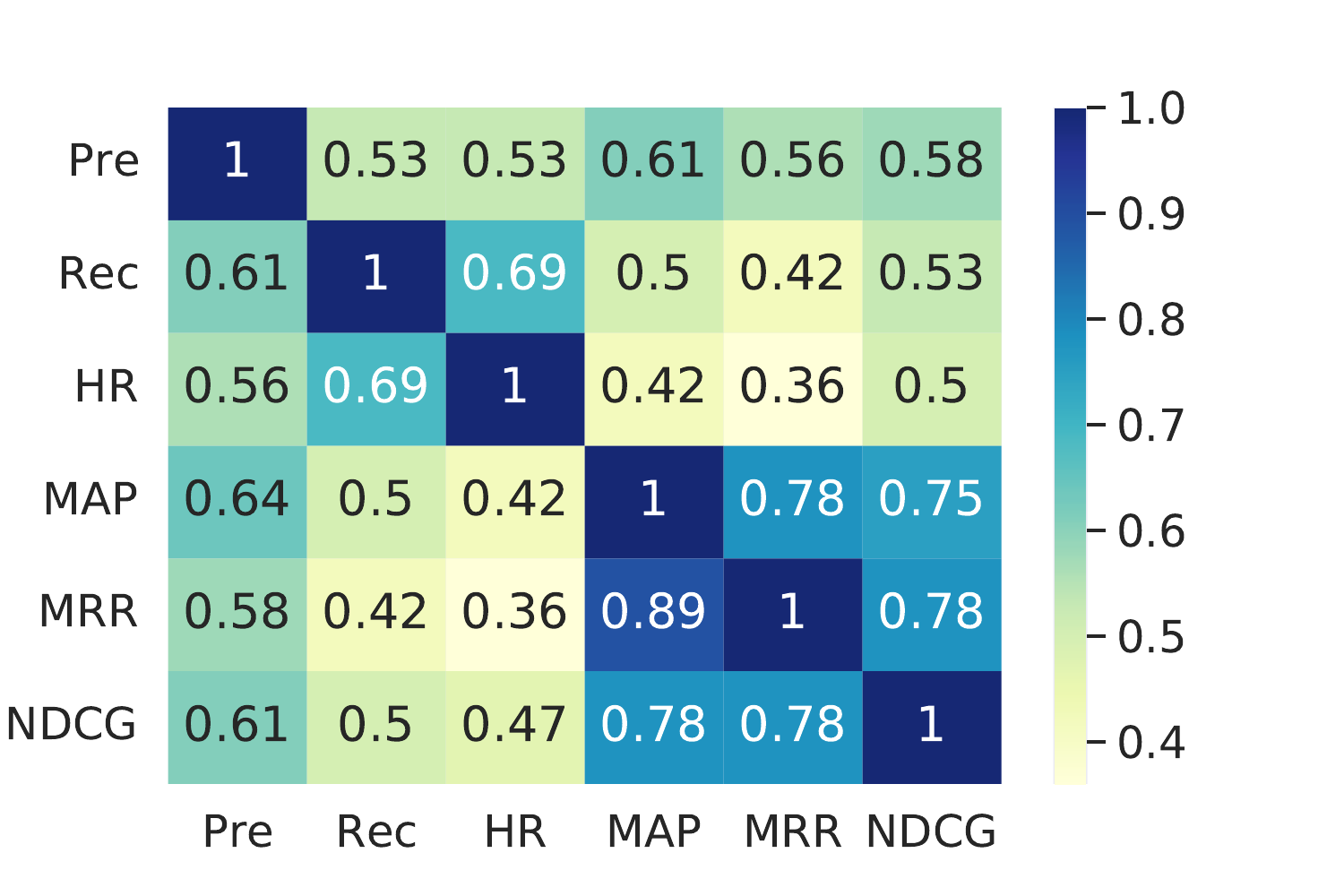}}
    \hspace{-0.1in}
    \subfigure[TLOO]{\includegraphics[width=0.25\textwidth]{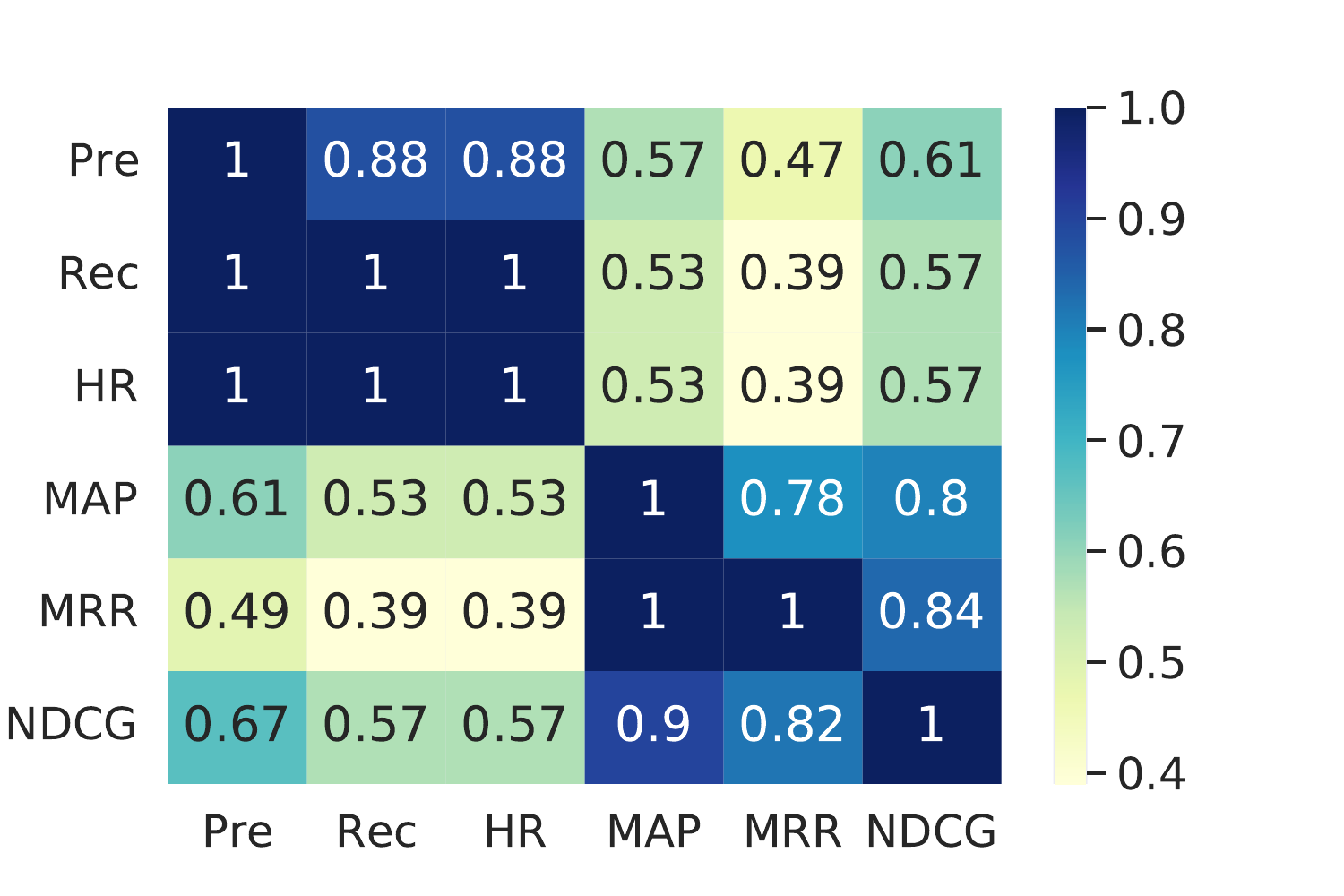}}
    \hspace{-0.1in}
    \subfigure[RLOO]{\includegraphics[width=0.25\textwidth]{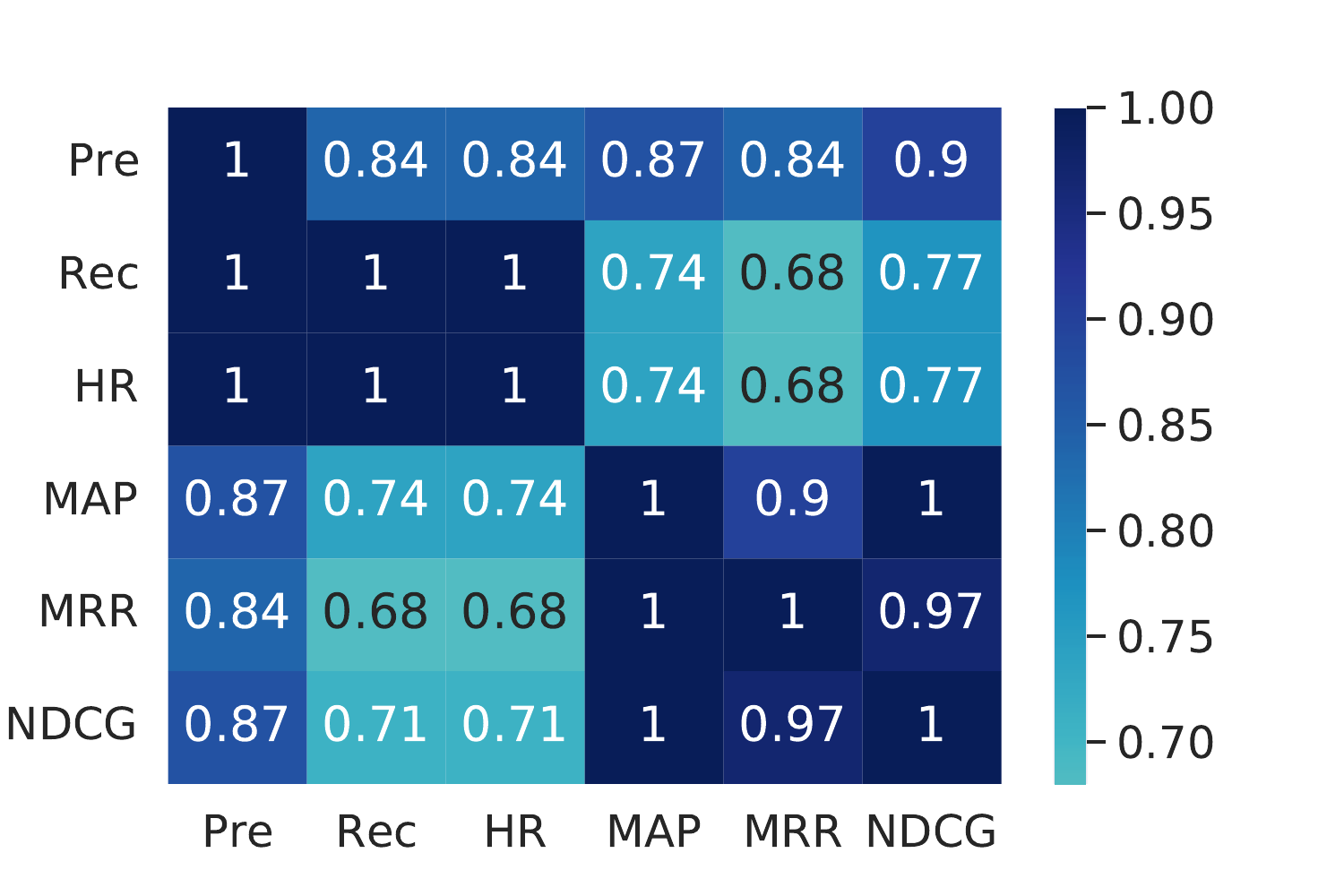}}
    \subfigure[TSBR]{\includegraphics[width=0.25\textwidth]{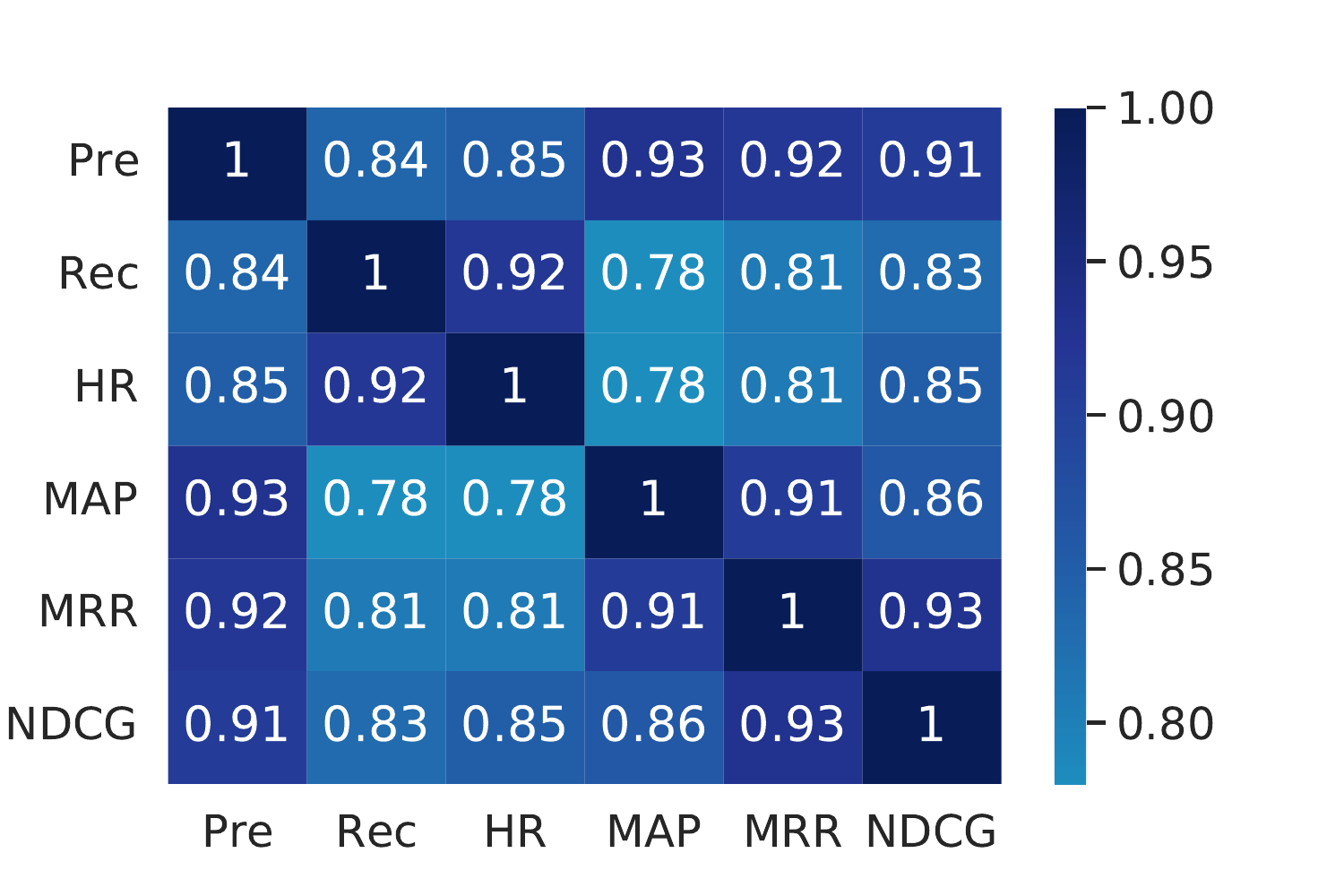}}
    \hspace{-0.1in}
    \subfigure[RSBR]{\includegraphics[width=0.25\textwidth]{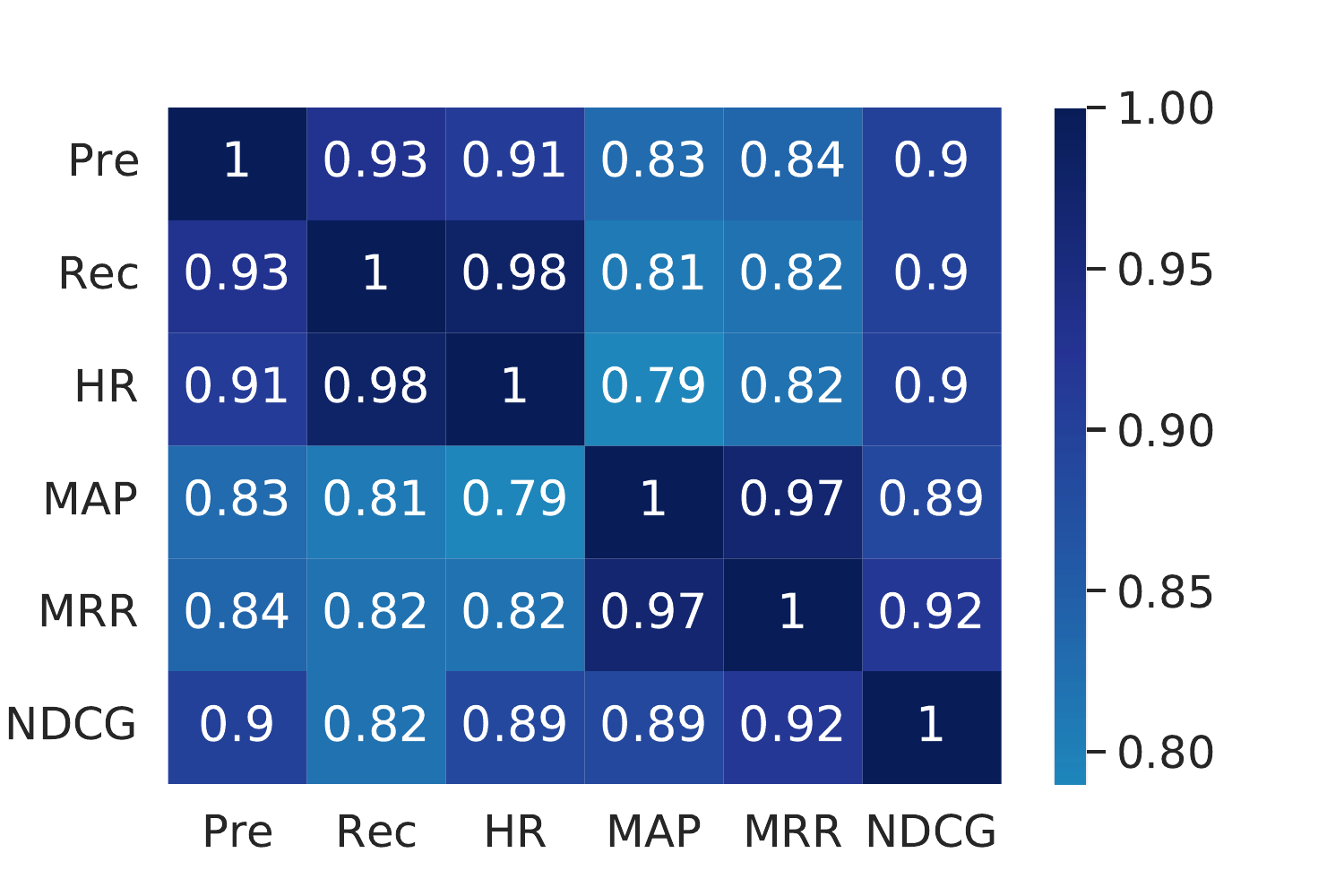}}
    \hspace{-0.1in}
    \subfigure[TLOO]{\includegraphics[width=0.25\textwidth]{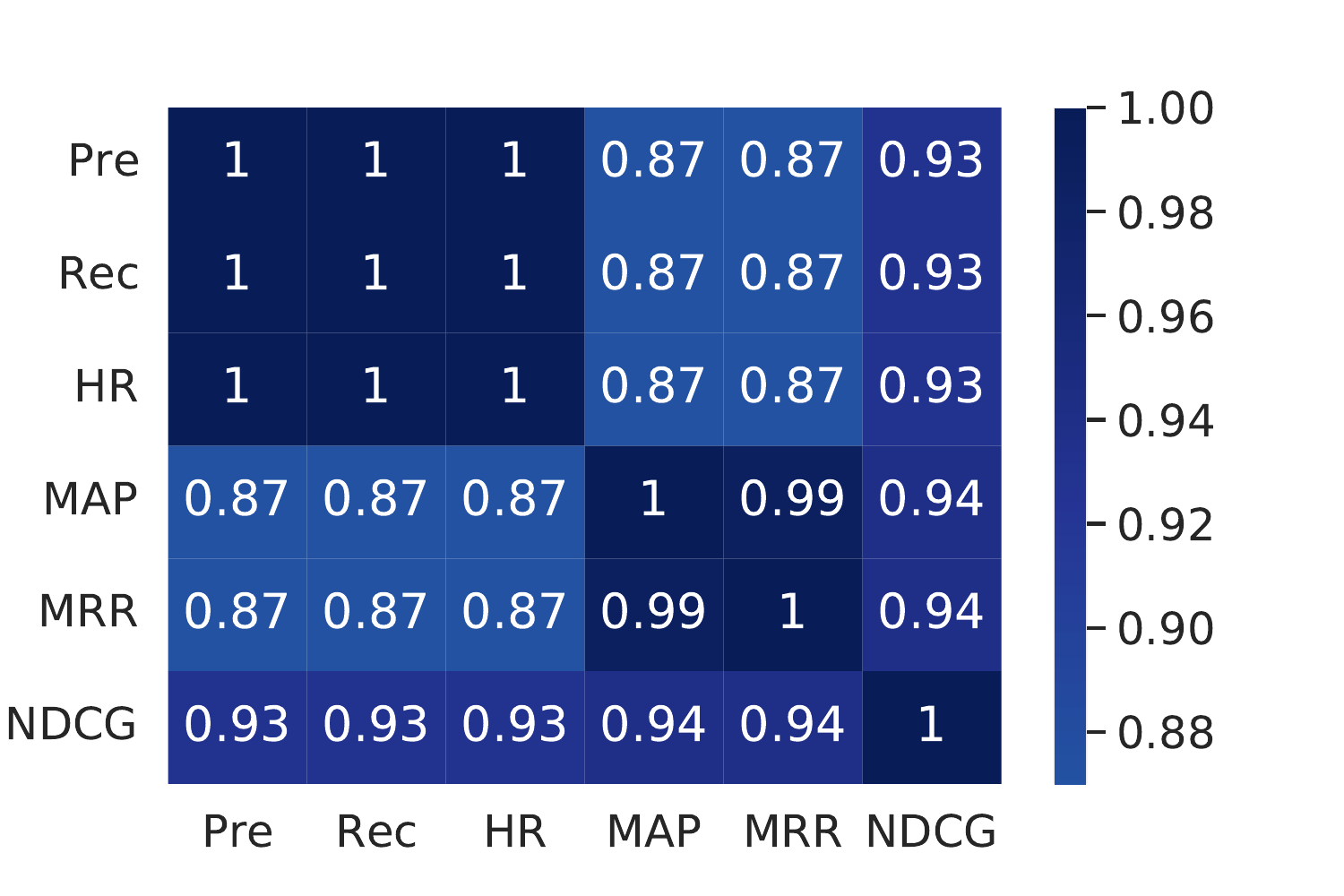}}
    \hspace{-0.1in}
    \subfigure[RLOO]{\includegraphics[width=0.25\textwidth]{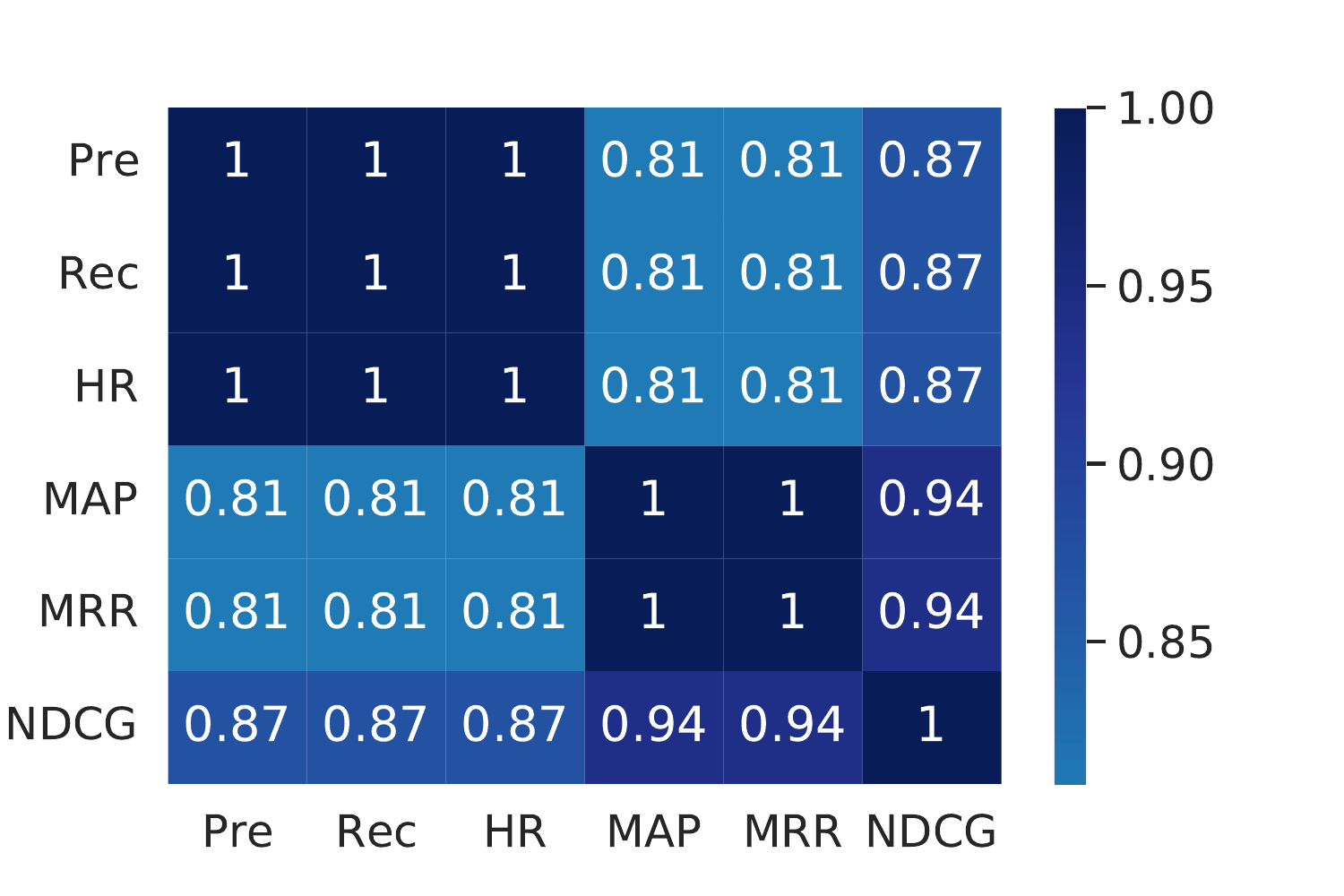}}
    \vspace{-0.1in}
    \caption{The correlations of evaluation metrics w.r.t. different data splitting methods on 10-filter. `Pre' and `Rec' are Precision and Recall, respectively. }
    \label{fig:correlation}
    \vspace{-0.15in}
\end{figure*}

\subsubsection{Complexity of Comparison Baselines}\label{subsec:baseline_selection} 
Table~\ref{tab:complexity_train} shows the training time for the ten baselines on the six datasets with optimal hyper-parameters found by Bayesian HyperOpt on 10-filter view via split-by-ratio. Note that, the optimal batch size may differ for different baselines, which may also affect the training time. 
All the experiments are executed on an Nvidia V100 GPU with 32 GiB memory, each running is paired with 11 Intel(R) Xeon(R) Platinum 8260 CPU (2.4GHz)  sharing 40 GiB memory. 

According to Table~\ref{tab:complexity_train}, we can note that \textit{MostPop} is the fastest one in training, as it merely ranks all the items by the calculated popularity. \textit{PureSVD} is the runner-up with time complexity $\mathcal{O}(\min \{m^2f, n^2f\})$, where $m, n, f$ are the number of users, items and singular values, respectively. Compared with other LFMs and DLMs, it achieves a better balance between time complexity and ranking accuracy. Particularly, it performs comparably and sometimes even better than, e.g., BPRMF and NeuMF on LastFM, as depicted by Figure~\ref{fig:data-preprocessing}, while its training time is hundreds or thousands times less than that of BPRMF and NeuMF as shown in Table~\ref{tab:complexity_train}. Although the training efficiency of ItemKNN ranks third among all baselines with 10-filter setting, the time cost quadratically increases with origin setting due to its time complexity $\mathcal{O}(mn^2)$.   
Besides, the similarity matrix also takes up huge memory, for example, on the original AMZe ($n \approx 10^6$), it will cost $(64 \text{ bit} * 10^6 * 10^6) /10^{12} = 64 T$ to save the similarity matrix. To ease this issue, we only keep the top-$100$ similar items for each target item in the memory. 

The training time of BPRMF and BPRFM is comparable, where the time complexity for both methods is $\mathcal{O}(\vert \mathcal{R}\vert d)$, where $\mathcal{R}$ is the total number of observed feedback and $d$ is the dimension of latent factors. Similar to ItemKNN, the time cost of SLIM with 10-filter setting is acceptable, while it tremendously increases with origin setting due to its time complexity $\mathcal{O}(\vert R\vert n)$. Even with the 10-filter view, it takes the longest training time among all baselines on the two large datasets (i.e., Yelp and AMZe).
Meanwhile, it also suffers from the huge memory cost issue because of the learned item similarity matrix. Hence, both ItemKNN and SLIM are not scalable for large-scale datasets. 
Regarding the four DLMs (i.e., NeuMF, NFM, NGCF and Multi-VAE), NFM and Multi-VAE are usually more efficient than NeuMF and NGCF regarding time complexity. Although DLMs yield comparable performance with LFMs, they generally cost much more training time, especially on larger datasets. For example, on AMZe, the training time of NeuMF and NGCF is around 20 times larger than that of BPRMF.

\subsubsection{Correlations of Evaluation Metrics}
As discussed in Section~\ref{subsubsec:data_preprocessing}, we adopt Bayesian HyperOpt to perform hyper-parameter optimization for 30 trials via optimizing NDCG@10. However, six metrics are used in our study, namely Precision, Recall, HR, MAP, MRR and NDCG. The best hyper-parameter settings for optimal NDCG does not necessarily guarantee optimums w.r.t. the other five metrics.
Hence, we study the correlation of different metrics
when their respective optimums are achieved.
In particular, for each baseline 
on each dataset with 10-filter view, the Bayesian HyperOpt executes 30 trails; we thus have 30 entries for the validation performance of the baseline
correspondingly, where each entry includes the results on the six metrics, e.g., [Precision: 0.24; Recall: 0.07; HR: 0.57; MAP: 0.17; MRR: 0.76; NDCG: 0.42]. 
Due to the optimal results for the six metrics may not achieve simultaneously, we select the optimal one among the 30 entries for each metric, and ultimately obtain six entries, where each entry records the best result on the corresponding metric. 

Based on the six selected entries of each method per dataset, we pair-wisely calculate and record the times that any two of them (e.g., NDCG and HR) can achieve their best results simultaneously entry by entry. For example, given the optimal entry for NDCG, we will check whether the rest five metrics (e.g., HR) in this entry are optimal or not. If yes, we will add one at the corresponding position (NDCG, HR) of the correlation matrix; otherwise 0. The same rule is applied to the optimal entries for the other five metrics. Except MostPop, as it does not have any hyper-parameters, we accumulate the results of nine baselines across the six datasets (9*6=54), and
ultimately obtain their correlation matrices regarding time-/random-aware split-by-ratio and leave-one-out
as illustrated by Figures~\ref{fig:correlation}(a-d), where all values are normalized into the range of $[0,1]$ (divided by 54), and a darker color indicates a stronger correlation, that is, a higher probability of two metrics achieving their best results in the meanwhile.  

The results help verify our argument that best hyper-parameter settings for optimal NDCG cannot ensure optimal results for all the other five metrics.
Several detailed findings can be noted.
\textbf{(1)} The correlation matrix is asymmetrical, for instance, the correlation for (NDCG, HR, 0.69) is higher than (HR, NDCG, 0.61) as shown in Figure~\ref{fig:correlation}(a). That is to say, the probability of a model with best NDCG to achieve the best HR is higher than that of a model with best HR to reap the optimal NDCG. \textbf{(2)} Similar trends can be noticed within a same base data splitting method (i.e., TSBR and RSBR; TLOO and RLOO) no matter whether the timestamp information is considered or not; whilst the patterns are different across different base splitting fashions (i.e., TSBR/RSBR and TLOO/RLOO). \textbf{(3)} Regarding TSBR/RSBR, the best MRR and MAP are more easily to be guaranteed concurrently, e.g., (MRR, MAP, 0.81) and (MAP, MRR, 0.80) in Figure~\ref{fig:correlation}(a); and (MRR, MAP, 0.89) and (MAP, MRR, 0.78) in Figure~\ref{fig:correlation}(b). \textbf{(4)} For TLOO/RLOO, Precision, Recall and HR are more likely to be optimized simultaneously; and MAP, MRR and NDCG have a higher probability to reach their peaks together. This is mainly due to only one positive item inside the test set for each user; consequently, Recall and HR are equivalent, which is positively correlated with Precision; meanwhile, MAP, MRR and NDCG are also positively correlated with each other. 

Additionally, we examine the Kendall's correlation~\cite{valcarce2018robustness} among metrics in terms of indicating recommendation performance on the ten baselines across the six datasets under 10-filter view with different data splitting methods. The results are depicted in Figures~\ref{fig:correlation}(e-h), where a darker color (a stronger correlation) implies that the metrics produce more identical ranking. We find that \textbf{(1)} different from Figures~\ref{fig:correlation}(a-d), the Kendall's correlation matrix is symmetrical; \textbf{(2)} similarly, the trends are consistent within a same base data splitting method, e.g.,  Figures~\ref{fig:correlation}(g-h), while vary slightly across different base splitting ways, e.g., Figures~\ref{fig:correlation}(e) and \ref{fig:correlation}(g); and \textbf{(3)}
a common observation across Figures~\ref{fig:correlation}(e-h) is that MAP, MRR and NDCG are more likely to generate consistent ranking. Besides, for TSBR/RSBR, (Precision, NDCG) and (Recall, HR) show a fairly strong correlation, while w.r.t. TLOO/RLOO, (Precision, Recall, HR) exhibits obvious correlation, which is also caused by the single positive item inside the test set for each user as explained previously. 
In summary, a convincing and solid evaluation should be performed w.r.t. more diverse metrics.

\begin{figure}[t]
\centering
\subfigure[ML-1M]{
\begin{tikzpicture}[scale=0.4]
\pgfplotsset{%
    width=0.55\textwidth,
    height=0.4\textwidth
}
\begin{axis}[
    ybar,
    bar width=5pt,
    ylabel={NDCG@10},
    ylabel style ={font = \Large},
    xlabel style ={font = \Large},
    enlarge x limits={abs=0.7cm},
    scaled ticks=false,
    tick label style={/pgf/number format/fixed, font=\Large},
    ymin=0, ymax=0.8,
    symbolic x coords={MF, FM, NeuMF, NFM, NGCF},
    xtick=data,
    ytick={0,0.2,0.4,0.6,0.8},
    legend style={at={(0.5,0.98)}, anchor=north,legend columns=4, column sep=0.2cm, draw=none, font=\large},
]
\addplot coordinates {
(MF,0.4601) (FM,0.6466)
(NeuMF, 0.6278) (NFM,0.6232) (NGCF,0.5494)};
\addplot coordinates {
(MF,0.4399) (FM,0.6312)
(NeuMF, 0.4322) (NFM,0.5973) (NGCF,0.5974)};  
\addplot coordinates {
(MF,0.4629) (FM,0.6489)
(NeuMF, 0.6295) (NFM,0.6223) (NGCF,0.5942)};
\addplot coordinates {
(MF,0.1889) (FM,0.6290)
(NeuMF, 0.6009) (NFM,0.6240) (NGCF,0.4630)};
\legend{BPR, CE, Hinge, Top1}
\end{axis}
\end{tikzpicture}}
\subfigure[LastFM]{
\begin{tikzpicture}[scale=0.4]
\pgfplotsset{%
    width=0.55\textwidth,
    height=0.4\textwidth
}
\begin{axis}[
    ybar,
    bar width=5pt,
    ylabel style ={font = \Large},
    xlabel style ={font = \Large},
    enlarge x limits={abs=0.7cm},
    scaled ticks=false,
    tick label style={/pgf/number format/fixed, font=\Large},
    ymin=0, ymax=0.65,
    symbolic x coords={MF, FM, NeuMF, NFM, NGCF},
    xtick=data,
    ytick={0,0.2,0.4,0.6},
    legend style={at={(0.5,0.98)}, anchor=north,legend columns=4, column sep=0.2cm, draw=none, font=\large},
]
\addplot coordinates {
(MF,0.5592) (FM,0.5513)
(NeuMF, 0.4626) (NFM,0.2532) (NGCF,0.5349)};
\addplot coordinates {
(MF,0.4387) (FM,0.5431)
(NeuMF, 0.5044) (NFM,0.2833) (NGCF,0.4400)};  
\addplot coordinates {
(MF,0.5382) (FM,0.5107)
(NeuMF, 0.4547) (NFM,0.2588) (NGCF,0.4501)};
\addplot coordinates {
(MF,0.1378) (FM,0.2875)
(NeuMF, 0.1414) (NFM,0.2798) (NGCF,0.4819)};
\legend{BPR, CE, Hinge, Top1}
\end{axis}
\end{tikzpicture}}
\subfigure[Book-X]{
\begin{tikzpicture}[scale=0.4]
\pgfplotsset{%
    width=0.55\textwidth,
    height=0.4\textwidth
}
\begin{axis}[
    ybar,
    bar width=5pt,
    ylabel={NDCG@10},
    ylabel style ={font = \Large},
    xlabel style ={font = \Large},
    enlarge x limits={abs=0.7cm},
    scaled ticks=false,
    tick label style={/pgf/number format/fixed, font=\Large},
    ymin=0, ymax=0.35,
    symbolic x coords={MF, FM, NeuMF, NFM, NGCF},
    xtick=data,
    ytick={0,0.1,0.2,0.3},
    legend style={at={(0.5,0.98)}, anchor=north,legend columns=4, column sep=0.2cm, draw=none, font=\large},
]
\addplot coordinates {
(MF,0.2983) (FM,0.2779)
(NeuMF, 0.2333) (NFM,0.2380) (NGCF,0.2533)};
\addplot coordinates {
(MF,0.2670) (FM,0.2629)
(NeuMF, 0.2852) (NFM,0.2407) (NGCF,0.2314)};  
\addplot coordinates {
(MF,0.2813) (FM,0.2652)
(NeuMF, 0.0786) (NFM,0.2404) (NGCF,0.2173)};
\addplot coordinates {
(MF,0.0328) (FM,0.2389)
(NeuMF, 0.0329) (NFM,0.2369) (NGCF,0.2628)};
\legend{BPR, CE, Hinge, Top1}
\end{axis}
\end{tikzpicture}}
\subfigure[Epinions]{
\begin{tikzpicture}[scale=0.4]
\pgfplotsset{%
    width=0.55\textwidth,
    height=0.4\textwidth
}
\begin{axis}[
    ybar,
    bar width=5pt,
    ylabel style ={font = \Large},
    xlabel style ={font = \Large},
    enlarge x limits={abs=0.7cm},
    scaled ticks=false,
    tick label style={/pgf/number format/fixed, font=\Large},
    ymin=0, ymax=0.13,
    symbolic x coords={MF, FM, NeuMF, NFM, NGCF},
    xtick=data,
    ytick={0,0.04,0.08,0.12},
    legend style={at={(0.5,0.98)}, anchor=north,legend columns=4, column sep=0.2cm, draw=none, font=\large},
]
\addplot coordinates {
(MF,0.0820) (FM,0.0979)
(NeuMF, 0.0724) (NFM,0.0756) (NGCF,0.1052)};
\addplot coordinates {
(MF,0.0689) (FM,0.0815)
(NeuMF, 0.1084) (NFM,0.0725) (NGCF,0.0820)};  
\addplot coordinates {
(MF,0.0790) (FM,0.0941)
(NeuMF, 0.0522) (NFM,0.0759) (NGCF,0.0800)};
\addplot coordinates {
(MF,0.0452) (FM,0.0754)
(NeuMF, 0.0335) (NFM,0.0747) (NGCF,0.1005)};
\legend{BPR, CE, Hinge, Top1}
\end{axis}
\end{tikzpicture}}
\subfigure[Yelp]{
\begin{tikzpicture}[scale=0.4]
\pgfplotsset{%
    width=0.55\textwidth,
    height=0.4\textwidth
}
\begin{axis}[
    ybar,
    bar width=5pt,
    ylabel={NDCG@10},
    ylabel style ={font = \Large},
    xlabel style ={font = \Large},
    enlarge x limits={abs=0.7cm},
    scaled ticks=false,
    tick label style={/pgf/number format/fixed, font=\Large},
    ymin=0, ymax=0.5,
    symbolic x coords={MF, FM, NeuMF, NFM, NGCF},
    xtick=data,
    ytick={0,0.1,0.2,0.3,0.4},
    legend style={at={(0.5,0.98)}, anchor=north,legend columns=4, column sep=0.2cm, draw=none, font=\large},
]
\addplot coordinates {
(MF,0.3698) (FM,0.3701)
(NeuMF, 0.2900) (NFM,0.2028) (NGCF,0.4187)};
\addplot coordinates {
(MF,0.3572) (FM,0.3777)
(NeuMF, 0.4032) (NFM,0.1932) (NGCF,0.3321)};  
\addplot coordinates {
(MF,0.3607) (FM,0.3630)
(NeuMF, 0.2568) (NFM,0.1998) (NGCF,0.3405)};
\addplot coordinates {
(MF,0.1328) (FM,0.2041)
(NeuMF, 0.0333) (NFM,0.1996) (NGCF,0.3882)};
\legend{BPR, CE, Hinge, Top1}
\end{axis}
\end{tikzpicture}}
\subfigure[AMZe]{
\begin{tikzpicture}[scale=0.4]
\pgfplotsset{%
    width=0.55\textwidth,
    height=0.4\textwidth
}
\begin{axis}[
    ybar,
    bar width=5pt,
    ylabel style ={font = \Large},
    xlabel style ={font = \Large},
    enlarge x limits={abs=0.7cm},
    scaled ticks=false,
    tick label style={/pgf/number format/fixed, font=\Large},
    ymin=0, ymax=0.38,
    symbolic x coords={MF, FM, NeuMF, NFM, NGCF},
    xtick=data,
    ytick={0,0.1,0.2,0.3},
    legend style={at={(0.5,0.98)}, anchor=north,legend columns=4, column sep=0.2cm, draw=none, font=\large},
]
\addplot coordinates {
(MF,0.3117) (FM,0.3264)
(NeuMF, 0.3106) (NFM,0.2858) (NGCF,0.2664)};
\addplot coordinates {
(MF,0.3113) (FM,0.3282)
(NeuMF, 0.3244) (NFM,0.1942) (NGCF,0.2604)};  
\addplot coordinates {
(MF,0.2714) (FM,0.3222)
(NeuMF, 0.2852) (NFM,0.2375) (NGCF,0.2455)};
\addplot coordinates {
(MF,0.0200) (FM,0.3282)
(NeuMF, 0.0201) (NFM,0.3046) (NGCF,0.2659)};
\legend{BPR, CE, Hinge, Top1}
\end{axis}
\end{tikzpicture}}
\vspace{-0.1in}
\caption{Performance of baselines w.r.t. time-aware split-by-ratio on 10-filter view across the six datasets with different loss functions.}\label{fig:loss-function}
\vspace{-0.15in}
\end{figure}
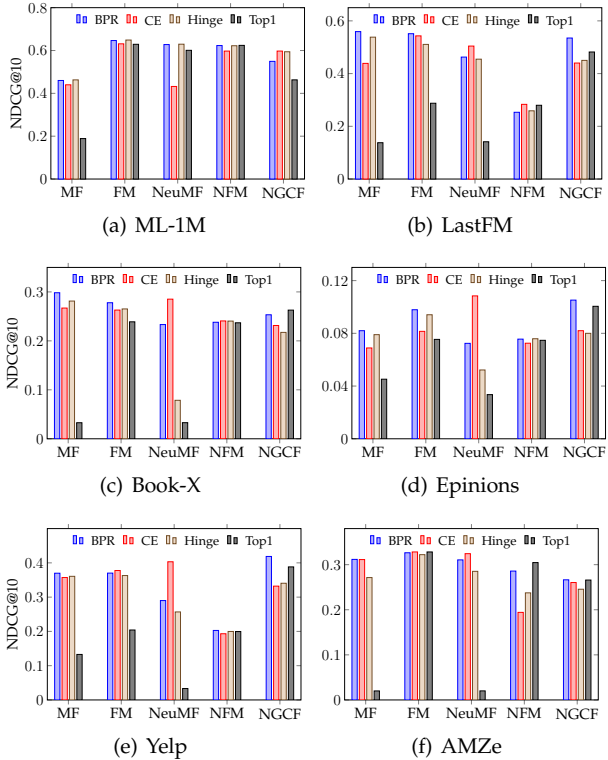
\begin{figure}[t]
\centering
\subfigure[ML-1M]{
\begin{tikzpicture}[scale=0.4]
\pgfplotsset{%
    width=0.55\textwidth,
    height=0.4\textwidth
}
\begin{axis}[
    ybar,
    bar width=5pt,
    ylabel={NDCG@10},
    ylabel style ={font = \Large},
    xlabel style ={font = \Large},
    enlarge x limits={abs=0.8cm},
    scaled ticks=false,
    tick label style={/pgf/number format/fixed, font=\Large},
    ymin=0.2, ymax=0.8,
    symbolic x coords={BPRMF, BPRFM, NeuMF, NFM, NGCF},
    xtick=data,
    ytick={0.2,0.4,0.6},
    legend style={at={(0.5,0.98)}, anchor=north,legend columns=5, column sep=0.1cm, draw=none, font=\large},
]
\addplot coordinates {
(BPRMF,0.4601) (BPRFM,0.6466)
(NeuMF, 0.5102) (NFM,0.6232) (NGCF,0.5494)};
\addplot coordinates {
(BPRMF,0.4732) (BPRFM,0.6501)
(NeuMF, 0.4114) (NFM,0.6219) (NGCF,0.4967)};  
\addplot coordinates {
(BPRMF,0.4847) (BPRFM,0.6477)
(NeuMF, 0.4311) (NFM,0.6233) (NGCF,0.5342)};
\addplot coordinates {
(BPRMF,0.5853) (BPRFM,0.6169)
(NeuMF, 0.4041) (NFM,0.5500) (NGCF,0.4354)};
\addplot coordinates {
(BPRMF,0.4864) (BPRFM,0.6218)
(NeuMF, 0.4083) (NFM,0.6262) (NGCF,0.4159)};
\legend{U, HP, LP, U+HP, U+LP}
\end{axis}
\end{tikzpicture}}
\subfigure[LastFM]{
\begin{tikzpicture}[scale=0.4]
\pgfplotsset{%
    width=0.55\textwidth,
    height=0.4\textwidth
}
\begin{axis}[
    ybar,
    bar width=5pt,
    ylabel style ={font = \Large},
    xlabel style ={font = \Large},
    enlarge x limits={abs=0.8cm},
    scaled ticks=false,
    tick label style={/pgf/number format/fixed, font=\Large},
    ymin=0, ymax=0.65,
    symbolic x coords={BPRMF, BPRFM, NeuMF, NFM, NGCF},
    xtick=data,
    ytick={0,0.2,0.4,0.6},
    legend style={at={(0.5,0.98)}, anchor=north,legend columns=5, column sep=0.1cm, draw=none, font=\large},
]
\addplot coordinates {
(BPRMF,0.5592) (BPRFM,0.5513)
(NeuMF, 0.5044) (NFM,0.2532) (NGCF,0.5349)};
\addplot coordinates {
(BPRMF,0.5591) (BPRFM,0.5371)
(NeuMF, 0.4232) (NFM,0.2323) (NGCF,0.5178)};  
\addplot coordinates {
(BPRMF,0.5562) (BPRFM,0.5382)
(NeuMF, 0.4179) (NFM,0.1164) (NGCF,0.5129)};
\addplot coordinates {
(BPRMF,0.4081) (BPRFM,0.4297)
(NeuMF, 0.3841) (NFM,0.1948) (NGCF,0.4382)};
\addplot coordinates {
(BPRMF,0.4602) (BPRFM,0.4936)
(NeuMF, 0.4100) (NFM,0.2385) (NGCF,0.4842)};
\legend{U, HP, LP, U+HP, U+LP}
\end{axis}
\end{tikzpicture}}
\subfigure[Book-X]{
\begin{tikzpicture}[scale=0.4]
\pgfplotsset{%
    width=0.55\textwidth,
    height=0.4\textwidth
}
\begin{axis}[
    ybar,
    bar width=5pt,
    ylabel={NDCG@10},
    ylabel style ={font = \Large},
    xlabel style ={font = \Large},
    enlarge x limits={abs=0.8cm},
    scaled ticks=false,
    tick label style={/pgf/number format/fixed, font=\Large},
    ymin=0.1, ymax=0.35,
    symbolic x coords={BPRMF, BPRFM, NeuMF, NFM, NGCF},
    xtick=data,
    ytick={0.1,0.2,0.3},
    legend style={at={(0.5,0.98)}, anchor=north,legend columns=5, column sep=0.1cm, draw=none, font=\large},
]
\addplot coordinates {
(BPRMF,0.2983) (BPRFM,0.2779)
(NeuMF, 0.2852) (NFM,0.2380) (NGCF,0.2533)};
\addplot coordinates {
(BPRMF,0.2984) (BPRFM,0.2767)
(NeuMF, 0.2867) (NFM,0.2337) (NGCF,0.2342)};  
\addplot coordinates {
(BPRMF,0.3006) (BPRFM,0.2750)
(NeuMF, 0.2881) (NFM,0.2399) (NGCF,0.2363)};
\addplot coordinates {
(BPRMF,0.2254) (BPRFM,0.2175)
(NeuMF, 0.2554) (NFM,0.2124) (NGCF,0.2195)};
\addplot coordinates {
(BPRMF,0.2445) (BPRFM,0.2360)
(NeuMF, 0.2712) (NFM,0.2403) (NGCF,0.2355)};
\legend{U, HP, LP, U+HP, U+LP}
\end{axis}
\end{tikzpicture}}
\subfigure[Epinions]{
\begin{tikzpicture}[scale=0.4]
\pgfplotsset{%
    width=0.55\textwidth,
    height=0.4\textwidth
}
\begin{axis}[
    ybar,
    bar width=5pt,
    ylabel style ={font = \Large},
    xlabel style ={font = \Large},
    enlarge x limits={abs=0.8cm},
    scaled ticks=false,
    tick label style={/pgf/number format/fixed, font=\Large},
    ymin=0.03, ymax=0.125,
    symbolic x coords={BPRMF, BPRFM, NeuMF, NFM, NGCF},
    xtick=data,
    ytick={0.03,0.06,0.09},
    legend style={at={(0.5,0.98)}, anchor=north,legend columns=5, column sep=0.1cm, draw=none, font=\large},
]
\addplot coordinates {
(BPRMF,0.0820) (BPRFM,0.0979)
(NeuMF, 0.1084) (NFM,0.0756) (NGCF,0.1052)};
\addplot coordinates {
(BPRMF,0.0821) (BPRFM,0.0945)
(NeuMF, 0.0997) (NFM,0.0759) (NGCF,0.1045)};  
\addplot coordinates {
(BPRMF,0.0819) (BPRFM,0.0959)
(NeuMF, 0.0986) (NFM,0.0749) (NGCF,0.1000)};
\addplot coordinates {
(BPRMF,0.0697) (BPRFM,0.0816)
(NeuMF, 0.0973) (NFM,0.0597) (NGCF,0.0694)}; 
\addplot coordinates {
(BPRMF,0.0741) (BPRFM,0.0805)
(NeuMF, 0.0956) (NFM,0.0771) (NGCF,0.0721)};
\legend{U, HP, LP, U+HP, U+LP}
\end{axis}
\end{tikzpicture}}
\subfigure[Yelp]{
\begin{tikzpicture}[scale=0.4]
\pgfplotsset{%
    width=0.55\textwidth,
    height=0.4\textwidth
}
\begin{axis}[
    ybar,
    bar width=5pt,
    ylabel={NDCG@10},
    ylabel style ={font = \Large},
    xlabel style ={font = \Large},
    enlarge x limits={abs=0.8cm},
    scaled ticks=false,
    tick label style={/pgf/number format/fixed, font=\Large},
    ymin=0.1, ymax=0.5,
    symbolic x coords={BPRMF, BPRFM, NeuMF, NFM, NGCF},
    xtick=data,
    ytick={0.1,0.2,0.3,0.4,0.5},
    legend style={at={(0.5,0.98)}, anchor=north,legend columns=5, column sep=0.1cm, draw=none, font=\large},
]
\addplot coordinates {
(BPRMF,0.3698) (BPRFM,0.3701)
(NeuMF, 0.4032) (NFM,0.2028) (NGCF,0.4187)};
\addplot coordinates {
(BPRMF,0.3697) (BPRFM,0.3718)
(NeuMF, 0.2682) (NFM,0.2038) (NGCF,0.3942)};  
\addplot coordinates {
(BPRMF,0.3678) (BPRFM,0.3720)
(NeuMF, 0.2465) (NFM,0.1990) (NGCF,0.3943)};
\addplot coordinates {
(BPRMF,0.3103) (BPRFM,0.3165)
(NeuMF, 0.3413) (NFM,0.1991) (NGCF,0.3854)};
\addplot coordinates {
(BPRMF,0.3113) (BPRFM,0.3104)
(NeuMF, 0.3328) (NFM,0.2040) (NGCF,0.3862)};
\legend{U, HP, LP, U+HP, U+LP}
\end{axis}
\end{tikzpicture}}
\subfigure[AMZe]{
\begin{tikzpicture}[scale=0.4]
\pgfplotsset{%
    width=0.55\textwidth,
    height=0.4\textwidth
}
\begin{axis}[
    ybar,
    bar width=5pt,
    ylabel style ={font = \Large},
    xlabel style ={font = \Large},
    enlarge x limits={abs=0.8cm},
    scaled ticks=false,
    tick label style={/pgf/number format/fixed, font=\Large},
    ymin=0, ymax=0.39,
    symbolic x coords={BPRMF, BPRFM, NeuMF, NFM, NGCF},
    xtick=data,
    ytick={0,0.1,0.2,0.3},
    legend style={at={(0.5,0.98)}, anchor=north,legend columns=5, column sep=0.1cm, draw=none, font=\large},
]
\addplot coordinates {
(BPRMF,0.3117) (BPRFM,0.3264)
(NeuMF, 0.3244) (NFM,0.2858) (NGCF,0.2664)};
\addplot coordinates {
(BPRMF,0.3120) (BPRFM,0.3265)
(NeuMF, 0.3251) (NFM,0.2959) (NGCF,0.2620)};  
\addplot coordinates {
(BPRMF,0.3103) (BPRFM,0.3279)
(NeuMF, 0.3262) (NFM,0.2772) (NGCF,0.2665)};
\addplot coordinates {
(BPRMF,0.3036) (BPRFM,0.3128)
(NeuMF, 0.3163) (NFM,0.2120) (NGCF,0.2565)};
\addplot coordinates {
(BPRMF,0.3145) (BPRFM,0.3269)
(NeuMF, 0.3263) (NFM,0.2337) (NGCF,0.2688)};
\legend{U, HP, LP, U+HP, U+LP}
\end{axis}
\end{tikzpicture}}
\vspace{-0.1in}
\caption{Performance of baselines w.r.t. time-aware split-by-ratio on 10-filter view across the six datasets with different sampling strategies.}\label{fig:negative-sample}
\vspace{-0.15in}
\end{figure}
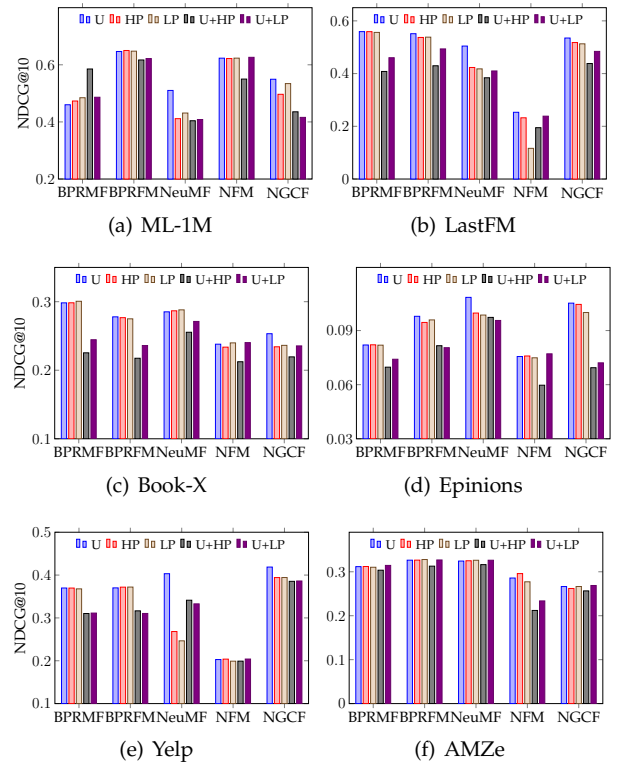

\subsection{Analysis on Model-dependent Hyper-Factors}

\subsubsection{Impacts of Loss Functions}\label{subsubsec:loss-function}
To examine the impacts of different objective functions, we adopt the optimal parameters for the baselines found on 10-filter view with time-aware split-by-ratio in Section~\ref{subsubsec:data_preprocessing}, and only vary objective functions for MF, FM, NeuMF, NFM and NGCF. 
The results are depicted in Figure~\ref{fig:loss-function}, where BPR (pair-wise log loss), CE (point-wise cross entropy loss), Hinge (pair-wise hinge loss) and Top1 (pair-wise top1 loss) correspond to $\mathcal{L}_{pai} + f_{ll}, \mathcal{L}_{poi} + f_{cl}, \mathcal{L}_{pai} + f_{hl}$ and $\mathcal{L}_{pai} + f_{tl}$
in Table~\ref{tab:summary_objective_functions}, respectively. 
Several conclusions can be drawn. As a whole, for different baselines on the six datasets, (1) BPR loss  generally achieves the best performance; (2) CE loss and Hinge loss perform comparably; and (3) Top1 loss possesses the largest performance variation. From the perspective of different baselines, (1) MF and FM usually achieve the best performance with BPR loss; (2) NeuMF performs better with CE loss in most cases; (3) NFM is relatively less sensitive to different losses;  and (4) NGCF generally obtains better accuracy with either BPR or Top1 loss.

\subsubsection{Impacts of Negative Sampling Strategies}
We now explore the impact of different negative samplers, i.e., uniform (U), high-popularity (HP), low-popularity (LP), uniform+high-popularity (U+HP) and uniform+low-popularity (U+LP) on BPRMF, BPRFM, NeuMF, NFM and NGCF across the six datasets under 10-filter view with time-aware split-by-ratio. To this end, we only vary negative samplers for the baselines while keeping other parameters fixed. First, the uniform sampler, though simple, achieves comparable performance in comparison with popularity samplers (HP and LP) as illustrated in Figure~\ref{fig:negative-sample}.
Intuitively, users may not tend to buy the less popular items, that is, the items with low popularity are more likely to be the negative items for users. However, it is overturned by the empirical results. Second, U+HP and U+LP samplers are generally defeated by U/HP/LP samplers. However, there are some exceptions, e.g., BPRMF on ML-1M and NeuMF on Yelp. 
Lastly, U+LP exceeds U+HP in most cases, indicating that generally the popular items have a lower probability to be negative items than the less popular ones. 

\begin{figure}[t]
\centering
\subfigure[ML-1M]{
\begin{tikzpicture}[scale=0.4]
\pgfplotsset{%
    width=0.55\textwidth,
    height=0.4\textwidth
}
\begin{axis}[
    ybar,
    bar width=10pt,
    ylabel={NDCG@10},
    ylabel style ={font = \Large},
    xlabel style ={font = \Large},
    xticklabel style={rotate=20},
    enlarge x limits={abs=0.7cm},
    scaled ticks=false,
    tick label style={/pgf/number format/fixed, font=\Large},
    ymin=0.2, ymax=0.72,
    symbolic x coords={BPRMF, BPRFM, NeuMF, NFM, NGCF, VAE},
    xtick=data,
    ytick={0.2,0.4,0.6,0.8},
    legend style={at={(0.6,0.98)}, anchor=north,legend columns=2, column sep=0.2cm, draw=none, font=\large},
]
\addplot coordinates {
(BPRMF,0.6100) (BPRFM,0.6208)
(NeuMF, 0.4194) (NFM,0.6261) (NGCF,0.5494) (VAE,0.6294)};
\addplot coordinates {
(BPRMF,0.4601) (BPRFM,0.6466)
(NeuMF, 0.4135) (NFM,0.6300) (NGCF,0.5092) (VAE,0.6287)}; \legend{uniform, normal}
\end{axis}
\end{tikzpicture}}
\subfigure[LastFM]{
\begin{tikzpicture}[scale=0.4]
\pgfplotsset{%
    width=0.55\textwidth,
    height=0.4\textwidth
}
\begin{axis}[
    ybar,
    bar width=10pt,
    ylabel style ={font = \Large},
    xlabel style ={font = \Large},
    xticklabel style={rotate=20},
    enlarge x limits={abs=0.7cm},
    scaled ticks=false,
    tick label style={/pgf/number format/fixed, font=\Large},
    ymin=0, ymax=0.6,
    symbolic x coords={BPRMF, BPRFM, NeuMF, NFM, NGCF, VAE},
    xtick=data,
    ytick={0,0.2,0.4,0.6},
    legend style={at={(0.5,0.98)}, anchor=north,legend columns=1, column sep=0.2cm, draw=none, font=\large}
]
\addplot coordinates {
(BPRMF,0.3705) (BPRFM,0.3951)
(NeuMF, 0.4304) (NFM,0.2077) (NGCF,0.5349) (VAE,0.2897)};
\addplot coordinates {
(BPRMF,0.5592) (BPRFM,0.5513)
(NeuMF, 0.4322) (NFM,0.1049) (NGCF,0.5112) (VAE,0.2901)}; \legend{uniform, normal}
\end{axis}
\end{tikzpicture}}
\subfigure[Book-X]{
\begin{tikzpicture}[scale=0.4]
\pgfplotsset{%
    width=0.55\textwidth,
    height=0.4\textwidth
}
\begin{axis}[
    ybar,
    bar width=10pt,
    ylabel={NDCG@10},
    ylabel style ={font = \Large},
    xlabel style ={font = \Large},
    xticklabel style={rotate=20},
    enlarge x limits={abs=0.7cm},
    scaled ticks=false,
    tick label style={/pgf/number format/fixed, font=\Large},
    ymin=0.1, ymax=0.33,
    symbolic x coords={BPRMF, BPRFM, NeuMF, NFM, NGCF, VAE},
    xtick=data,
    ytick={0.1,0.2,0.3},
    legend style={at={(0.6,0.98)}, anchor=north,legend columns=2, column sep=0.2cm, draw=none, font=\large},
]
\addplot coordinates {
(BPRMF,0.2061) (BPRFM,0.2021)
(NeuMF, 0.2840) (NFM,0.2366) (NGCF,0.2533) (VAE,0.2390)};
\addplot coordinates {
(BPRMF,0.2983) (BPRFM,0.2779)
(NeuMF, 0.2868) (NFM,0.2383) (NGCF,0.2434) (VAE,0.2378)}; \legend{uniform, normal}
\end{axis}
\end{tikzpicture}}
\subfigure[Epinions]{
\begin{tikzpicture}[scale=0.4]
\pgfplotsset{%
    width=0.55\textwidth,
    height=0.4\textwidth
}
\begin{axis}[
    ybar,
    bar width=10pt,
    ylabel style ={font = \Large},
    xlabel style ={font = \Large},
    xticklabel style={rotate=20},
    enlarge x limits={abs=0.7cm},
    scaled ticks=false,
    tick label style={/pgf/number format/fixed, font=\Large},
    ymin=0.06, ymax=0.115,
    symbolic x coords={BPRMF, BPRFM, NeuMF, NFM, NGCF, VAE},
    xtick=data,
    ytick={0.06,0.08,0.10},
    legend style={at={(0.6,0.98)}, anchor=north,legend columns=2, column sep=0.2cm, draw=none, font=\large},
]
\addplot coordinates {
(BPRMF,0.0755) (BPRFM,0.0709)
(NeuMF, 0.1020) (NFM,0.0752) (NGCF,0.1052) (VAE,0.0745)};
\addplot coordinates {
(BPRMF,0.0820) (BPRFM,0.0979)
(NeuMF, 0.1027) (NFM,0.0757) (NGCF,0.1018) (VAE,0.0749)};
\legend{uniform, normal}
\end{axis}
\end{tikzpicture}}
\subfigure[Yelp]{
\begin{tikzpicture}[scale=0.4]
\pgfplotsset{%
    width=0.55\textwidth,
    height=0.4\textwidth
}
\begin{axis}[
    ybar,
    bar width=10pt,
    ylabel={NDCG@10},
    ylabel style ={font = \Large},
    xlabel style ={font = \Large},
    xticklabel style={rotate=20},
    enlarge x limits={abs=0.7cm},
    scaled ticks=false,
    tick label style={/pgf/number format/fixed, font=\Large},
    ymin=0.1, ymax=0.45,
    symbolic x coords={BPRMF, BPRFM, NeuMF, NFM, NGCF, VAE},
    xtick=data,
    ytick={0.1,0.2,0.3,0.4},
    legend style={at={(0.4,0.98)}, anchor=north,legend columns=2, column sep=0.2cm, draw=none, font=\large},
]
\addplot coordinates {
(BPRMF,0.2686) (BPRFM,0.2256)
(NeuMF, 0.2770) (NFM,0.2040) (NGCF,0.4187) (VAE,0.2041)};
\addplot coordinates {
(BPRMF,0.3698) (BPRFM,0.3701)
(NeuMF, 0.2656) (NFM,0.2034) (NGCF,0.3948) (VAE,0.2041)}; \legend{uniform, normal}
\end{axis}
\end{tikzpicture}}
\subfigure[AMZe]{
\begin{tikzpicture}[scale=0.4]
\pgfplotsset{%
    width=0.55\textwidth,
    height=0.4\textwidth
}
\begin{axis}[
    ybar,
    bar width=10pt,
    ylabel style ={font = \Large},
    xlabel style ={font = \Large},
    xticklabel style={rotate=20},
    enlarge x limits={abs=0.7cm},
    scaled ticks=false,
    tick label style={/pgf/number format/fixed, font=\Large},
    ymin=0.2, ymax=0.35,
    symbolic x coords={BPRMF, BPRFM, NeuMF, NFM, NGCF, VAE},
    xtick=data,
    ytick={0.2,0.3,0.4},
    legend style={at={(0.6,0.98)}, anchor=north,legend columns=2, column sep=0.2cm, draw=none, font=\large},
]
\addplot coordinates {
(BPRMF,0.2831) (BPRFM,0.3037)
(NeuMF, 0.3241) (NFM,0.2893) (NGCF,0.2664) (VAE,0.3281)};
\addplot coordinates {
(BPRMF,0.3117) (BPRFM,0.3264)
(NeuMF, 0.3224) (NFM,0.2763) (NGCF,0.2657) (VAE,0.3285)};
\legend{uniform, normal}
\end{axis}
\end{tikzpicture}}
\vspace{-0.1in}
\caption{Performance of baselines w.r.t. time-aware split-by-ratio on 10-filter across the six datasets with different initializers.}\label{fig:initializer}
\vspace{-0.15in}
\end{figure}
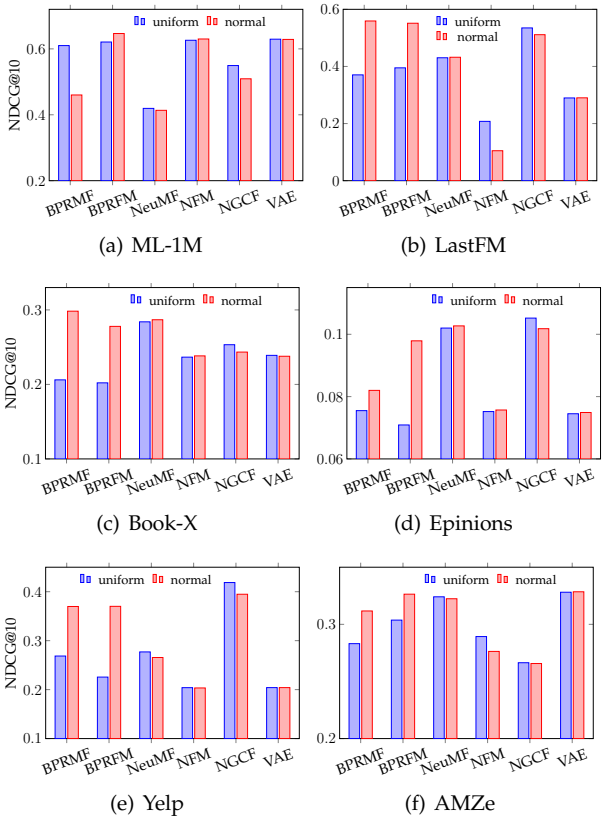
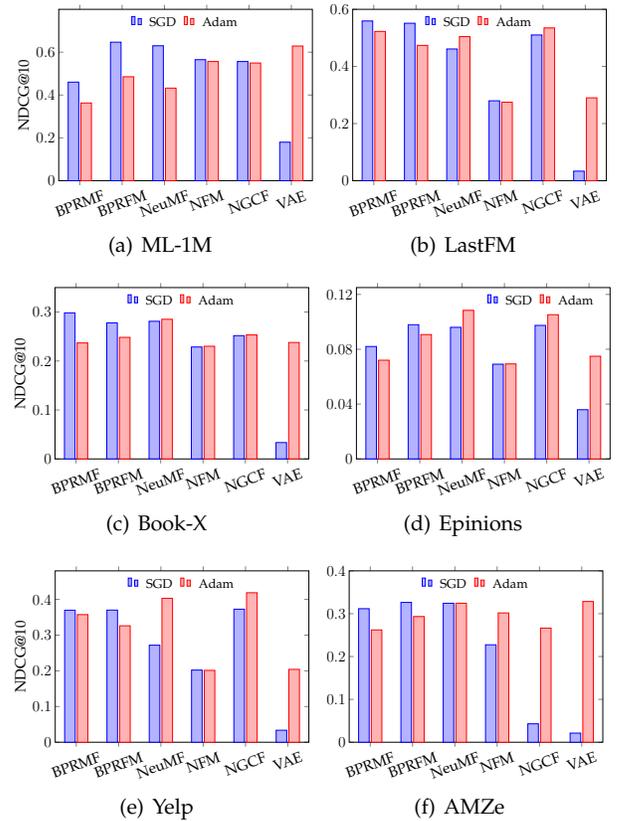
\begin{figure}[t]
\centering
\subfigure[ML-1M]{
\begin{tikzpicture}[scale=0.4]
\pgfplotsset{%
    width=0.55\textwidth,
    height=0.4\textwidth
}
\begin{axis}[
    ybar,
    bar width=10pt,
    ylabel={NDCG@10},
    ylabel style ={font = \Large},
    xlabel style ={font = \Large},
    xticklabel style={rotate=20},
    enlarge x limits={abs=0.7cm},
    scaled ticks=false,
    tick label style={/pgf/number format/fixed, font=\Large},
    ymin=0, ymax=0.8,
    symbolic x coords={BPRMF, BPRFM, NeuMF, NFM, NGCF, VAE},
    xtick=data,
    ytick={0,0.2,0.4,0.6},
    legend style={at={(0.5,0.98)}, anchor=north,legend columns=2, column sep=0.2cm, draw=none, font=\large},
]
\addplot coordinates {
(BPRMF,0.4601) (BPRFM,0.6466) 
(NeuMF, 0.6299) (NFM, 0.5650) (NGCF,0.5564) (VAE,0.1797)};
\addplot coordinates {
(BPRMF,0.3628) (BPRFM,0.4852) 
(NeuMF, 0.4322) (NFM, 0.5565) (NGCF,0.5494) (VAE,0.6287)}; 
\legend{SGD, Adam}
\end{axis}
\end{tikzpicture}}
\subfigure[LastFM]{
\begin{tikzpicture}[scale=0.4]
\pgfplotsset{%
    width=0.55\textwidth,
    height=0.4\textwidth
}
\begin{axis}[
    ybar,
    bar width=10pt,
    ylabel style ={font = \Large},
    xlabel style ={font = \Large},
    xticklabel style={rotate=20},
    enlarge x limits={abs=0.7cm},
    scaled ticks=false,
    tick label style={/pgf/number format/fixed, font=\Large},
    ymin=0, ymax=0.6,
    symbolic x coords={BPRMF, BPRFM, NeuMF, NFM, NGCF, VAE},
    xtick=data,
    ytick={0,0.2,0.4,0.6},
    legend style={at={(0.5,0.98)}, anchor=north,legend columns=2, column sep=0.2cm, draw=none, font=\large},
]
\addplot coordinates {
(BPRMF,0.5592) (BPRFM,0.5513) 
(NeuMF, 0.4612) (NFM, 0.2797) (NGCF,0.5102) (VAE,0.0333)};
\addplot coordinates {
(BPRMF,0.5226) (BPRFM,0.4737) 
(NeuMF, 0.5044) (NFM, 0.2748) (NGCF,0.5349) (VAE,0.2901)}; 
\legend{SGD, Adam}
\end{axis}
\end{tikzpicture}}
\subfigure[Book-X]{
\begin{tikzpicture}[scale=0.4]
\pgfplotsset{%
    width=0.55\textwidth,
    height=0.4\textwidth
}
\begin{axis}[
    ybar,
    bar width=10pt,
    ylabel={NDCG@10},
    ylabel style ={font = \Large},
    xlabel style ={font = \Large},
    xticklabel style={rotate=20},
    enlarge x limits={abs=0.7cm},
    scaled ticks=false,
    tick label style={/pgf/number format/fixed, font=\Large},
    ymin=0, ymax=0.35,
    symbolic x coords={BPRMF, BPRFM, NeuMF, NFM, NGCF, VAE},
    xtick=data,
    ytick={0,0.1,0.2,0.3},
    legend style={at={(0.5,0.98)}, anchor=north,legend columns=2, column sep=0.2cm, draw=none, font=\large},
]
\addplot coordinates {
(BPRMF,0.2983) (BPRFM,0.2779) 
(NeuMF, 0.2811) (NFM, 0.2286) (NGCF,0.2518) (VAE,0.0334)};
\addplot coordinates {
(BPRMF,0.2369) (BPRFM,0.2483) 
(NeuMF, 0.2852) (NFM, 0.2301) (NGCF,0.2533) (VAE,0.2378)}; 
\legend{SGD, Adam}
\end{axis}
\end{tikzpicture}}
\subfigure[Epinions]{
\begin{tikzpicture}[scale=0.4]
\pgfplotsset{%
    width=0.55\textwidth,
    height=0.4\textwidth
}
\begin{axis}[
    ybar,
    bar width=10pt,
    ylabel style ={font = \Large},
    xlabel style ={font = \Large},
    xticklabel style={rotate=20},
    enlarge x limits={abs=0.7cm},
    scaled ticks=false,
    tick label style={/pgf/number format/fixed, font=\Large},
    ymin=0, ymax=0.125,
    symbolic x coords={BPRMF, BPRFM, NeuMF, NFM, NGCF, VAE},
    xtick=data,
    ytick={0,0.04,0.08,0.12},
    legend style={at={(0.73,0.98)}, anchor=north,legend columns=2, column sep=0.2cm, draw=none, font=\large},
]
\addplot coordinates {
(BPRMF,0.0820) (BPRFM,0.0979) 
(NeuMF, 0.0960) (NFM, 0.0691) (NGCF,0.0974) (VAE,0.0359)};
\addplot coordinates {
(BPRMF,0.0720) (BPRFM,0.0907) 
(NeuMF, 0.1084) (NFM, 0.0693) (NGCF,0.1052) (VAE,0.0749)}; 
\legend{SGD, Adam}
\end{axis}
\end{tikzpicture}}
\subfigure[Yelp]{
\begin{tikzpicture}[scale=0.4]
\pgfplotsset{%
    width=0.55\textwidth,
    height=0.4\textwidth
}
\begin{axis}[
    ybar,
    bar width=10pt,
    ylabel={NDCG@10},
    ylabel style ={font = \Large},
    xlabel style ={font = \Large},
    xticklabel style={rotate=20},
    enlarge x limits={abs=0.7cm},
    scaled ticks=false,
    tick label style={/pgf/number format/fixed, font=\Large},
    ymin=0, ymax=0.48,
    symbolic x coords={BPRMF, BPRFM, NeuMF, NFM, NGCF, VAE},
    xtick=data,
    ytick={0,0.1,0.2,0.3,0.4},
    legend style={at={(0.5,0.98)}, anchor=north,legend columns=2, column sep=0.2cm, draw=none, font=\large},
]
\addplot coordinates {
(BPRMF,0.3698) (BPRFM,0.3701) 
(NeuMF, 0.2718) (NFM, 0.2021) (NGCF,0.3727) (VAE,0.0336)};
\addplot coordinates {
(BPRMF,0.3575) (BPRFM,0.3261) 
(NeuMF, 0.4032) (NFM, 0.2013) (NGCF,0.4187) (VAE,0.2041)}; 
\legend{SGD, Adam}
\end{axis}
\end{tikzpicture}}
\subfigure[AMZe]{
\begin{tikzpicture}[scale=0.4]
\pgfplotsset{%
    width=0.55\textwidth,
    height=0.4\textwidth
}
\begin{axis}[
    ybar,
    bar width=10pt,
    ylabel style ={font = \Large},
    xlabel style ={font = \Large},
    xticklabel style={rotate=20},
    enlarge x limits={abs=0.7cm},
    scaled ticks=false,
    tick label style={/pgf/number format/fixed, font=\Large},
    ymin=0, ymax=0.4,
    symbolic x coords={BPRMF, BPRFM, NeuMF, NFM, NGCF, VAE},
    xtick=data,
    ytick={0,0.1,0.2,0.3,0.4},
    legend style={at={(0.5,0.98)}, anchor=north,legend columns=2, column sep=0.2cm, draw=none, font=\large},
]
\addplot coordinates {
(BPRMF,0.3117) (BPRFM,0.3264) 
(NeuMF, 0.3244) (NFM, 0.2274) (NGCF,0.0430) (VAE,0.0213)};
\addplot coordinates {
(BPRMF,0.2621) (BPRFM,0.2934) 
(NeuMF, 0.3244) (NFM, 0.3018) (NGCF,0.2664) (VAE,0.3285)}; 
\legend{SGD, Adam}
\end{axis}
\end{tikzpicture}}
\vspace{-0.1in}
\caption{Performance of baselines w.r.t. time-aware split-by-ratio on 10-filter across the six datasets with different optimizers.}\label{fig:optimizer}
\vspace{-0.15in}
\end{figure}
\begin{figure}[t]
\centering
\subfigure[ML-1M]{
\begin{tikzpicture}[scale=0.4]
\pgfplotsset{%
    width=0.55\textwidth,
    height=0.4\textwidth
}
\begin{axis}[
    ybar,
    bar width=5pt,
    ylabel={NDCG@10},
    ylabel style ={font = \Large},
    xlabel style ={font = \Large},
    xticklabel style={rotate=20},
    enlarge x limits={abs=0.7cm},
    scaled ticks=false,
    tick label style={/pgf/number format/fixed, font=\Large},
    ymin=0, ymax=0.8,
    symbolic x coords={BPRMF, BPRFM, NeuMF, NFM, NGCF, VAE},
    xtick=data,
    ytick={0,0.2,0.4,0.6,0.8},
    legend style={at={(0.5,0.98)}, anchor=north,legend columns=4, column sep=0.2cm, draw=none, font=\large},
]
\addplot coordinates {
(BPRMF,0.4601) (BPRFM,0.6466)
(NeuMF, 0.4322) (NFM,0.6232) (NGCF,0.5494) (VAE,0.6287)};
\addplot coordinates {
(BPRMF,0.4747) (BPRFM,0.6515)
(NeuMF, 0.4339) (NFM,0.6341) (NGCF,0.4777) (VAE,0.1930)};
\addplot coordinates {
(BPRMF,0.4686) (BPRFM,0.6438)
(NeuMF, 0.5349) (NFM,0.6231) (NGCF,0.4923) (VAE,0.6279)};
\addplot coordinates {
(BPRMF,0.0) (BPRFM,0.0)
(NeuMF, 0.4469) (NFM,0.5917) (NGCF,0.5149) (VAE,0.6352)};
\legend{+all, -L2, -ES, -dropout}
\end{axis}
\end{tikzpicture}}
\subfigure[LastFM]{
\begin{tikzpicture}[scale=0.4]
\pgfplotsset{%
    width=0.55\textwidth,
    height=0.4\textwidth
}
\begin{axis}[
    ybar,
    bar width=5pt,
    ylabel style ={font = \Large},
    xlabel style ={font = \Large},
    xticklabel style={rotate=20},
    enlarge x limits={abs=0.7cm},
    scaled ticks=false,
    tick label style={/pgf/number format/fixed, font=\Large},
    ymin=0, ymax=0.65,
    symbolic x coords={BPRMF, BPRFM, NeuMF, NFM, NGCF, VAE},
    xtick=data,
    ytick={0,0.2,0.4,0.6},
    legend style={at={(0.5,0.98)}, anchor=north,legend columns=4, column sep=0.2cm, draw=none, font=\large},
]
\addplot coordinates {
(BPRMF,0.5592) (BPRFM,0.5513)
(NeuMF, 0.5044) (NFM,0.2532) (NGCF,0.5349) (VAE,0.2901)};
\addplot coordinates {
(BPRMF,0.5634) (BPRFM,0.5573)
(NeuMF, 0.4474) (NFM,0.1826) (NGCF,0.5166) (VAE,0.2851)};  
\addplot coordinates {
(BPRMF,0.5642) (BPRFM,0.5396)
(NeuMF, 0.4167) (NFM,0.2211) (NGCF,0.5022) (VAE,0.2886)};
\addplot coordinates {
(BPRMF,0.0) (BPRFM,0.0)
(NeuMF, 0.4337) (NFM,0.0895) (NGCF,0.5163) (VAE,0.2926)};
\legend{+all, -L2, -ES, -dropout}
\end{axis}
\end{tikzpicture}}
\subfigure[Book-X]{
\begin{tikzpicture}[scale=0.4]
\pgfplotsset{%
    width=0.55\textwidth,
    height=0.4\textwidth
}
\begin{axis}[
    ybar,
    bar width=5pt,
    ylabel={NDCG@10},
    ylabel style ={font = \Large},
    xlabel style ={font = \Large},
    xticklabel style={rotate=20},
    enlarge x limits={abs=0.7cm},
    scaled ticks=false,
    tick label style={/pgf/number format/fixed, font=\Large},
    ymin=0, ymax=0.35,
    symbolic x coords={BPRMF, BPRFM, NeuMF, NFM, NGCF, VAE},
    xtick=data,
    ytick={0,0.1,0.2,0.3},
    legend style={at={(0.5,0.98)}, anchor=north,legend columns=4, column sep=0.2cm, draw=none, font=\large},
]
\addplot coordinates {
(BPRMF,0.2983) (BPRFM,0.2779)
(NeuMF, 0.2852) (NFM,0.2380) (NGCF,0.2533) (VAE,0.2378)};
\addplot coordinates {
(BPRMF,0.2980) (BPRFM,0.2790)
(NeuMF, 0.2216) (NFM,0.2387) (NGCF,0.2381) (VAE,0.0340)};  
\addplot coordinates {
(BPRMF,0.3002) (BPRFM,0.2779)
(NeuMF, 0.2894) (NFM,0.2377) (NGCF,0.2406) (VAE,0.2400)};
\addplot coordinates {
(BPRMF,0.0) (BPRFM,0.0)
(NeuMF, 0.2292) (NFM,0.1231) (NGCF,0.2410) (VAE,0.2396)};
\legend{+all, -L2, -ES, -dropout}
\end{axis}
\end{tikzpicture}}
\subfigure[Epinions]{
\begin{tikzpicture}[scale=0.4]
\pgfplotsset{%
    width=0.55\textwidth,
    height=0.4\textwidth
}
\begin{axis}[
    ybar,
    bar width=5pt,
    ylabel style ={font = \Large},
    xlabel style ={font = \Large},
    xticklabel style={rotate=20},
    enlarge x limits={abs=0.7cm},
    scaled ticks=false,
    tick label style={/pgf/number format/fixed, font=\Large},
    ymin=0, ymax=0.13,
    symbolic x coords={BPRMF, BPRFM, NeuMF, NFM, NGCF, VAE},
    xtick=data,
    ytick={0,0.04,0.08,0.12},
    legend style={at={(0.5,0.98)}, anchor=north,legend columns=4, column sep=0.2cm, draw=none, font=\large},
]
\addplot coordinates {
(BPRMF,0.0820) (BPRFM,0.0979)
(NeuMF, 0.1084) (NFM,0.0756) (NGCF,0.1052) (VAE,0.0749)};
\addplot coordinates {
(BPRMF,0.0822) (BPRFM,0.0955)
(NeuMF, 0.0523) (NFM,0.0759) (NGCF,0.0637) (VAE,0.0372)};  
\addplot coordinates {
(BPRMF,0.0831) (BPRFM,0.0973)
(NeuMF, 0.1060) (NFM,0.0749) (NGCF,0.0811) (VAE,0.0740)};
\addplot coordinates {
(BPRMF,0.0) (BPRFM,0.0)
(NeuMF, 0.0840) (NFM,0.0666) (NGCF,0.1032) (VAE,0.0761)};
\legend{+all, -L2, -ES, -dropout}
\end{axis}
\end{tikzpicture}}
\subfigure[Yelp]{
\begin{tikzpicture}[scale=0.4]
\pgfplotsset{%
    width=0.55\textwidth,
    height=0.4\textwidth
}
\begin{axis}[
    ybar,
    bar width=5pt,
    ylabel={NDCG@10},
    ylabel style ={font = \Large},
    xlabel style ={font = \Large},
    xticklabel style={rotate=20},
    enlarge x limits={abs=0.7cm},
    scaled ticks=false,
    tick label style={/pgf/number format/fixed, font=\Large},
    ymin=0, ymax=0.5,
    symbolic x coords={BPRMF, BPRFM, NeuMF, NFM, NGCF, VAE},
    xtick=data,
    ytick={0,0.1,0.2,0.3,0.4,0.5},
    legend style={at={(0.5,0.98)}, anchor=north,legend columns=4, column sep=0.2cm, draw=none, font=\large},
]
\addplot coordinates {
(BPRMF,0.3698) (BPRFM,0.3701)
(NeuMF, 0.4032) (NFM,0.2028) (NGCF,0.4187) (VAE,0.2041)};
\addplot coordinates {
(BPRMF,0.3682) (BPRFM,0.3703)
(NeuMF, 0.3530) (NFM,0.2035) (NGCF,0.3476) (VAE,0.0442)};  
\addplot coordinates {
(BPRMF,0.3675) (BPRFM,0.3715)
(NeuMF, 0.2868) (NFM,0.2031) (NGCF,0.3948) (VAE,0.2043)};
\addplot coordinates {
(BPRMF,0.0) (BPRFM,0.0)
(NeuMF, 0.2733) (NFM,0.1165) (NGCF,0.3923) (VAE,0.2044)};
\legend{+all, -L2, -ES, -dropout}
\end{axis}
\end{tikzpicture}}
\subfigure[AMZe]{
\begin{tikzpicture}[scale=0.4]
\pgfplotsset{%
    width=0.55\textwidth,
    height=0.4\textwidth
}
\begin{axis}[
    ybar,
    bar width=5pt,
    ylabel style ={font = \Large},
    xlabel style ={font = \Large},
    xticklabel style={rotate=20},
    enlarge x limits={abs=0.7cm},
    scaled ticks=false,
    tick label style={/pgf/number format/fixed, font=\Large},
    ymin=0, ymax=0.4,
    symbolic x coords={BPRMF, BPRFM, NeuMF, NFM, NGCF, VAE},
    xtick=data,
    ytick={0,0.1,0.2,0.3,0.4},
    legend style={at={(0.5,0.98)}, anchor=north,legend columns=4, column sep=0.2cm, draw=none, font=\large},
]
\addplot coordinates {
(BPRMF,0.3117) (BPRFM,0.3264)
(NeuMF, 0.3244) (NFM,0.2858) (NGCF,0.2664) (VAE,0.3285)};
\addplot coordinates {
(BPRMF,0.3109) (BPRFM,0.3264)
(NeuMF, 0.1956) (NFM,0.2176) (NGCF,0.2668) (VAE,0.0916)};  
\addplot coordinates {
(BPRMF,0.3125) (BPRFM,0.3285)
(NeuMF, 0.3264) (NFM,0.2123) (NGCF,0.2643) (VAE,0.3273)};
\addplot coordinates {
(BPRMF,0.0) (BPRFM,0.0)
(NeuMF, 0.1918) (NFM,0.2759) (NGCF,0.1494) (VAE,0.3284)};
\legend{+all, -L2, -ES, -dropout}
\end{axis}
\end{tikzpicture}}
\vspace{-0.1in}
\caption{Performance of baselines w.r.t. time-aware split-by-ratio on 10-filter across the six datasets with different strategies to avoid over-fitting.}\label{fig:overfitting}
\vspace{-0.15in}
\end{figure}
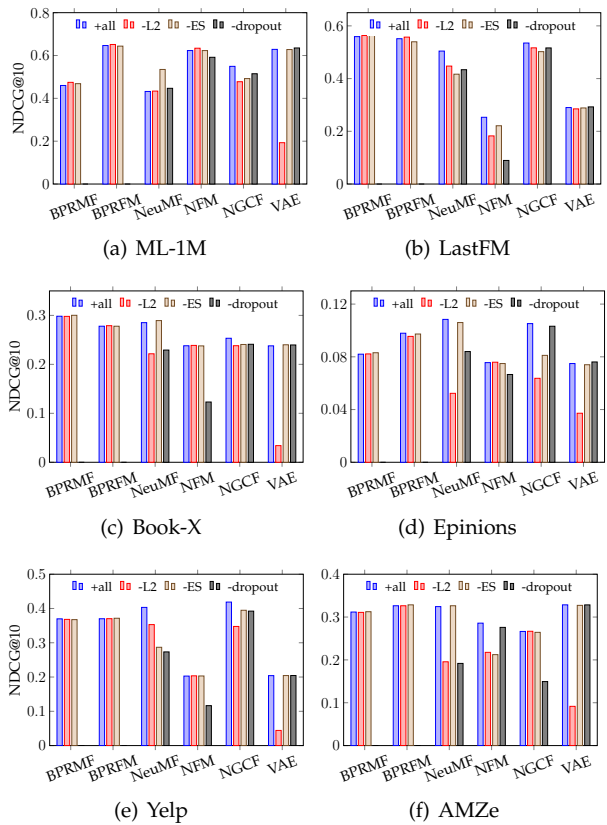

\subsubsection{Impacts of Parameter Initializers}\label{subsec:parameter-initializer}
To study the impact of different parameter initializers, we compare the results of six baselines (BPRMF, BPRFM, NeuMF, NFM, NGCF and Multi-VAE) across the six datasets under 10-filter view with time-aware split-by-ratio. Specifically, for BPRMF and BPRFM, we adopt uniform ($a=1$) and normal distribution ($\sigma=0.01$) for initialization; while for the rest four deep learning baselines, we utilize Xavier uniform and normal distribution for initialization. As depicted in Figure~\ref{fig:initializer}, we can note that (1) for the two LFMs, i.e., BPRMF and BPRFM, initializer with normal distribution dramatically beats that with uniform distribution; and (2) for the four DLMs, some baselines (e.g., NGCF) gains better accuracy with uniform distribution than normal distribution on the six datasets; while some perform comparably with the two types of initializers, e.g., Multi-VAE, across the six datasets. In a nutshell, different parameter initializers produce different recommendation performance. With a proper initializer, the LFMs may easily beat DLMs, for example, BPRMF defeats DLMs on LastFM and Book-X.

\subsubsection{Impacts of Model Optimizers}\label{subsec:model-optimizer}
We further investigate the impacts of different optimiziers on the final recommendation performance. In particular, we vary optimizers (i.e., SGD, and Adam) for the six baselines (i.e., BPRMF, BPRFM, NeuMF, NFM, NGCF and Multi-VAE) on the six datasets under 10-filter view with time-aware split-by-ratio. The results are presented in Figure~\ref{fig:optimizer}, where we observe that a better performance is achieved via SGD in comparison with Adam for LFMs (i.e., BPRMF and BPRFM); whereas Adam generally outperforms SGD regarding DLMs (i.e., NeuMF, NFM, NGCF and Multi-VAE).

\subsubsection{Impacts of Strategies to Avoid Over-fitting}
As illustrated in Section~\ref{subsubsec:over-fitting}, regularization term, dropout and early-stop mechanism are widely adopted to avoid over-fitting. To verify their impacts, we compare the results of six baselines (BPRMF, BPRFM, NeuMF, NFM, NGCF and Multi-VAE) across the six datasets under 10-fiter view with time-aware split-by-ratio by removing these strategies. In particular, +all, -L2, -dropout and -ES respectively indicate the baseline with all over-fitting prevention strategies, variant without L2 regularization term, variant without 
dropout (only for deep learning baseline), and variant without early-stop. The results are displayed in Figure~\ref{fig:overfitting}, where several observations can be noted. First, the overfitting prevention strategies generally facilitate to enhance the recommendation accuracy to some extent for all baselines across the six datasets. However, there are a few exceptions, e.g., NeuMF on ML-1M, indicating some of these strategies may also lead to the underfitting issue occasionally. Second, the impact of these strategies is more significant on DLMs (e.g., NeuMF) than LFMs (e.g., BPRMF). Lastly, the performance of DLMs may be remarkably affected by a certain strategy, e.g., NFM is heavily affected by dropout, whilst a major impact of L2 regularization term on Multi-VAE can be observed.  

\subsubsection{Impacts of Hyper-parameter Tuning Strategies}
\begin{figure}[t]
\centering
\subfigure[ML-1M]{
\begin{tikzpicture}[scale=0.4]
\pgfplotsset{%
    width=0.55\textwidth,
    height=0.4\textwidth
}
\begin{axis}[
    ybar,
    bar width=10pt,
    ylabel={NDCG@10},
    ylabel style ={font = \Large},
    xlabel style ={font = \Large},
    enlarge x limits={abs=0.7cm},
    scaled ticks=false,
    tick label style={/pgf/number format/fixed, font=\Large},
    ymin=0.2, ymax=0.67,
    symbolic x coords={ItemKNN, BPRMF, NeuMF, NGCF},
    xtick=data,
    ytick={0.2,0.4,0.6},
    legend style={at={(0.3,0.98)}, anchor=north,legend columns=1, column sep=0.2cm, draw=none, font=\large},
]
\addplot coordinates {
(ItemKNN,0.4233) (BPRMF,0.4601)
(NeuMF, 0.4322) (NGCF,0.5494)};
\addplot coordinates {
(ItemKNN,0.3657) (BPRMF,0.5085)
(NeuMF, 0.6475) (NGCF,0.6028)};  
\legend{validation, test}
\end{axis}
\end{tikzpicture}}
\hspace{0.05in}
\subfigure[LastFM]{
\begin{tikzpicture}[scale=0.4]
\pgfplotsset{%
    width=0.55\textwidth,
    height=0.4\textwidth
}
\begin{axis}[
    ybar,
    bar width=10pt,
    ylabel style ={font = \Large},
    xlabel style ={font = \Large},
    enlarge x limits={abs=0.7cm},
    scaled ticks=false,
    tick label style={/pgf/number format/fixed, font=\Large},
    ymin=0.4, ymax=0.62,
    symbolic x coords={ItemKNN, BPRMF, NeuMF, NGCF},
    xtick=data,
    ytick={0.4,0.5,0.6},
    legend style={at={(0.5,0.98)}, anchor=north,legend columns=2, column sep=0.2cm, draw=none, font=\large},
]
\addplot coordinates {
(ItemKNN,0.5577) (BPRMF,0.5592)
(NeuMF, 0.5044) (NGCF,0.5349)};
\addplot coordinates {
(ItemKNN,0.5968) (BPRMF,0.5659)
(NeuMF, 0.5381) (NGCF,0.5671)};  
\legend{validation, test}
\end{axis}
\end{tikzpicture}}
\hspace{0.05in}
\subfigure[Book-X]{
\begin{tikzpicture}[scale=0.4]
\pgfplotsset{%
    width=0.55\textwidth,
    height=0.4\textwidth
}
\begin{axis}[
    ybar,
    bar width=10pt,
    ylabel={NDCG@10},
    ylabel style ={font = \Large},
    xlabel style ={font = \Large},
    enlarge x limits={abs=0.7cm},
    scaled ticks=false,
    tick label style={/pgf/number format/fixed, font=\Large},
    ymin=0.1, ymax=0.35,
    symbolic x coords={ItemKNN, BPRMF, NeuMF, NGCF},
    xtick=data,
    ytick={0.1,0.2,0.3},
    legend style={at={(0.5,0.98)}, anchor=north,legend columns=2, column sep=0.2cm, draw=none, font=\large},
]
\addplot coordinates {
(ItemKNN,0.1921) (BPRMF,0.2983)
(NeuMF, 0.2852) (NGCF,0.2533)};
\addplot coordinates {
(ItemKNN,0.3253) (BPRMF,0.3134)
(NeuMF, 0.2840) (NGCF,0.3221)};  
\legend{validation, test}
\end{axis}
\end{tikzpicture}}
\hspace{0.05in}
\subfigure[Epinions]{
\begin{tikzpicture}[scale=0.4]
\pgfplotsset{%
    width=0.55\textwidth,
    height=0.4\textwidth
}
\begin{axis}[
    ybar,
    bar width=10pt,
    ylabel style ={font = \Large},
    xlabel style ={font = \Large},
    enlarge x limits={abs=0.7cm},
    scaled ticks=false,
    tick label style={/pgf/number format/fixed, font=\Large},
    ymin=0.06, ymax=0.112,
    symbolic x coords={ItemKNN, BPRMF, NeuMF, NGCF},
    xtick=data,
    ytick={0.06,0.08,0.10},
    legend style={at={(0.3,0.98)}, anchor=north,legend columns=1, column sep=0.2cm, draw=none, font=\large},
]
\addplot coordinates {
(ItemKNN,0.0813) (BPRMF,0.0820)
(NeuMF, 0.1084) (NGCF,0.1052)};
\addplot coordinates {
(ItemKNN,0.0836) (BPRMF,0.0838)
(NeuMF, 0.1089) (NGCF,0.1098)};  
\legend{validation, test}
\end{axis}
\end{tikzpicture}}
\vspace{-0.1in}
\caption{Performance of baselines w.r.t. time-aware split-by-ratio on 10-filter across the six datasets by tuning on validation and test sets.}\label{fig:parameter-tuning}
\vspace{-0.15in}
\end{figure}
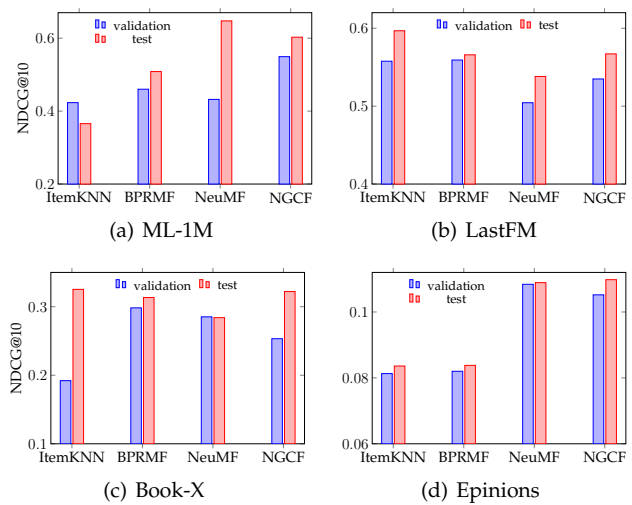

As illustrated in Section~\ref{subsubsec:parameter-tuning}, a validation set should be held out for hyper-parameter tuning to avoid data leakage. To investigate its impact, we compare with the results by directly tuning hyper-parameters on the test set under 10-filter view with time-aware split-by-ratio. For simplicity, we only select
four representative baselines on four datasets as displayed in Figure~\ref{fig:parameter-tuning}. Accordingly, we can easily find that in most cases directly tuning hyper-parameters on the test set indeed guarantees a better performance compared with tuning hyper-parameters on the validation set. This implies that the empirical results disclosed in previous studies without validation set might be overestimated.

\section{Benchmarking Recommendation}

\subsection{Standardized Procedures}
Section~\ref{sec:theoretical_study} shows the hyper-factors in recommendation evaluation, and their impacts are empirically analyzed in Section~\ref{sec:experiement}.
To achieve a rigorous evaluation, 
the \textbf{mixed mode} discussed in Section~\ref{subsec:categorization-evaluation-modes} is encouraged to be adopted.
Accordingly, 
we propose a series of standardized procedures 
and correspondingly call for endeavors of all researchers, aiming to effectively enhance the standardization of recommendation evaluation. Regarding model-independent hyper-factors, five procedures are recommended.
\begin{itemize}[leftmargin=*]
    \item It is impossible to evaluate recommenders on all public datasets covering each domain. However, at least one widely-used dataset discussed in Section~\ref{subsubsec:dataset} should be considered, especially for the papers evaluated on the private datasets (e.g., confidential data from commercial companies). Otherwise, the results could not be easily reproduced by the subsequent studies.
    \item Section~\ref{subsubsec:data_preprocessing} verifies that different data pre-processing strategies impact the performance. Besides origin view, 5- and 10-filter views are recommended to ease the data sparsity issue, and a clear description on data pre-processing details is indispensable.
    \item For data splitting methods, both time-aware split-by-ratio and time-aware leave-one-out are recommended. With timestamp, the real recommendation scenario will be better simulated. W.r.t. split-by-ratio, both global- and user-level work well and $\rho=80\%$ is recommended for a more feasible and convenient comparison.
    \item The representative baselines with different types (MMs, LFMs and DLMs) in Section~\ref{subsubsec:comparison_baselines} are recommended to be selected and compared. As shown in Section~\ref{subsubsec:data_preprocessing}, the performance of different types of baselines vary a lot in different scenarios, that is, the MMs (e.g., MostPop) and simple LFMs (e.g., PureSVD) sometimes even perform better than DLMs (e.g., NeuMF). The more diverse baselines are compared, the more comprehensive and reliable the evaluation is.
    \item At least two of the six discussed metrics in Section~\ref{subsubsec:evluation_metrics} should be adopted, where one (e.g., Precision) measures whether a test item is present on the top-N recommendation list, and the other (e.g., NDCG) measures the ranking positions of the recommended items.
\end{itemize} 

With respect to model-dependent hyper-factors, there are also five procedures recommended as below.
\begin{itemize}[leftmargin=*]
    \item For a fair comparison, it is better to evaluate all methods with a same type of objective functions and thus better positioning a proposed method's contributions.
    \item All the compared methods should adopt the same negative sampler, except the papers with the goal of proposing or studying different negative sampling strategies.
    \item The parameter initializer and model optimizer should be consistent across all compared methods as demonstrated in Section~\ref{subsec:parameter-initializer} and Section~\ref{subsec:model-optimizer}.
    \item The same basic overfitting prevention strategies should be applied to all compared methods, except the methods with specially-designed strategies, e.g., the message dropout in NGCF~\cite{wang2019neural}.
    \item With regards to the hyper-parameter tuning, a nested validation is mandatory, that is, retaining partial (e.g., $10\%$) training data as validation set. Bayesian HyperOpt, as a more intelligent parameter searching strategy, is recommended, and the search space should be kept the same for the shared parameters of different baselines. The number of trails (we set 30 in this study by following~\cite{dacrema2019we}) may be increased for further performance improvements. Most importantly, the optimal parameter settings should be well reported for reproduction.
\end{itemize}
Meanwhile, the source codes and datasets for each publication should be available for reproduction~\cite{raff2019step}. The conference venues could make them as necessities, measure the quality, and even require a short code demonstration along with each accepted paper during the conference.

\subsection{Performance of Baselines}
With the goal of providing a better reference for fair comparison, 
Tables~\ref{tab:ml-1m}-\ref{tab:amze} (Appendix) show the performance of ten baselines across six metrics on the six datasets under three different views (i.e., origin, 5-filter and 10-filter) with time-aware split-by-ratio ($N=10$). 
Due to space limitation, other results (e.g., $N=1, 5, 20, 30, 50$ and other data splitting methods) are on our GitHub.
All optimal hyper-parameters are found by Bayesian HyperOpt to optimize NDCG@10 for 30 trials (see Section~\ref{subsubsec:data_preprocessing}), and the corresponding detailed parameter settings are shown in Tables~\ref{tab:parameter-setting-origin}-\ref{tab:parameter-setting-10-filter}.

Based on the results, several major observations can be noted. \textbf{(1)} BPRFM achieves the best performance on ML-1M across all views. \textbf{(2)} Regarding LastFM, ItemKNN/NGCF performs the best on origin and 5-filter views, while SLIM achieves the best performance on 10-filter view. \textbf{(3)} For Book-X, BPRFM/NGCF and PureSVD/NGCF respectively beat other baselines on origin and 5-filter views, and PureSVD is the winner on 10-filter view. \textbf{(4)} W.r.t. Epinions, NeuMF obtains the highest accuracy on origin view; and NeuMF/NGCF helps reach the best performance on 5- and 10-filter views.
\textbf{(5)} On Yelp, the best performance on the origin, 5- and 10-filter views are respectively gained by BPRFM, NGCF and NGCF. \textbf{(6)} With regards to AMZe, NeuMF/Multi-VAE defeat other baselines on origin view; BPRFM/NFM obtains the optimal results on 5-filter view; and Multi-VAE is the top method on 10-filter view.

\section{Related Work}

While long been recognized as a key feature of scientific discoveries, reproducibility has been increasingly characterized as a crisis recently \cite{open2015estimating,baker2016reproducibility,munafo2017manifesto}. It is becoming a primary concern in computer and information science, evidenced by the recently developed ACM policy on Artifact Review and Badging\footnote{\url{www.acm.org/publications/policies/artifact-review-badging}; see also SIGIR's implementation of the policy \cite{ferro2018sigir}.} and emerging efforts including seminars \cite{freire2016report}, workshops \cite{clancy2019overview}, reproducibility checklist\footnote{\url{aaai.org/Conferences/AAAI-22/reproducibility-checklist/}.} \cite{pineau2021improving},
and focused tracks at major conferences, such as ECIR \cite{hanbury2015advances}, ACM MM \cite{mm2021}, SIGIR\footnote{\url{sigir.org/sigir2022/call-for-reproducibility-track-papers/.}}, and ISWC \cite{hotho2021semantic}. Specific to recommender systems research, besides the reproducibility track starting from 2020 on the premier conference for recommender systems -- RecSys \cite{recsys2020}, the discussions have been concentrated on the fairness of comparison between newly proposed and baseline methods \cite{dacrema2019we,rendle2019difficulty}. In very recent work, Dacrema et al. \cite{dacrema2019we} find 
neural models 
hardly outperform fine-tuned memory- and latent factor-based methods, a similar finding also discovered by Rendle et al. \cite{rendle2019difficulty}.

Despite the importance, improving reproducibility in recommender systems research is highly challenging due to the many influential evaluation factors for recommendation performance. Said et al. \cite{said2014comparative} find large differences in the effectiveness of recommendation methods across different implementation frameworks as well as across evaluation datasets and metrics. A companion toolkit RiVal \cite{said2014rival} was released to allow for the control of data splitting and evaluation metrics, while Elliot \cite{anelli2021elliot} further improves it by implementing more baselines and incorporating statistic significance tests. 
Beel et al. \cite{beel2016towards} find a similar phenomenon in news and research paper recommendation and identify influential factors such as user characteristics and time of recommendation. Valcarce et al. \cite{valcarce2018robustness} specifically study the properties of evaluation metrics for item ranking, marking precision as the most robust and NDCG presenting the highest discriminative power. More recently, Rendle et al. \cite{rendle2019difficulty} demonstrate the importance of hyperparameter search in baseline methods, e.g., matrix factorization, and stress the need for standardized benchmarks where methods should be extensively tuned for fair comparison. Sachdeva et al. \cite{sachdeva2022sampling} specifically examine the impact of dataset sampling strategies on model performance, and indicate that sampling methods, including the random ways do matter with regard to final performance of recommendation algorithms.

Existing benchmarks are, however, either restricted to pre-neural methods \cite{said2014comparative}, a single evaluation factor \cite{valcarce2018robustness}, or rating prediction \cite{rendle2019difficulty} which has been discouraged as a way to formulate the recommendation problem \cite{mcnee2006being}; besides, most of the existing benchmarks consider two or three datasets (including \cite{dacrema2019we}), ignoring the richness of available datasets often chosen by newly published work. 
The two most recent work, RecBole~\cite{zhao2021recbole} and Elliot~\cite{anelli2021elliot}, has partially alleviated the aforementioned issues by implementing more baselines (neural ones included), considering varied datasets and recommendation scenarios (e.g., temporal and context-aware ones), and incorporating hyper-parameter optimization strategies. However, similar to other recommender system libraries (e.g., Librec \cite{guo2015librec}, MyMediaLite \cite{gantner2011mymedialite}, and Surprise \cite{hug2020surprise}), they strive to provide a unified framework for developing and reproducing algorithms for different scenarios in terms of different evaluation metrics.

Instead, aimed for a full treatment of evaluation issues, our work takes a bottom-up approach analyzing an extensive amount of literature to search for and summarize important evaluation factors, denoted as \emph{hyper-factors} (categorized as model-dependent and model-independent ones), which might influence model performance in model evaluation, towards the goal of performing rigorous evaluation. We further present a benchmark supported by an empirical study at a bigger-than-ever scale with the hope of laying a strong foundation for future research.

\section{Conclusion}
This paper aims to benchmark recommendation for reproducible evaluation and fair comparison from the angles of both practical theory analysis and empirical study. Regarding theory analysis, 141 recommendation papers published in the four recent years (2017-2020) from eight top tier conferences have been systematically reviewed, whereby we define and extract the hyper-factors affecting recommendation evaluation, classied into model-independent (e.g., dataset splitting methods) and -dependent (e.g., loss function design) factors. Accordingly, different modes for rigorous evaluation are defined and discussed in-depth. To support the empirical study, a user-friendly Python toolkit -- DaisyRec 2.0 has been released and updated by seamlessly accommodating the extracted hyper-factors. Thereby, the impacts of different hyper-factors on evaluation are then empirically examined and comprehensively analyzed. Lastly, we create benchmarks for rigorous evaluation by proposing standardized procedures and providing the performance of ten well-tuned state-of-the-art algorithms on six widely-used datasets across six metrics as a reference for later study.
For the future work, we plan to deepen our investigation by, for example, diving into more diverse (e.g., session/sequential-aware) recommendation tasks, and more evaluation metrics (e.g., diversity, novelty and serendipity).

\IEEEdisplaynontitleabstractindextext
\bibliographystyle{IEEEtran}
\bibliography{reference}

\clearpage
\newpage
\appendix

\begin{table}[!ht]
\scriptsize
\addtolength{\tabcolsep}{1pt}
\centering
\caption{Representative datasets in various domains for recommendation, where `SNs' and `LBSNs' are short for social networks and the location-based social networks, respectively.
}\label{tab:summary_datasets}
\vspace{-0.15in}
\begin{tabular}{|l|l|}
    \specialrule{.15em}{.05em}{.05em}
    \textbf{Domain}    
    &\multicolumn{1}{c|}{\textbf{Representative Datasets}}\\
    \specialrule{.05em}{.05em}{.05em}
    \specialrule{.05em}{.05em}{.05em}
    \multirow{3}{*}{Movie}
    & MovieLens (100K/1M/10M/20M/25M/Latest), Netflix, \\ 
    & Amazon, FilmTrust, Douban, Yahoo!Movie, 
    Flixster, \\
    & CiaoDVD, EachMovie, IMDB, Watcha~\cite{chae2020ar}\\
    \specialrule{.05em}{.05em}{.05em}
    \multirow{3}{*}{Consumable} & Amazon (Clothing, Home, Office, Sports, Electronic, \\ 
    & Cell Phones, Beauty, Food, Health, Toy, Baby), 
    \\ &Taobao, Beibei, Tmall, Alibaba, Retailrocket, Flipkart~\cite{kang2019candidate} \\
    \specialrule{.05em}{.05em}{.05em} 
    \multirow{2}{*}{Music} & Last.fm, Yahoo!Music, Douban, Amazon-CDs, KKBox, \\
    & Kollect.fm, EchoNest, MSD~\cite{mcfee2012million}, TasteProfile~\cite{tran2018regularizing}
    \\\specialrule{.05em}{.05em}{.05em}
    \multirow{2}{*}{Book} &Amazon-Book/Kindle, Douban, Dianping, IntentBooks, \\
    & Book-Crossing, LibraryThing, GoodReads \\\specialrule{.05em}{.05em}{.05em}
    \multirow{2}{*}{News} &Bing-News, Baidu-News, Medium, Adressa-News~\cite{liu2020hypernews}, \\
    & Microsoft News~\cite{liu2020kred} 
    \\\specialrule{.05em}{.05em}{.05em}
    \multirow{2}{*}{SNs} & Epinions, Delicious,
    Douban, Last.fm, Yelp, Ciao, Xing, \\
    & FilmTrust, Twitter, Wechat, Weibo \\\specialrule{.05em}{.05em}{.05em}
    LBSNs   & Foursquare, Yelp, Gowalla, BrightKite 
    \\\specialrule{.05em}{.05em}{.05em}
    Paper &Aminer, CiteULike, PRSDataset~\cite{guo2020leveraging}
    \\\specialrule{.05em}{.05em}{.05em}
    Image & Pinterest, Flickr, Aesthetic Visual Analysis~\cite{yu2018aesthetic} \\
    \specialrule{.15em}{.05em}{.05em}
    \end{tabular}
    \vspace{-0.15in}
\end{table}
%


\begin{table}[!ht]
    \centering
    \scriptsize
    \addtolength{\tabcolsep}{-4.5pt}
	\caption{The abbreviation and full names of the eight conferences.}
	\label{tab:conference-name}
	\vspace{-0.15in}
    \begin{tabular}{|l|l|}
    \specialrule{.15em}{.05em}{.05em}
    Abbre. &Full Names \\
    \specialrule{.05em}{.05em}{.05em}
    AAAI & AAAI Conference on Artificial Intelligence\\
    CIKM & Conference on Information and Knowledge Management\\
    IJCAI & International Joint Conference on Artificial Intelligence \\
    KDD & SIGKDD Conference on Knowledge Discovery and Data Mining\\
    RecSys & ACM Conference on Recommender System\\
    SIGIR & SIGIR Conference on Research and Development in Information Retrieval\\
    WSDM & International Conference on Web Search and Data Mining\\
    WWW & The ACM Web Conference\\
    \specialrule{.15em}{.05em}{.05em}
    \end{tabular}
    \vspace{-0.1in}
\end{table}


\begin{table}[!ht]
\centering
\scriptsize
\caption{The equations for the six evaluation metrics.}\label{tab:evaluation_metrics}
\addtolength{\tabcolsep}{4pt}
\renewcommand{\arraystretch}{1.8}
\vspace{-0.15in}
\begin{tabular}{l}
\specialrule{.15em}{.05em}{.05em}
     $Precision@N =\frac{1}{\vert \mathcal{U}\vert}\sum_u \frac{1}{N}\sum_{j=1}^N rel_j \;$ \\
     $Recall@N = \frac{1}{\vert \mathcal{U}\vert}\sum_u\frac{1}{\vert T(u)\vert}\sum_{j=1}^N rel_j \;$ \\
     $HR@N  = \frac{1}{\vert \mathcal{U}\vert}\sum_u \delta(R(u)\cap T(u) \neq \emptyset) \;$  \\
     $MAP@N =\frac{1}{\vert \mathcal{U}\vert} \sum_{u}\frac{1}{N}\sum_{k=1}^N Precision@k \;$ \\
     $MRR@N =\frac{1}{\vert \mathcal{U}\vert}\sum_u{\sum_{j=1}^N j^{-1} rel_j}$  \\
     $NDCG@N  =\frac{1}{\vert \mathcal{U}\vert}\sum_u\frac{DCG@N}{IDCG@N}, \; DCG@N=\sum_{j=1}^N\frac{2^{rel_j} - 1}{log_2(j+1)}$ \\
     \specialrule{.15em}{.05em}{.05em}
\end{tabular}
\vspace{-0.15in}
\end{table}

\begin{table}[!ht]
    \centering
    \scriptsize
    \renewcommand{\arraystretch}{1.4}
    \addtolength{\tabcolsep}{-5pt}
    \caption{The update rules of commonly-used optimizers. 
    }\label{tab:optimizer}
    \vspace{-0.15in}
    \begin{tabular}{|l|l|}
    \specialrule{.15em}{.05em}{.05em}
    Optimizer   &  Update Rule \\
    \specialrule{.05em}{.05em}{.05em}
    GD    & $\boldsymbol{\theta}_{t+1} \leftarrow \boldsymbol{\theta}_t - \eta \nabla_{\theta}  \mathcal{L}(\boldsymbol{\theta}_t)$ \\
    SGD &$\boldsymbol{\theta}_{t+1} \leftarrow \boldsymbol{\theta}_t - \eta \nabla_{\theta}  \mathcal{L}(\boldsymbol{\theta}_t; (u, i))$ \\
    MB-SGD &$\boldsymbol{\theta}_{t+1} \leftarrow \boldsymbol{\theta}_t - \eta \nabla_{\theta}  \mathcal{L}(\boldsymbol{\theta}_t; B(u, i))$\\
    AdaGrad &$\boldsymbol{\theta}_{t+1} \leftarrow \boldsymbol{\theta}_t - \frac{\eta}{\sqrt{\boldsymbol{G}_t+\epsilon}}\cdot \boldsymbol{g}_t; \; \boldsymbol{g}_t = \nabla_{\theta} \mathcal{L}(\boldsymbol{\theta}_t)$ \\
    RMSProp &$\boldsymbol{\theta}_{t+1} \leftarrow \boldsymbol{\theta}_t - \frac{\eta}{\sqrt{\mathbb{E}[g^2]_t+\epsilon}} \cdot \boldsymbol{g}_t; \; \mathbb{E}(\boldsymbol{g}^2)_t = \gamma\mathbb{E}[\boldsymbol{g}^2]_{t-1} + (1-\gamma) \boldsymbol{g}_t^2$\\
    \multirow{2}{*}{Adam} &$\boldsymbol{\theta}_{t+1} \leftarrow \boldsymbol{\theta}_t - \frac{\eta}{\sqrt{\boldsymbol{\hat{v}}_t+\epsilon}}\cdot \boldsymbol{\hat{m}}_t; \; \boldsymbol{\hat{m}}_t = \frac{\boldsymbol{m}_t}{1-\beta_1^t}\; \boldsymbol{\hat{v}}_t = \frac{\boldsymbol{v}_t}{1-\beta_2^t}$ \\
    &$\boldsymbol{m}_t = \beta_1 \boldsymbol{m}_{t-1} + (1-\beta_1)\boldsymbol{g}_t; \; \boldsymbol{v}_t = \beta_2 \boldsymbol{v}_{t-1} + (1-\beta_2)\boldsymbol{g}_t^2;$ \\
\end{tabular}
    \vspace{-0.15in}
\end{table}

\begin{table}[!ht]
    \centering
    \scriptsize
    \caption{The performance of ten baselines on ML-1M ($N=10$) across the six metrics, where the best performance is highlighted in bold.}
    \label{tab:ml-1m}
    \vspace{-0.15in}
    \begin{tabular}{|c|c|c|c|c|c|c|}
    \specialrule{.15em}{.05em}{.05em}
    \textbf{Origin} &Pre &Rec &HR &MAP &MRR &NDCG  \\
    \specialrule{.05em}{.05em}{.05em}
    MostPop &0.2554 &0.0530 &0.6997 &0.1734 &0.8290 &0.4952\\
    ItemKNN &0.2218 &0.0676 &0.6521 &0.1428 &0.6804 &0.4261\\
    PureSVD &0.2389 &0.0760 &0.6840 &0.1558 &0.7430 &0.4562\\
    SLIM &-- &-- &-- &-- &-- &--\\
    BPRMF &0.2854 &0.0819 &0.7585 &0.1924 &0.8932 &0.5275\\
    BPRFM &\textbf{0.4379} &\textbf{0.1156} &\textbf{0.8667} &\textbf{0.3360} &\textbf{1.3718} &\textbf{0.6745}\\
    NeuMF &0.4210 &0.1023 &0.8515 &0.3215 &1.3105 &0.6528\\
\end{tabular}
    \vspace{-0.1in}
\end{table}

\begin{table}[!ht]
    \centering
    \scriptsize
    \caption{The performance of ten baselines on LastFM ($N=10$) across the six metrics, where the best performance is highlighted in bold.}
    \label{tab:yelp}
    \vspace{-0.15in}
    \begin{tabular}{|c|c|c|c|c|c|c|}
    \specialrule{.15em}{.05em}{.05em}
    \textbf{Origin} &Pre &Rec &HR &MAP &MRR &NDCG  \\
    \specialrule{.05em}{.05em}{.05em}
    MostPop &0.0712 &0.0716 &0.3978 &0.0338 &0.2509 &0.2311\\
    ItemKNN &\textbf{0.4219} &\textbf{0.4332} &0.9580 &\textbf{0.3324} &\textbf{1.5841} &\textbf{0.8026}\\
    PureSVD &0.4173 &0.4304 &0.9480 &0.3321 &1.5807 &0.7989\\
    SLIM    &-- &-- &-- &-- &-- &--\\
    BPRMF &0.3796 &0.3906 &0.9411 &0.2867 &1.4303 &0.7624\\
    BPRFM &0.3367 &0.3465 &0.8970 &0.2415 &1.2322 &0.6897\\
    NeuMF &0.3561 &0.3663 &0.9320 &0.2552 &1.3112 &0.7279\\
\end{tabular}
    \vspace{-0.1in}
\end{table}

\begin{table}[!ht]
    \centering
    \scriptsize
    \caption{The performance of ten baselines on Book-X ($N=10$) across the six metrics, where the best performance is highlighted in bold.}
    \label{tab:lastfm}
    \vspace{-0.15in}
    \begin{tabular}{|c|c|c|c|c|c|c|}
    \specialrule{.15em}{.05em}{.05em}
    \textbf{Origin} &Pre &Rec &HR &MAP &MRR &NDCG  \\
    \specialrule{.05em}{.05em}{.05em}
    MostPop &0.0048 &0.0150 &0.0411 &0.0023 &0.0214 &0.0249\\
    ItemKNN &0.0214 &0.0283 &0.1027 &0.0159 &0.1010 &0.0782\\
    PureSVD &0.0776 &0.1761 &0.3247 &0.0560 &0.3178 &0.2344\\
    SLIM  &-- &-- &-- &-- &-- &--\\
    BPRMF &0.0868 &0.2309 &0.3924 &0.0586 &0.3450 &0.2687\\
    BPRFM &0.0954 &\textbf{0.2970} &\textbf{0.4603} &0.0636 &\textbf{0.3836} &\textbf{0.3115}\\
    NeuMF &0.0693 &0.2067 &0.3578 &0.0421 &0.2655 &0.2289\\
\end{tabular}
    \vspace{-0.1in}
\end{table}

\begin{table}[!ht]
    \centering
    \scriptsize
    \caption{The performance of ten baselines on Epinions ($N=10$) across the six metrics, where the best performance is highlighted in bold.}
    \label{tab:epinions}
    \vspace{-0.15in}
    \begin{tabular}{|c|c|c|c|c|c|c|}
    \specialrule{.15em}{.05em}{.05em}
    \textbf{Origin} &Pre &Rec &HR &MAP &MRR &NDCG  \\
    \specialrule{.05em}{.05em}{.05em}
    MostPop &0.0016 &0.0009 &0.0157 &0.0005 &0.0052 &0.0076\\
    ItemKNN &0.0480 &0.0193 &0.1599 &0.0365 &0.1753 &0.1176\\
    PureSVD &0.0681 &0.0413 &0.2462 &0.0490 &0.2424 &0.1675\\
    SLIM &-- &-- &-- &-- &-- &--\\
    BPRMF &0.0721 &0.0467 &0.2861 &0.0483 &0.2523 &0.1844\\
    BPRFM &0.0774 &0.0448 &0.3018 &0.0494 &0.2576 &0.1884\\
    NeuMF &\textbf{0.0882} &\textbf{0.0573} &\textbf{0.3519} &\textbf{0.0588} &\textbf{0.3176} &\textbf{0.2329}\\
\end{tabular}
    \vspace{-0.1in}
\end{table}

\begin{table}[!ht]
    \centering
    \scriptsize
    \caption{The performance of ten baselines on Yelp ($N=10$) across the six metrics, where the best performance is highlighted in bold.}
    \label{tab:bookx}
    \vspace{-0.15in}
    \begin{tabular}{|c|c|c|c|c|c|c|}
    \specialrule{.15em}{.05em}{.05em}
    \textbf{Origin} &Pre &Rec &HR &MAP &MRR &NDCG  \\
    \specialrule{.05em}{.05em}{.05em}
    MostPop &0.0012 &0.0058 &0.0113 &0.0005 &0.0048 &0.0061\\
    ItemKNN &0.0128 &0.0302 &0.0699 &0.0094 &0.0628 &0.0524\\
    PureSVD &0.0365 &0.1168 &0.1938 &0.0218 &0.1417 &0.1222\\
    SLIM  &-- &-- &-- &-- &-- &--\\
    BPRMF &0.0425 &0.1573 &0.2447 &0.0234 &0.1594 &0.1471\\
    BPRFM &\textbf{0.0508} &\textbf{0.2144} &\textbf{0.3141} &0.0271 &\textbf{0.1925} &\textbf{0.1859}\\
    NeuMF &0.0500 &0.2092 &0.3048 &\textbf{0.0272} &0.1916 &0.1827\\
\end{tabular}
    \vspace{-0.1in}
\end{table}

\begin{table}[!ht]
    \centering
    \scriptsize
    \caption{The performance of ten baselines on AMZe ($N=10$) across the six metrics, where the best performance is highlighted in bold.}
    \label{tab:amze}
    \vspace{-0.15in}
    \begin{tabular}{|c|c|c|c|c|c|c|}
    \specialrule{.15em}{.05em}{.05em}
    \textbf{Origin} &Pre &Rec &HR &MAP &MRR &NDCG  \\
    \specialrule{.05em}{.05em}{.05em}
    MostPop &0.0025 &0.0169 &0.0245 &0.0006 &0.0063 &0.0104\\
    ItemKNN &0.0018 &0.0114 &0.0176 &0.0008 &0.0074 &0.0096\\
    PureSVD &0.0202 &0.1095 &0.1474 &0.0117 &0.0971 &0.0969\\
    SLIM  &-- &-- &-- &-- &-- &--\\
    BPRMF &0.0323 &0.2060 &0.2642 &0.0170 &0.1510 &0.1655\\
    BPRFM &0.0518 &0.3530 &0.4249 &0.0286 &0.2549 &0.2744\\
    NeuMF &\textbf{0.0518} &\textbf{0.3531} &\textbf{0.4250} &\textbf{0.0286} &0.2550 &\textbf{0.2745} \\
\end{tabular}
    \vspace{-0.1in}
\end{table}

\begin{table*}[htbp]
    \centering
    \scriptsize
    \renewcommand{\arraystretch}{1.0}
    \addtolength{\tabcolsep}{-1pt}
    \caption{The opitmal hyper-parameter settings found by Bayesian HyperOpt for different baselines 
    with time-aware split-by-ratio under origin view.}\label{tab:parameter-setting-origin}
    \vspace{-0.15in}
    \begin{tabular}{|l|l|l|l|l|l|l|l|l|l|}
    \specialrule{.15em}{.05em}{.05em}
    \textit{\textbf{\textcolor{blue}{Origin}}}
    &\multicolumn{1}{c|}{\textbf{Parameter}}
    &\multicolumn{1}{c|}{\textbf{ML-1M}} &\multicolumn{1}{c|}{\textbf{LastFM}} &\multicolumn{1}{c|}{\textbf{Book-X}} &\multicolumn{1}{c|}{\textbf{Epinions}} &\multicolumn{1}{c|}{\textbf{Yelp}} &\multicolumn{1}{c|}{\textbf{AMZe}} 
     &\multicolumn{1}{c|}{\textbf{Searching space}}
     &\multicolumn{1}{c|}{\textbf{Description}}
     \\
    \specialrule{.05em}{.05em}{.05em}
    \textbf{ItemKNN}
    &--maxk
    &8 &49 &83 &14 &31 &9 
    &$[1, 100]$
    &the number of neighbors \\
    \specialrule{.05em}{.05em}{.05em}
    \textbf{PureSVD}
    &--factors
    &39 &24 &10 &99 &96 &6
    &$[1, 100]$
    &the number of singular values\\
    \specialrule{.05em}{.05em}{.05em}
    \multirow{6}{*}{\textbf{BPRMF}} 
    &--num\_ng
    &4 &5 &5 &5
    &5 &4
    &$[1, 5]$
    &the number of negative items\\
    &--factors
    &83 &92 &48 &97
    &72 &74
    &$[1, 100]$
    &the dimension of latent factors\\
    &--epochs
    &50 &50 &50 &50
    &50 &50
    &--
    &the number of epochs\\
    &--lr
    &0.0056 &0.0080 &0.0098 &0.0095
    &0.0089 &0.0047
    &$[10^{-4}, 10^{-2}]$
    &learning rate\\
    &--reg\_2
    &0.0002 &0.0012 &0.0008 &0.0002 &0.0088 &0.0016
    &$[10^{-4}, 10^{-2}]$
    &L2 regularization coefficient\\
\end{tabular}
\vspace{-0.1in}
\end{table*}

\begin{table*}[htbp]
    \centering
    \scriptsize
    \renewcommand{\arraystretch}{1.0}
    \addtolength{\tabcolsep}{-1pt}
    \caption{The opitmal hyper-parameter settings found by Bayesian HyperOpt for different baselines 
    with time-aware split-by-ratio under 5-filter view.}\label{tab:parameter-setting-5-filter}
    \vspace{-0.15in}
    \begin{tabular}{|l|l|l|l|l|l|l|l|l|l|}
    \specialrule{.15em}{.05em}{.05em}
    \textit{\textbf{\textcolor{blue}{5-filter}}}
    &\multicolumn{1}{c|}{\textbf{Parameter}}
    &\multicolumn{1}{c|}{\textbf{ML-1M}} &\multicolumn{1}{c|}{\textbf{LastFM}} &\multicolumn{1}{c|}{\textbf{Book-X}} &\multicolumn{1}{c|}{\textbf{Epinions}} &\multicolumn{1}{c|}{\textbf{Yelp}} &\multicolumn{1}{c|}{\textbf{AMZe}} 
     &\multicolumn{1}{c|}{\textbf{Searching space}}
     &\multicolumn{1}{c|}{\textbf{Description}}
     \\
    \specialrule{.05em}{.05em}{.05em}
    \textbf{ItemKNN} &--maxk
    &73 &55 &65 &30 &75 &41
    &$[1, 100]$
    &the number of neighbors \\
    \specialrule{.05em}{.05em}{.05em}
    \textbf{PureSVD} 
    &--factors
    &2 &33 &80 &70 &100 &2 
    &$[1, 100]$
    &the number of singular values\\
    \specialrule{.05em}{.05em}{.05em}
    \multirow{6}{*}{\textbf{BPRMF}} 
    &--num\_ng 
    &1 &5 &4 &5 &5 &4
    &$[1, 5]$
    &the number of negative items\\
    &--factors
    &100 &56 &99 &70 &92 &74
    &$[1, 100]$
    &the dimension of latent factors\\
    &--epochs
    &50 &50 &50 &50
    &50 &50 &--
    &the number of epochs\\
    &--lr
    &0.0018 &0.0060 &0.0093 &0.0098
    &0.0051 &0.0047
    &$[10^{-4}, 10^{-2}]$
    &learning rate\\
    &--reg\_2
    &0.0007 &0.0003 &0.0014 &0.0029 &0.0001 &0.0016
    &$[10^{-4}, 10^{-2}]$
    &L2 regularization coefficient\\
\end{tabular}
    \vspace{-0.1in}
\end{table*}

\begin{table*}[htbp]
    \centering
    \scriptsize
    \renewcommand{\arraystretch}{1.0}
    \addtolength{\tabcolsep}{-1pt}
    \caption{The opitmal hyper-parameter settings found by Bayesian HyperOpt for different baselines 
    with time-aware split-by-ratio under 10-filter view.}\label{tab:parameter-setting-10-filter}
    \vspace{-0.15in}
    \begin{tabular}{|l|l|l|l|l|l|l|l|l|l|}
    \specialrule{.15em}{.05em}{.05em}
    \textit{\textbf{\textcolor{blue}{10-filter}}}
    &\multicolumn{1}{c|}{\textbf{Parameter}}
    &\multicolumn{1}{c|}{\textbf{ML-1M}} &\multicolumn{1}{c|}{\textbf{LastFM}} &\multicolumn{1}{c|}{\textbf{Book-X}} &\multicolumn{1}{c|}{\textbf{Epinions}} &\multicolumn{1}{c|}{\textbf{Yelp}} &\multicolumn{1}{c|}{\textbf{AMZe}} 
     &\multicolumn{1}{c|}{\textbf{Searching space}}
     &\multicolumn{1}{c|}{\textbf{Description}}
     \\
    \specialrule{.05em}{.05em}{.05em}
    \textbf{ItemKNN}
    &--maxk
    &100 &28 &71 &14 &73 &68 
    &$[1, 100]$
    &the number of neighbors \\
    \specialrule{.05em}{.05em}{.05em}
    \textbf{PureSVD} 
    &--factors
    &39 &12 &98 &98 &93 &9 
    &$[1, 100]$
    &the number of singular values\\
    \specialrule{.05em}{.05em}{.05em}
    \multirow{5}{*}{\textbf{BPRMF}} 
    &--num\_ng
    &3 &5 &3 &5 &5 &3
    &$[1, 5]$
    &the number of negative items\\
    &--factors
    &22 &92 &88 &49 &98 &97
    &$[1, 100]$
    &the dimension of latent factors\\
    &--epochs
    &50 &50 &50 &50
    &50 &50 &--
    &the number of epochs\\
    &--lr
    &0.0004 &0.0057 &0.0098 &0.0097 &0.0099 &0.0078
    &$[10^{-4}, 10^{-2}]$
    &learning rate\\
    &--reg\_2
    &0.0002 &0.0088 &0.0001 &0.0050 &0.0007 &0.0004
    &$[10^{-4}, 10^{-2}]$
    &L2 regularization coefficient\\
\end{tabular}
    \vspace{-0.1in}
\end{table*}

\begin{table*}[htbp]
    \centering
    \scriptsize
    \renewcommand{\arraystretch}{1.0}
    \addtolength{\tabcolsep}{-1pt}
    \caption{The opitmal hyper-parameter settings found by Bayesian HyperOpt for different baselines 
    with time-aware leave-one-out under 10-filter view.}\label{tab:parameter-setting-10-filter}
    \vspace{-0.15in}
    \begin{tabular}{|l|l|l|l|l|l|l|l|l|l|}
    \specialrule{.15em}{.05em}{.05em}
    \textit{\textbf{\textcolor{blue}{10-filter}}}
    &\multicolumn{1}{c|}{\textbf{Parameter}}
    &\multicolumn{1}{c|}{\textbf{ML-1M}} &\multicolumn{1}{c|}{\textbf{LastFM}} &\multicolumn{1}{c|}{\textbf{Book-X}} &\multicolumn{1}{c|}{\textbf{Epinions}} &\multicolumn{1}{c|}{\textbf{Yelp}} &\multicolumn{1}{c|}{\textbf{AMZe}} 
     &\multicolumn{1}{c|}{\textbf{Searching space}}
     &\multicolumn{1}{c|}{\textbf{Description}}
     \\
    \specialrule{.05em}{.05em}{.05em}
    \textbf{ItemKNN}
    &--maxk
    &22 &70 &100 &51 &52 &40 
    &$[1, 100]$
    &the number of neighbors \\
    \specialrule{.05em}{.05em}{.05em}
    \textbf{PureSVD} 
    &--factors
    &45 &26 &89 &96 &97 &11
    &$[1, 100]$
    &the number of singular values\\
    \specialrule{.05em}{.05em}{.05em}
    \multirow{5}{*}{\textbf{BPRMF}} 
    &--num\_ng
    &5 &4 &5 &5 &5 &5
    &$[1, 5]$
    &the number of negative items\\
    &--factors
    &49 &66 &76 &56 &61 &100
    &$[1, 100]$
    &the dimension of latent factors\\
    &--epochs
    &50 &50 &50 &50
    &50 &50 &--
    &the number of epochs\\
    &--lr
    &0.0040 &0.0082 &0.0098 &0.0098 &0.0047 &0.0051
    &$[10^{-4}, 10^{-2}]$
    &learning rate\\
    &--reg\_2
    &0.0014 &0.0007 &0.0013 &0.0016 &0.0012 &0.0006
    &$[10^{-4}, 10^{-2}]$
    &L2 regularization coefficient\\
\end{tabular}
    \vspace{-0.1in}
\end{table*}






\end{document}